\documentclass[iop]{emulateapj}

\usepackage{natbib}
\usepackage{graphicx}
\usepackage{amsmath}
\usepackage{hyperref}
\usepackage{afterpage}

\newcommand{\vh}{v_{\rm helio}}
\newcommand{\vsys}{v_{\rm sys}}
\newcommand{\LCDM}{$\Lambda$CDM}
\newcommand{\mvir}{M_{\rm vir}}
\newcommand{\vmax}{V_{\rm max}}

\newcommand{\Reff}{R_{\rm eff}}
\newcommand{\slos}{\sigma_{\rm LOS}}
\newcommand{\kps}{km s$^{-1}$}
\newcommand{\ndwarfs}{15}
\newcommand{\nmasks}{33}

\newcommand{ \comppapp}{(R.L. Beaton et al., in preparation)}
\newcommand{ \comppapalt}{R.L. Beaton et al., in preparation}

\newcommand{\satsrad}{139}
\newcommand{\andmassnum}{8.0^{+4.1}_{-3.7} \times 10^{11} M_{\odot}}
\newcommand{\andmass}{M_{\rm M31}(<\satsrad\;{\rm kpc}) = \andmassnum}
\newcommand{\andmvirnum}{1.2^{+0.9}_{-0.7} \times 10^{12} M_\odot}
\newcommand{\andmvir}{\mvir = \andmvirnum}

\renewcommand{\S}{Section }
\newcommand{\Ss}{Sections }

\begin{document}

\title{The SPLASH Survey: Spectroscopy of \ndwarfs{} M31 Dwarf Spheroidal Satellite Galaxies\footnotemark[*]} 
\shorttitle{SPLASH dSphs}

\keywords{ Local Group --- galaxies: dwarf --- galaxies: individual (And I, And III, And V, And VII, And IX, And X, And XI, And XII, And XIII,  And XIV, And XV, And XVI,  And XVIII, And XXI, And XXII) --- galaxies: kinematics and dynamics ---  galaxies: fundamental parameters ---  dark matter}

\author{Erik J. Tollerud\altaffilmark{1}, Rachael L. Beaton\altaffilmark{2,11}, Marla C. Geha\altaffilmark{3}, James S. Bullock\altaffilmark{1}, Puragra Guhathakurta\altaffilmark{4},  Jason S. Kalirai\altaffilmark{5,6}, Steve R. Majewski\altaffilmark{2,11}, Evan N. Kirby\altaffilmark{7,12}, Karoline M. Gilbert\altaffilmark{8,12}, Basilio Yniguez\altaffilmark{1}, Richard J. Patterson\altaffilmark{2,11}, James C. Ostheimer\altaffilmark{2,11}, Jeff Cooke\altaffilmark{9}, Claire E. Dorman\altaffilmark{4}, Abrar Choudhury\altaffilmark{10}, and Michael C. Cooper\altaffilmark{1,12}}
\altaffiltext{1}{Center for Cosmology, Department of Physics and Astronomy, The University of California, Irvine, Irvine, CA, 92697, USA; etolleru@uci.edu, bullock@uci.edu, byniguez@uci.edu; m.cooper@uci.edu}
\altaffiltext{2}{Department of Astronomy, University of Virginia, P.O. Box 3818, Charlottesville, VA 22903, USA; rlb9n@virginia.edu, srm4n@mail.astro.virginia.edu, ricky@virginia.edu, james.ostheimer@gmail.com}
\altaffiltext{3}{Astronomy Department, Yale University, New Haven, CT 06510, USA; marla.geha@yale.edu}
\altaffiltext{4}{University of California Observatories/Lick Observatory, University of California at Santa Cruz, Santa Cruz, CA 95064, USA; raja@ucolick.org, cdorman@ucolick.org}
\altaffiltext{5}{Space Telescope Science Institute, 3700 San Martin Drive, Baltimore, MD 21218, USA; jkalirai@stsci.edu}
\altaffiltext{6}{Center for Astrophysical Sciences, Johns Hopkins University, Baltimore, MD, 21218}
\altaffiltext{7}{Department of Astronomy, California Institute of Technology, 1200 East California Boulevard, MC 249-17, Pasadena, CA 91125, USA; enk@astro.caltech.edu}
\altaffiltext{8}{Department of Astronomy, University of Washington, Box 351580, Seattle, WA 98195, USA; kmgilber@u.washington.edu}
\altaffiltext{9}{Centre for Astrophysics and Supercomputing, Swinburne University of Technology, Mail H39, PO Box 218, Hawthorn, VIC 3122, Australia; jcooke@astro.swin.edu.au}
\altaffiltext{10}{Bellarmine College Preparatory, 960 West Hedding Street, San Jose, CA 95126, USA}
\altaffiltext{11}{Visiting Astronomer, Kitt Peak National Observatory, National Optical Astronomy Observatory, which is operated by the Association of Universities for Research in Astronomy under cooperative agreement with the National Science Foundation.}
\altaffiltext{12}{Hubble Fellow}

\begin{abstract}

We present a resolved-star spectroscopic survey of  \ndwarfs{} dwarf spheroidal (dSph) satellites of the Andromeda Galaxy (M31). We filter foreground contamination from Milky Way (MW) stars, noting that MW substructure is evident in this contaminant sample. We also filter M31 halo field giant stars, and identify the remainder as probable dSph members. We then use these members to determine the kinematical properties of the dSphs. For the first time, we confirm that And XVIII, XXI, and XXII show kinematics consistent with bound, dark matter-dominated galaxies. From the velocity dispersions for the full sample of dSphs we determine masses, which we combine with the size and luminosity of the galaxies to produce mass-size-luminosity scaling relations. With these scalings we determine that the M31 dSphs are fully consistent with the MW dSphs, suggesting that the well-studied MW satellite population provides a fair sample for broader conclusions. We also estimate dark matter halo masses of the satellites, and find that there is no sign that the luminosity of these galaxies depends on  their dark halo mass, a result consistent with what is seen for MW dwarfs.  Two of the M31 dSphs (And XV, XVI) have estimated maximum circular velocities smaller than $12$ km/s (to 1$\sigma$), which likely places them within the lowest mass dark matter halos known to host stars (along with Bo\"{o}tes I of the MW).  Finally, we use the systemic velocities of the M31 satellites to estimate the mass of the M31 halo, obtaining a virial mass consistent with previous results.

\end{abstract}

\section{Introduction}
\label{sec:intro}

Dwarf spheroidal (dSph) galaxies are among the most extreme objects in the pantheon of galaxies.  Their low luminosities ($10^3 < L/L_\odot <  10^8$), 
lack of significant gas \citep{grcevich09}, and low numbers compared to \LCDM{} expectations  \citep{kl99ms,moo99ms} are all puzzles that remain to be 
solved.  The difficulty in understanding the count of dSph galaxies around the Milky Way (MW) and M31 is known as the missing satellites problem, an issue that 
has prompted a flurry of activity modeling these galaxies \citep[recently,][and references therein]{krav10satrev,bullock10msp,kaz11stirr,font11}.  Most 
models rely heavily on feedback scenarios that are tied directly to the masses of the dark matter halos that presumably host dSph galaxies. In this sense, 
mass determinations for dwarfs are among the most important diagnostic measurements for testing theoretical predictions at the frontier of galaxy formation.

{ \let\thefootnote\relax\footnotetext{* The data presented herein were obtained at the W.M. Keck Observatory, which is operated as a scientific partnership among the California Institute of Technology, the University of California and the National Aeronautics and Space Administration. The Observatory was made possible by the generous financial support of the W.M. Keck Foundation} }
\addtocounter{footnote}{-1}

While the brightest dSphs can be detected at the distance of nearby clusters \citep[e.g.,][]{hilker03,durrell07},  faint, diffuse  dSphs can only be detected via resolved star counts, which limits detection to the Local Group (LG).  Kinematics of these galaxies thus requires resolved star spectroscopy at extragalactic distances.   
Thus, despite the motivations outlined above, detailed study of a large population of dSphs has been limited to the MW satellites \citep[e.g.,][]{mateo98,sandg,walker09,simon11seg1}.   

Studies of the MW dSph population have resulted in puzzles that have provided interesting challenges to \LCDM{} and galaxy formation models. The  missing satellite problem noted above is the classic example, for which a variety of solutions have surfaced \citep[e.g,][]
{bullock00,stri07msp,tollerud08,bovill09,kop09,krav10satrev}.   However, there remain other questions such as the cause of their low gas fractions \citep
{grcevich09,nichols11}, their morphologies \citep{kaz11stirr}, or the curiously small number with the high densities expected by \LCDM{} \citep{BKBK11}.  
These studies are based entirely on the MW dSph population, as this has been the only available data set.   Yet there is evidence that the MW has had an 
unusual merger history relative to similarly bright galaxies such as M31 \citep{guhathakurta06,hammer07}.  
Furthermore, there are hints that the dSph populations of M31 and the MW exhibit different scaling relations \citep{mcc05andbig,kalirai10}.  
Hence, expanding the sample of satellite systems is crucial to generalizing the information dSphs provide about  galaxy formation.

Fortunately, the past few years have seen much growth in the known satellite population of M31.  This is primarily due to the advent of deep surveys of the region surrounding M31 specifically designed to search for substructure like dSphs or their remnants \citep[e.g.,][]{ibata07,pandas09nat}.   While the distance to M31 means that detection limits do not reach those of the MW's ultra-faint dSph system,  smaller angular coverage is needed to survey M31's environs.
  Thus, the full population of known M31 satellites (shown in Figure \ref{fig:overview}) now includes 27 dSphs \citep{zucker04,martin06,zucker07,maj07and14,irwin08,mc08,martin09,richardson11,Slater11And28,Bell11And29}.  These are further supplemented by 4 dwarf Elliptical (dE) satellites, which are similar in morphology to the dSph, but have somewhat higher luminosities.

  The imaging surveys that detected these dSphs provide data sets that allow characterization of the photometric properties of M31 satellites, but do not provide the kinematics of these dSphs necessary for characterizing their masses and dark matter content.  These kinematical data sets are also crucial for confirming the candidates' status as self-bound galaxies.  This is illustrated by two  M31 dSphs (And IV and VIII) that were originally identified as dSph 
  satellites, but were later shown by spectroscopic follow-up to be non-satellites \citep{ferg00no4,ibata04and8,merrett06and8}.  Beyond this, understanding the kinematics of M31 satellite system as a whole also provides insight into M31, its dark matter halo, and its accretion history \citep[e.g.,][]{evans00,vdm08}.  
  While kinematics exist for some of the M31 dSphs 
  \citep{cote99and2,guhathakurta00,chapman05and9,Chapman07,chapman08,Letarte09,collins10,kalirai10}, the large number of recent discoveries leave 
  many yet to be spectroscopically confirmed, and a homogeneously observed and reduced sample of a significant fraction of these satellites is necessary to
  properly determine characteristics of the satellites system \emph{as a whole}.

With these ends in mind, we report here on kinematics of \ndwarfs{} dSphs from the Spectroscopic and Photometric Landscape of Andromeda's Stellar Halo (SPLASH) Survey.  This ongoing survey of the environs of M31 aims to characterize the stellar halo of M31 and its satellite population via resolved star spectroscopy.  A companion paper on structural parameters and photometric properties is forthcoming \comppapp.

This paper is organized as follows. In \S \ref{sec:obs}, we describe the observations performed for this data set. In \S \ref{sec:analysis}, we describe the reduction and membership analysis performed homogeneously across the spectroscopic data set, as well as our method for  estimating total velocity dispersions for each satellite in the sample.  In \S \ref{sec:data}, we present the results of our full spectroscopic sample, and consider each satellite in turn, describing the results of our analysis and unique aspects of each galaxy. In \S \ref{sec:scaling}, we consider the scaling relations of M31 dSphs and compare them to the MW.  In \S \ref{sec:M31mass}, we use the galaxies' $\vsys$ to estimate the mass of M31.  Finally, we present our conclusions in \S \ref{sec:conc}.

\section{Observations}
\label{sec:obs}

\begin{figure*}[htbp!]
  \epsscale{1.15}
 \plotone{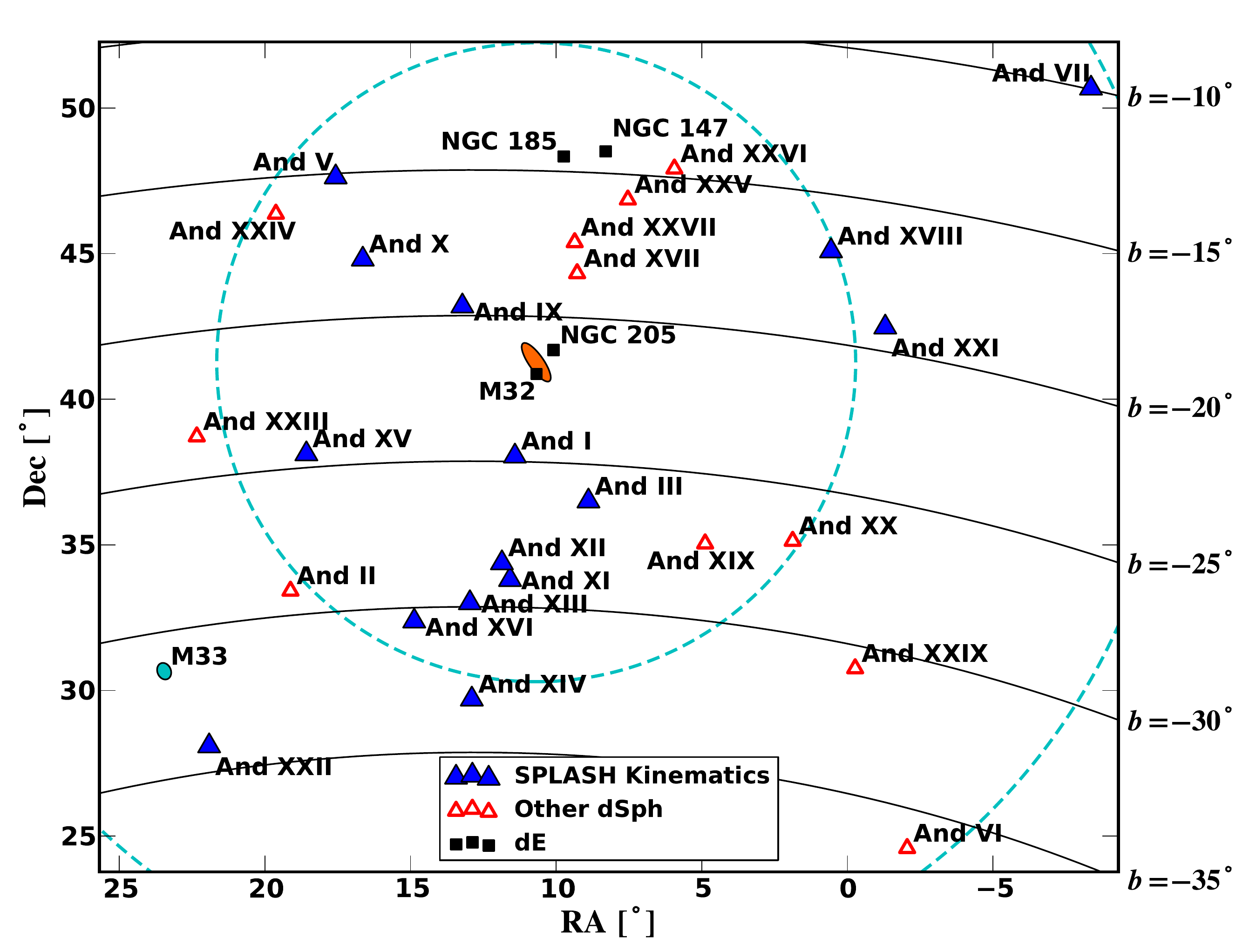}
 \caption{M31 and its satellites.  M31's disk is indicated by the (orange) ellipse in the center, and M33's disk is the (cyan) ellipse at the lower left.  The dSph sample presented here are filled (blue) triangles, and the outlined (red) triangles are the other dSphs. Squares (black) are M31's dE population.  The two (cyan) dashed circles centered on M31, at projected radii of 150 and 300 kpc, represent the approximate extent of the PAndAS survey \citep{pandas09nat} and the virial radius of M31's expected dark matter halo \citep{klypin02}, respectively. The (black) solid lines are lines of constant Galactic latitude. Note that And IV and VIII are absent from this figure because follow-up after discovery revealed that they were probably not dwarf galaxies, and And XXVIII is missing because it is much further to the west of M31 than any other satellites.}
 \label{fig:overview}
 \end{figure*}

We provide an overview of the M31 satellite system in Figure \ref{fig:overview}.  M31 is represented as the (orange) ellipse near the center, while the other symbols are likely M31 satellites.   The \ndwarfs{} satellites presented here with Keck/DEIMOS spectroscopy of resolved stars  are shown as filled (blue) triangles, while the remainder of the dSphs are shown as outlined (red) triangles (excepting the newly discovered And XXVIII, which is much further out than the other dSphs).  Also shown are M33 (cyan ellipse at the lower left), and the dE satellites (black squares). From this census it is apparent that our sample includes over half of the dSphs in the M31 system, with the spatial unevenness only due to the very recent discovery of some of the dSphs.  Also shown are lines of Galactic latitude, indicating significant variation in the MW foreground across M31's environs.

We present \nmasks{} DEIMOS slitmasks covering \ndwarfs{} of M31's dSphs,
with mask details provided in Table \ref{tab:masks}.  Eight of these masks have been presented elsewhere: the And I and III masks were previously presented in  \citet{kalirai10}, the And X data set was originally published by \citet{kalirai09andx}, and 2 masks of And XIV were described by \citet{maj07and14}.  There is an additional series of SPLASH slitmasks for And II that will be described in a forthcoming paper (Ho et al., in preperation). All masks were reduced and analyzed homogeneously using the procedure describe in the following sections, including those noted above that have been previously published.

 \begin{deluxetable*}{ccccccc}[h!]
 \tablecolumns{7}
 \tablecaption{Observed DEIMOS Slitmasks. }
 \tablehead{
   \colhead{Target} &
   \colhead{Mask Name} &
   \colhead{MJD} &
   \colhead{UTC Date} &
   \colhead{Total Exposure Time (s)} &
   \colhead{No. of Slits} &
   \colhead{No. of Successful Velocities\tablenotemark{a}}
 }
  \startdata

And I & d1\_1 & 53679.31 & 11/5/2005 & 4055 & 150 & 71 \\
And I & d1\_2 & 53994.62 & 9/16/2006 & 3600 & 145 & 88 \\
And III & d3\_1 & 53621.34 & 9/8/2005 & 3600 & 119 & 88 \\
And III & d3\_2 & 53621.39 & 9/8/2005 & 3600 & 117 & 83 \\
And III & d3\_3 & 55066.56 & 8/23/2009 & 3600 & 123 & 30 \\
And V & d5\_1 & 54739.45 & 9/30/2008 & 3000 & 101 & 52 \\
And V & d5\_2 & 54739.55 & 9/30/2008 & 2250 & 105 & 64 \\
And V & d5\_3 & 54740.54 & 10/1/2008 & 2250 & 75 & 37 \\
And VII & d7\_1 & 54682.43 & 8/4/2008 & 3000 & 152 & 110 \\
And VII & d7\_2 & 54682.46 & 8/4/2008 & 1800 & 147 & 74 \\
And IX & d9\_1 & 55069.56 & 8/26/2009 & 2700 & 115 & 74 \\
And IX & d9\_2 & 55069.60 & 8/26/2009 & 2400 & 101 & 58 \\
And X & d10\_1 & 53618.39 & 9/5/2005 & 3600 & 82 & 51 \\
And X & d10\_2 & 53618.44 & 9/5/2005 & 3600 & 93 & 54 \\
And XI & d11 & 55450.60 & 9/11/2010 & 3600 & 71 & 22 \\
And XII & d12 & 55451.00 & 9/12/2010 & 3600 & 62 & 22 \\
And XIII & d13\_1 & 54739.37 & 9/30/2008 & 6000 & 66 & 18 \\
And XIII & d13\_2 & 54740.48 & 10/1/2008 & 4800 & 65 & 15 \\
And XIII & d13\_3 & 55450.31 & 9/11/2010 & 3600 & 107 & 18 \\
And XIII & d13\_4 & 55450.36 & 9/11/2010 & 5400 & 114 & 12 \\
And XIII & d13\_5 & 55450.44 & 9/11/2010 & 5400 & 115 & 11 \\
And XIV & A170\_1 & 54059.20 & 11/20/2006 & 3600 & 93 & 39 \\
And XIV & A170\_2 & 54060.20 & 11/21/2006 & 3600 & 88 & 36 \\
And XIV & d14\_3 & 55065.55 & 8/22/2009 & 3600 & 99 & 33 \\
And XV & d15\_1 & 55068.36 & 8/25/2009 & 3600 & 105 & 82 \\
And XV & d15\_2 & 55068.43 & 8/25/2009 & 4800 & 120 & 60 \\
And XVI & d16\_1 & 55068.50 & 8/25/2009 & 4800 & 92 & 29 \\
And XVI & d16\_2 & 55068.56 & 8/25/2009 & 3600 & 77 & 16 \\
And XVIII & d18\_1 & 55069.40 & 8/26/2009 & 10800 & 76 & 49 \\
And XXI & d21\_1 & 55710.56 & 5/29/2011 & 6905 & 96 & 12 \\
And XXII & d22\_1 & 55450.55 & 9/11/2010 & 3600 & 140 & 10 \\
And XXII & d22\_2 & 55451.56 & 9/12/2010 & 4800 & 76 & 7 \\
And XXII & d22\_3 & 55476.59 & 10/7/2010 & 3955 & 125 & 6 

 \enddata
 
 \tablenotetext{a}{While for some masks there is an apparently low percentage of successful velocities, this is due to higher fractions of filler targets 
 in the sparser fields that tend to have lower odds of being RGB stars.  This is discussed in detail \S \ref{sec:specreduct}.}
 
 \label{tab:masks}
 \end{deluxetable*}

The details of the imaging used for target selection will be described in detail in the companion paper \comppapp.  Here we summarize the particulars relevant for spectroscopic target selection.  Our imaging is primarily in the Washington system (specifically, the $M$ and $T_2$ filters), which we convert for some target selection purposes to Johnson-Cousins $V$, $I$ using the relations of \citet{maj00}.    We obtained additional imaging of the same fields with the DDO51 intermediate-band filter \citep{maj00}.  This filter is centered near the surface-gravity dependent Mgb and MgH stellar absorption features, allowing it to discriminate between M31 giant stars and MW foreground dwarf stars.  Thus, selecting targets in e.g., the $M-DDO51$ versus $M-T_2$ color--color diagram allows for far more efficient selection of spectroscopic targets that have high probabilities of being giants \citep{kalirai06halo,guhathakurta06,gilbert06}.

\begin{figure}[htbp!]
  \epsscale{1}
 \plotone{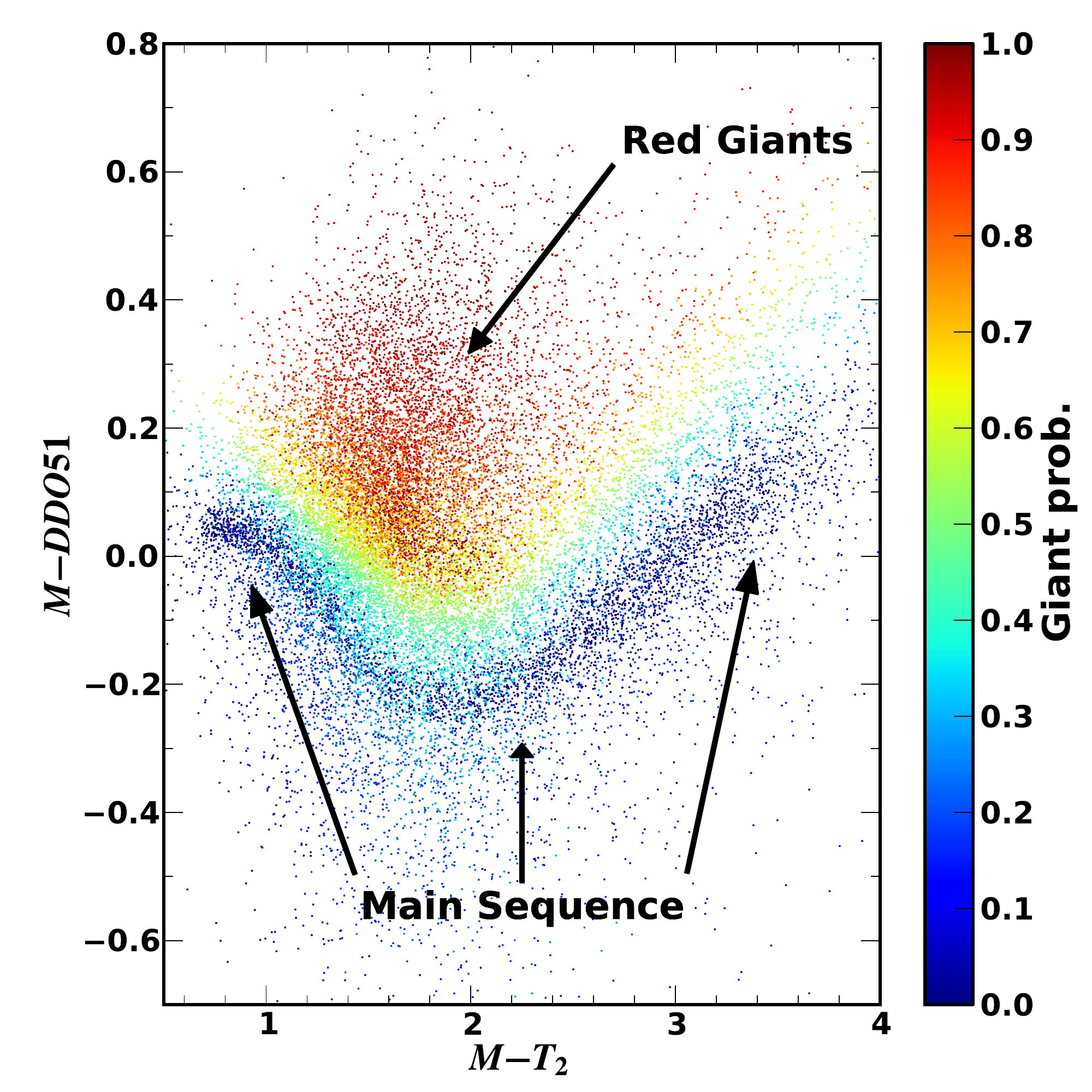}
 \caption{$M-T_2$, $M-DDO51$ color--color diagram used for pre-selection of likely giant stars, for a representative field near And I. Points are colored by red giant probability determined by the method of \citet{ost03}, which assigns high probabilities to stars near the indicated locus for stars with low surface gravity, while avoiding probable dwarf stars in the ``swoosh'' feature . Stars with high probability are selected for spectroscopic follow-up. }
 \label{fig:mdmtgprob}
 \end{figure}

We selected stars for spectroscopic follow-up from catalogs generated from this imaging.  Our candidates for each mask were selected by assigning priorities based on how far a given star was from a fiducial isochrone in the $M-T_2$, $T_2$ color--magnitude diagram (CMD).  This is similar to the membership selection described in \S \ref{sec:membership} and illustrated in the plots of \S \ref{sec:data}, but with looser restrictions.   We also assigned priorities based on how close a star was to the giant star locus in the $M-DDO51$ versus $M-T_2$  color--color diagram \citep{ost03}.  We demonstrate this part of the selection in Figure \ref{fig:mdmtgprob}, showing stars for a representative field and indicating the loci of red giant  stars (the targets) and main sequence stars (the contaminants), with points colored according to an estimate of the probability of the star being a giant.  Because most M31 dSphs are smaller than the footprint of a DEIMOS slitmask ($\approx 16' \times 5'$), in most cases we aligned the mask such that the dSph was centered to one side of the slitmask.  This  allows for measurement at larger radii of M31 halo structure within which a satellite might be embedded \citep[e.g., And I as described by ][]{kalirai10}.

Note that the above description is valid for all masks except those targeted at And X, And XV, And XVI, And XVIII, and And XXII.  For And X, Sloan Digital Sky Survey (SDSS) imaging was used for target selection \citep{sdssdr4}, for And XV and XVI selection was made using Canada France Hawaii Telescope (CFHT) archival imaging,  while or the last two dSph we made use of $B$ and $V$ band imaging from the Large Binocular Telescope \comppapp.   For these, distance from a fiducial isochrone in a CMD was used without DDO51 pre-selection.

\begin{figure}[htbp!]
  \epsscale{1}
 \plotone{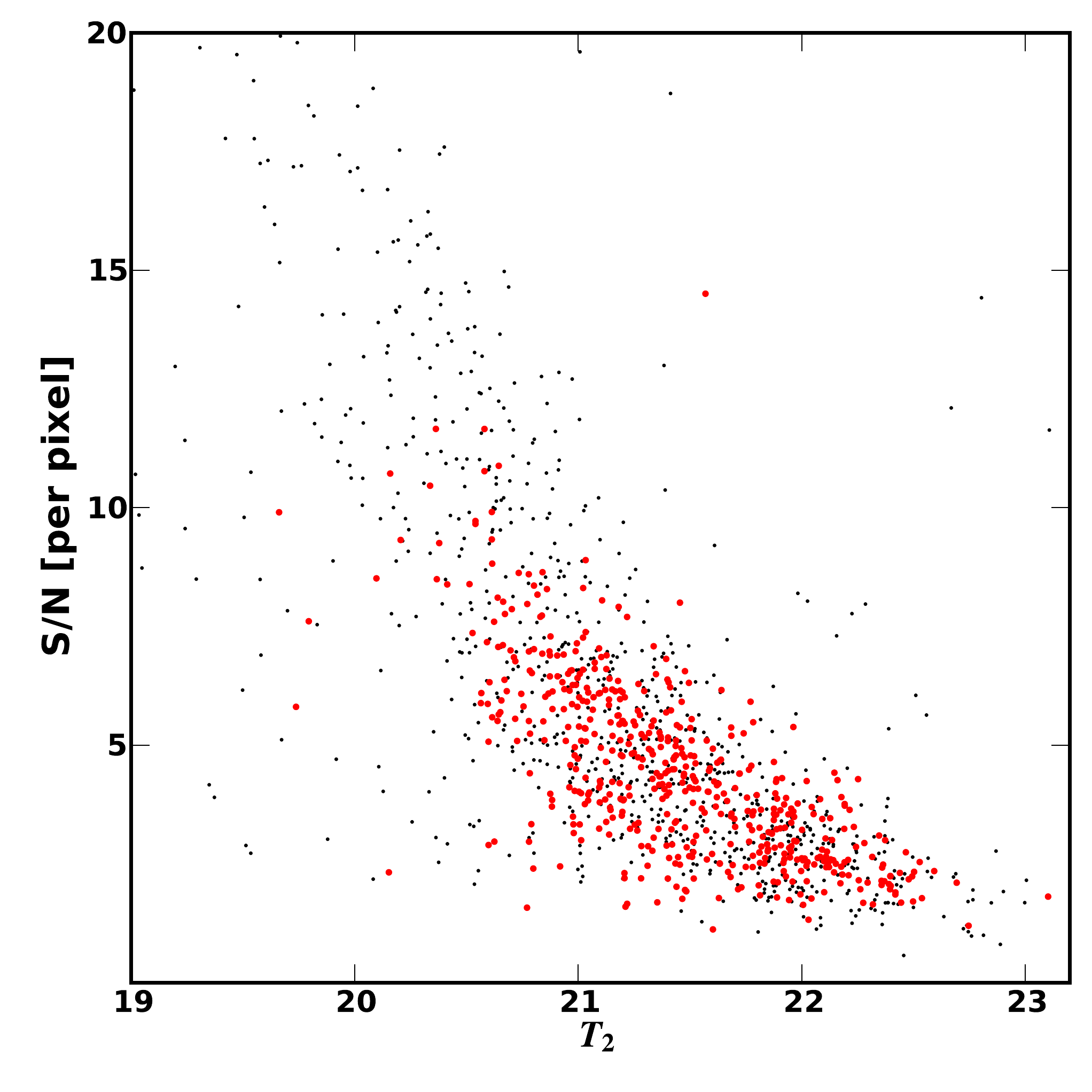}
 \caption{Per pixel signal-to-noise versus $T_2$-band magnitude for stars in this survey with successful spectra. Larger (red) circles are stars identified
 as likely dSph members following the prescription described in \S \ref{sec:membership}, while smaller (black) points are non-member stars.}
 \label{fig:snrvsmag}
 \end{figure}

The spectroscopic setup for the DEIMOS observations  used the 1200 lines mm$^{-1}$ grating with a central wavelength of 7800 \AA.  This provides spectral coverage over a range of 6400 - 9100 \AA{} (for objects centered in the mask along the dispersion direction), with a FWHM resolution of $\approx 1.3$ \AA.  This provides coverage of the Calcium triplet (CaT) stellar absorption feature near 8500 \AA{} and (depending on the slit location) H$\alpha$ to facilitate identification and accurate radial velocity measurements .  A typical total integration time of 3600 s per mask provides a mean signal-to-noise ratio per pixel of $ \sim 7$ for our entire sample.   Signal-to-noise  depends strongly on the target star's magnitude, so we plot in Figure \ref{fig:snrvsmag} the variation of signal-to-noise varies with magnitude.

\section{Spectroscopic Data Analysis}
\label{sec:analysis}

Here we describe the procedure applied homogeneously to each slitmask (Table \ref{tab:masks}) to extract radial velocities from the spectroscopic data, and to determine each satellite's kinematics.  

\subsection{Keck/DEIMOS Data Reductions}
\label{sec:specreduct}

We make use of the spec2d DEIMOS reduction pipeline\footnote{http://deep.berkeley.edu/spec2d/} and modified versions of the spec1d and zspec analysis codes developed for the DEEP2 survey \citep{davis03deep2,spec2dascl,newman12deep2}.  To summarize, the pipeline first rectifies each slit spectrum along the spatial direction and corrects for the slit function and fringing using quartz lamp flats. It then determines a two-dimensional wavelength solution from NeArKrXe arc lamp exposures taken for each mask. Science exposures are then combined with inverse variance weighting including cosmic ray rejection.  Finally, one-dimensional (1D) spectra are extracted using \citet{horne86} optimal extraction.

 \begin{figure*}[htbp!]
  \epsscale{1}
 \plotone{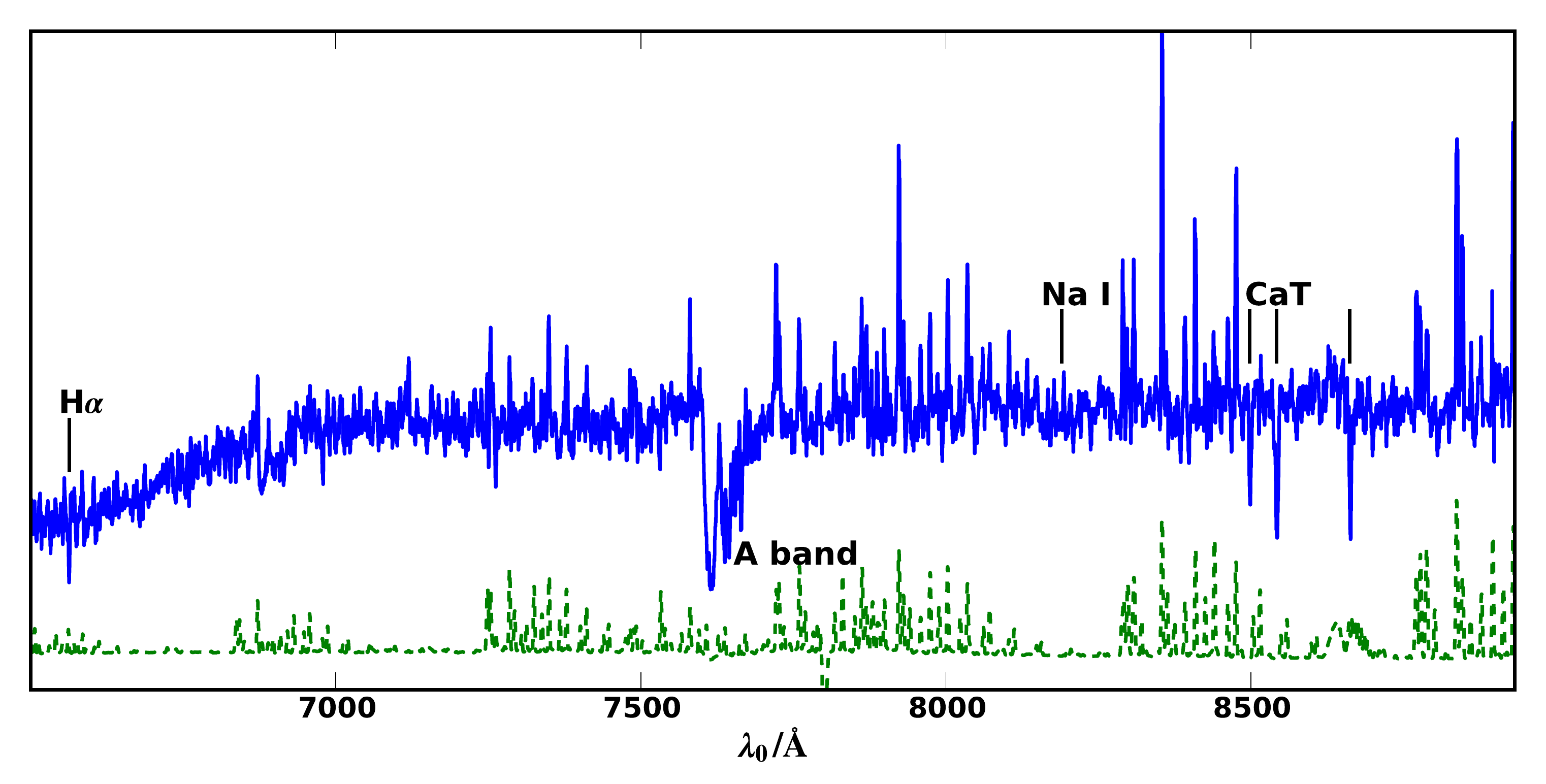}
 \caption{Example spectrum of a star with a high probability of dSph membership.  Solid (blue) line is the spectrum (smoothed with a 4 pixel boxcar filter), while dashed (green) shows the error spectrum, peaking at strong sky emission lines.  For this star, I$=20.7$, and un-smoothed S/N$=6$ per pixel. }
 \label{fig:specex}
 \end{figure*}

The above procedure thus yields 1D spectra for each slit, an example of which is given in Figure \ref{fig:specex}.  These spectra are then further analyzed to determine heliocentric radial velocities, following the same general procedure described by \citet{sandg} (See also \citealt{guhathakurta06} and \citealt{kalirai10}).  The spectra are first re-binned into equally-spaced logarithmic bins of 15 \kps.  These spectra are then cross-correlated with high signal-to-noise templates taken with DEIMOS using an identical spectroscopic setup.  These templates include dwarf, subgiant, giant, and asymptotic giant branch stars.  The largest subset is the giant templates, which range from spectral type F8 to M8, although most of our likely members best match early-K and late-G templates.  

We then manually examine each spectrum and remove contaminant spectra that are clearly galaxies, as well as those for which no absorption features are visible, i.e., low signal-to-noise spectra.   This process results in very different success rates from mask to mask, because masks with a low density of good 
M31 red giant branch (RGB) candidates will include a large number of ``filler'' targets.  These targets tend to be fainter by necessity, meaning the photometric errors are larger, 
and the fraction of background galaxies that overlap with the M31 RGB is much higher, increasing the contamination rate.  Furthermore, even true RGB targets are fainter, and thus identification of absorption lines becomes more difficult, further lowering the success rate.  Together, these effects are the main
reasons for the high variability in mask-to-mask success rates (i.e. the ratio of the last two columns in Table \ref{tab:masks}).

We also correct for mis-centering of stars in slits by matching to the A-band telluric feature as described by \citet{sandg} and \citet{sohn07}.  This correction is crucial, as mis-centered of stars in slits can result in velocity scatter comparable to that of the internal velocity dispersion of the dSphs for typical astrometric errors \citep[see the discussion of ][Section 2.3]{kalirai10}.  Finally, we correct the observed radial velocities to the heliocentric frame.

Understanding the errors on the radial velocity computed above is crucial to determining the internal velocity dispersion of the dSphs, and hence we 
determine the errors on each star by Monte Carlo  simulation.  For each spectrum that passes the inspection procedure described above, we generate 1000 
realizations of the 1D  spectrum, assuming independent, Poisson errors on each pixel based on the observed variance. 
 For each of these realizations we compute the telluric correction and the cross-correlation  radial velocity, and visually examine the histogram of radial 
 velocities for these simulations.  We reject any spectra that show multi-modal 
character in radial velocity (e.g., a secondary cross-correlation peak that is of comparable magnitude to the primary peak).  We then fit a Gaussian to the 
primary peak and use the best-fit parameters as the radial velocity $v_i$ and random error $\sigma_r$ for each spectrum.  Repeat measurements of stars for 
this spectroscopic setup are described in detail by \citet{sandg} and  \citet[][Section 2.4]{kalirai10}, and our sample includes the same repeat measurements 
as the latter.  Both sets of repeat measurements showed an additional systematic error component  of 2.2 \kps.   Thus, we adopt $\sigma_t \equiv \sqrt{\sigma_r^2 + (2.2 \, {\rm km} \, {\rm s}^{-1})^2}$ as our final radial velocity error estimate for each spectrum.

We also measure equivalent widths (EW) for the Na I $\lambda8190$ doublet $(\lambda \lambda 8183,8195)$, a surface-gravity dependent feature useful for dwarf/giant discrimination.  We adopt a blue pseudo-continuum from $8155 - 8175$ \AA, red $8203 - 8230$ \AA, and a line bandpass of $8178 - 8200.5$ \AA \citep{schiavon97}. For these measurements, we determine errors using the same Monte Carlo realizations and Gaussian fitting procedure described above for radial velocities.

This procedure thus provides a sample of stellar spectra with heliocentric radial velocities (and corresponding errors estimates), as well as indices for spectral features that are typically detectable for red giants at the distance of M31.  
We match these spectra to the corresponding photometric measurements to create a catalog of candidate M31 dSph stars.  This provides the input to our membership selection scheme, described below.

\subsection{Membership}
\label{sec:membership}

After determining the radial velocity for each
star in our sample, we next determine whether it is a member of a dSph
galaxy or a contaminant.  There are two distinct populations of stellar  contaminants: MW red dwarfs stars (foreground), and M31 red giants (halo field stars).   We consider each of these in turn and then describe our method for assigning membership probabilities to individual stars.

Foreground contamination from MW stars is typically worse than the M31 component, particularly for dwarfs that lie closer to the Galactic plane than M31.  \citet{gilbert06} describes a step-by-step approach for empirically determining the probability that a star is a MW contaminant or an M31 RGB star. While such an approach is valuable for searching for low density M31 halo field stars, the dSphs we consider here have much higher surface densities in their centers, and thus these diagnostics are less crucial near the centers of the satellite.  Furthermore, given that the dSphs studied here exhibit low-dispersion kinematical ``cold spikes'' at velocities closer to M31's systemic velocity than that of the MW, even a wide  velocity window that extends into the tails of the satellite velocity distribution will overlap negligibly with the MW stars.  This is borne out in the velocity histograms shown in \S \ref{sec:data}, as the distributions from the \citet{besancon} model of the MW 
do not overlap with the velocity windows we adopt.  

Halo field and spheroid stars from M31 also provide a significant source of contamination.  While these are generally lower surface density than MW foreground,
 M31 contamination is more problematic for selecting dSphs due to two effects. First, the DDO51 pre-selection technique described above is only 
effective at eliminating foreground dwarf stars --- it cannot distinguish M31 dSph red giants from M31 halo field red giants. 
Second, while the velocity distribution of the M31 halo is much broader than that of the dSphs, it  overlaps with the 
dSph  distributions for many of the satellites.  Hence, there is no definitive way to exclude M31 halo field stars from the dSphs with radial velocity measurements.  Indeed, 
given the hypothesis that  the outer reaches of galactic halos or spheroids are formed from dissolved satellites \citep[e.g.,][]{bj05,guhathakurta06}, it is 
unsurprising that the halo's stellar properties overlap with those of the satellites.  

Fortunately,  the surface densities of M31 halo field stars in the environs of these satellites are quite low relative to the dSphs.   M31's spheroid (non-disk) component can be decomposed into two sub-components: an ``inner spheroid'' that extends to $\sim 30$ kpc and features a de Vaucouleurs  profile, and a power-law ``halo'' component with a slope of $\sim-2.3$ extending out to at least $160$ kpc \citep{guhathakurta05}.  All of our dSph fields, aside  from And I and IX (see \S \ref{ssec:andi} and \ref{ssec:andix}), lie firmly in the halo regime, and the distances imply at most a few halo field stars per DEIMOS slitmask  \citep[][]{guhathakurta05,guhathakurta06,gilbert07}.  Hence, the probability of finding such a star near the center of a dSph with a similar velocity to that of the member stars is low.  Additionally, some of the dSphs have systemic velocities well separated from the M31 spheroid -  the M31 halo distribution is centered at $\sim -300$ \kps{} with a dispersion of $\sim 100$ \kps{} \citep{guhathakurta05,guhathakurta06,chapman06,gilbert07}. Furthermore, the metallicity distribution functions of the M31 halo and the dSphs are partially disjoint.  M31 dSphs examined here tend to have metallicities in the range [Fe/H]$\sim -1.4$ to $-2.0$ \citep{martin06,mc08,martin09,kalirai10}, while the M31 spheroid is more concentrated in the range [Fe/H]$\sim -1.5$ to $-0.1$ \citep{guhathakurta05,guhathakurta06,chapman06}.  While there is overlap, particularly in the outer halo where M31 metallicities are lower, the distributions are reasonably separated in the mean.  Thus, metallicity-based criteria have hope of approximately separating M31 halo structure from dSph stars in the outskirts of the satellites.  While we do not make use of spectroscopic metallicities due to the lower signal-to-noise for some of our spectra, our selection depends on this metallicity distinction implicitly through CDM selection.

With these two sources of contamination in mind, we adopt a method for determining membership probabilities that depends only on distance from the center of the target satellite and location in the ($T_2$,$M-T_2$) CMD.  Explicitly, the membership probability for each star in the survey is calculated as

\begin{equation}
\begin{split}
P_{\rm memb} = 
\exp \left(-\frac{\Delta \alpha^2+\Delta \delta^2}{2 \eta \Reff^2} -\frac{\Delta (M-T_2)^2}{2 \sigma_c} - \frac{\Delta T^2}{2 \sigma_m}\right)   \\
\times \left(\frac{\arctan[-10 (\Sigma_{\rm Na} - 2)]}{\pi} + \frac{1}{2}  \right)  \; ,
\end{split}
\label{eq:memb}
\end{equation}

\noindent where $\Delta \alpha$ and $\Delta \delta$ are the circular distance from the star to the center of the dSph (in RA and Dec, respectively), $\Delta (M-T_2)$ and $\Delta T$ are distances from fiducial isochrones in the  ($T_2$,$M-T_2$) CMD, $\Reff$ is the dSph's half-light radius, $\Sigma_{\rm Na}$ is the Na I equivalent width, and $\eta$, $\sigma_c$, and $\sigma_m$ are free parameters. 

The first term of this expression simply assigns a circular Gaussian spatial distribution of probabilities, weighting the region inside the half-light radius. While some of the dSphs have elliptical light distributions, here we use a circular distribution to avoid introducing biases for the cases where the ellipticity is not well measured.  In most cases, we adopt $\eta=1.5$, except in cases where halo contamination is much different from typical as indicated by the velocity distribution at large distances from the satellite.  In these cases we use a slightly different $\eta$ value to down- or up-weight the membership probability for stars beyond $\Reff$.  The satellites for which this is done are explicitly mentioned in \S \ref{sec:data}.  

The latter two terms inside the exponential define Gaussian acceptance regions around the fiducial isochrone.  We make use of \citet{girardi02cmd} isochrones\footnote{http://stev.oapd.inaf.it/cgi-bin/cmd} in the Washington photometric system for a range of metallicities from [Fe/H]=$-2.2$ to $-1.0$ with 12 Gyr ages, offset by the distance modulus of the satellite.     We then correct the $M$ and $T_2$ photometry for Galactic extinction/reddening using the \citet{sfd98dust} dust maps, and use photometry of the $DDO51$-selected stars within each dwarf's half light radius to empirically choose the best-fitting isochrone. With these  likely members as a guide, we adopt fiducial values of $\sigma_c=0.1$ and $\sigma_m=0.5$, although these parameters are adjusted for some dwarfs, as described in \S \ref{sec:data}.

Given that the only stars we can reach spectroscopically at the distance of M31 are red giant stars, which form a nearly vertical sequence in the CMD, the range of accepted colors, encoded in $\sigma_c$, contains contributions from both intrinsic metallicity scatter and photometric errors.  Meanwhile, $\sigma_m$ contains both distance and photometric uncertainties.  Uncertainty in the age of the stellar population induce further degeneracy in metallicity and distance.  This motivates the choice of fitting isochrones empirically rather than performing a detailed analysis of the isochrone-inferred metallicities, leaving such discussion to the companion paper \comppapp. 
We instead use the isochrones primarily to reject M31 halo field stars (which have a disjoint metallicity distribution, as described above).  This membership approach is thus not sensitive to specific assumptions regarding metallicity or age of the stellar populations, and has been found to be successful for both M31 dSphs \citep{kalirai10,collins10} and MW dSphs \citep{sandg,simon11seg1}.  As the plots in \S \ref{sec:data} show, these isochrones closely follow the spectroscopically apparent cold velocity spikes present for most of the satellite fields, validating the approach.  

The $\arctan$ term serves to favor stars that have Na I $\lambda8190$ values more consistent with RGB stars than those on the main sequence.  This feature is surface gravity sensitive and hence is usually very weak or undetectable in M31 RGB stars, while clearly apparent even on visual inspection of a spectrum of a cool (foreground) dwarf star.  This feature becomes weaker for hotter stars, however, and hence only rejects the cooler MW dwarf stars.  
The particular functional form is chosen to smoothly transition from 0 to 1 with an adjustable center and sharpness of transition.  While the center (here chosen as 2) and sharpness (10) are nominally free parameters,  they should not depend on the particulars of any dSph, but rather on the details of the observations and choice of bandpasses for measuring equivalent widths.  Hence, we keep them constant throughout this data set.   The precise choice of values  is informed by \citet[][Figure 4]{gilbert06},  as this choice rejects the main locus of MW foreground dwarfs, while only rejecting a small number of giants.

Equation \ref{eq:memb} thus defines a membership probability metric that is independent of velocity.  
Filtering velocity outliers could be warranted because inclusion of a few outlier stars as dSph members that are actually M31 contamination may artificially inflate the velocity dispersion.  However, 
filtering outliers that are true members in the tails of the dSph velocity distribution will serve to incorrectly 
\emph{decrease} the inferred dispersions\footnote{Numerical simulations assuming Gaussian distributions with sizes like those of our dSph samples imply this effect can bias the velocity dispersion at the few-percent level.}.  Particularly in the presence of contamination that overlaps all of our observable parameters, it is  impossible to completely correct for this effect with the information available.  Despite this, we do test applying an iterative $3\sigma$ clipping filter.  We find that it does not affect  our membership determination for any dSph other than And I, and defer the discussion for that particular dSph to \S \ref{ssec:andi}. Thus, we adopt a relatively agnostic approach of not using velocities explicitly in our membership formulation, aside from a wide selection window around the dSph's systemic velocity.

The aforementioned membership formulation leads to our adopted definition of ``member''  stars for the discussion below: those for which $P_{\rm memb}  > 0.1$.  As is apparent from the velocity
 histograms in \S \ref{sec:data}, this choice is conservative in the sense that it excludes some stars that are likely members based on their radial velocities. However, it also means that those that \emph{are} selected as members are generally rather secure members on the basis of all the information available aside from radial velocity. This serves to decrease the previously noted biases in the kinematical parameters when contamination is small relative to the member population (true in the inner regions  for most of the dSph we examine here).  

For cross-checking purposes, we considered alternate selection methods.  For And I, III, V, VII, IX, X, and XIV, we considered an alternate member selection 
method based on velocity, spatial proximity, and explicit photometric metallicity estimates (rather than the implicit metallicity dependence based on CMD location described above).  We find that the dispersions reported below all agree within $1\sigma$.  In addition, for a few dSphs, we compare the likelihood
method of \citealt{gilbert06} (adjusted to the distance of the dSph instead of the M31 mean).  We find that the method used in this paper rejects as MW stars 
nearly the same set of stars as those labeled probable dwarf stars by the \citet{gilbert06} likelihood method.
 
\subsection{dSph Kinematical Modeling}
\label{ssec:modeling}

With heliocentric radial velocities and membership probabilities, we are now prepared to describe the kinematics of the dSphs of our sample.  We model the velocities of member stars in each dSph as a Gaussian distribution with a systematic velocity $\mu$ and dispersion $\sigma$. This assumption of Gaussianity would be violated if there is significant contamination from unresolved binaries \citep[e.g.,][]{minor10bin}.  However, for the galaxies we present with more than two members, a Shapiro--Wilk test \citep{shapirotest} reveals that the null hypothesis that the radial velocity distributions are Gaussian cannot be rejected at the $p<0.05$ level for any of the galaxies.  This also holds if we apply the test to our entire data set after an offset to the systematic velocity and scaling by the dispersion. Thus, our assumption of Gaussianity is plausible for these data. 

We estimate parameters for the radial velocity distribution using a maximum likelihood estimator similar to that of \citet{kalirai10} and \citet{walker07}, but with a factor that includes our method of assigning membership probabilities.  The likelihood we adopt has the form

\begin{equation}
\begin{split}
\log L(\sigma,\mu | v_i,\sigma_{t,i},p_i) = -\frac{1}{2}\sum_{i=1}^N \Bigm[ p_i \log(\sigma^2+\sigma_{t,i}^2)  \\
+
 p_i \frac{(v_i-\mu)^2}{\sigma^2+\sigma_{t,i}^2} + p_i \log(2 \pi)\Bigm]
\end{split}
\label{eqn:loglike}
\end{equation}

\noindent where the $p_i$s are membership probabilities for each star computed as per Equation \ref{eq:memb}, and $\sigma_{t,i}$ is the per-star velocity error (including the 2.2 \kps{} floor).  This likelihood is numerically maximized,  the Hessian is computed at this maximum, and inverted to obtain the covariance matrix.  We adopt as  the error on $\sigma$ and $\mu$  the square root of the diagonal elements of this matrix.

As described in the previous subsection, for the final systemic velocity and velocity dispersion, we only accept stars with membership probabilities $P_{\rm memb}  > 0.1$.  We  thus filter out all of the stars with a high probability of being non-members, because even with low weights from Equation \ref{eqn:loglike},  large outliers strongly bias the parameter estimates.  For the same reason, we also reject stars with velocities far from the cold spike for each dwarf (limits determined on a case-by-case basis as detailed in \S \ref{sec:data}).  This serves to filter out a small number of MW foreground stars that happen to lie near a satellite spatially and near the fiducial isochrone in the CMD.  The precise range of accepted velocities for each satellite are specified in \S \ref{sec:data}.  We emphasize, however, that the ranges selected are significantly broader than the velocity peaks for each satellite, so as not to bias the kinematical parameter estimates.

\section{SPLASH M31 dSphs}
 \label{sec:data}

\begin{deluxetable*}{cccccccccc}
 \tablecolumns{9}
 \tablecaption{SPLASH M31 dSph Stars}
 \tablehead{
   \colhead{Mask\tablenotemark{1}} &
   \colhead{Star ID\tablenotemark{2}} &
   \colhead{RA\tablenotemark{3}} &
   \colhead{Dec\tablenotemark{4}} &
   \colhead{$T_2$\tablenotemark{5}} &
   \colhead{$M-T_2$\tablenotemark{6}} &
      \colhead{$\vh$ (km/s)\tablenotemark{7}} &
      \colhead{Na EW\tablenotemark{8}} &
      \colhead{$P_{\rm memb}$ \tablenotemark{9}} 
 }
  \startdata
  d1\_1 &  1005484 & $0^h45^m49.^s.6$ & $+38^{\circ}11'51''$ & $22.21 \pm 0.07$ & $1.67  \pm 0.11$ & $-396.1 \pm 8.6$ & $2.0 \pm 1.3$ & 0.0 \\
  d1\_1 &  1005334 & $0^h45^m51.^s.7$ & $+38^{\circ}8'57''$ & $20.73 \pm 0.02$ & $1.83 \pm 0.03$  & $-381.9 \pm 2.2$ & $0.7 \pm 0.4$ & 0.007  \\
  d1\_1 &  1005398 & $0^h45^m50.^s.8$ & $+38^{\circ}13'44''$ & $20.53 \pm 0.02$ & $1.76 \pm 0.03$  & $-371.3 \pm 3.5$ & $0.3 \pm 0.4$ & 0.0 \\
  ... & ...& ...& ...& ...& ...& ...& ...& ...

 \enddata
 
\tablenotetext{1}{Name of the mask on which this star was observed (see Table \ref{tab:masks}). }
\tablenotetext{2}{ID number of the star (unique within a dSph field).}
\tablenotetext{3}{Right ascension of the star (J2000).}
\tablenotetext{4}{Declination number of the star (J2000).}
\tablenotetext{5}{ Washington $T_2$ magnitude (extinction corrected).}
\tablenotetext{6}{Washington $M-T_2$ color (extinction corrected).}
\tablenotetext{7}{Heliocentric radial velocity of this star.}
\tablenotetext{8}{Equivalent width of Na I $\lambda 8190$ feature.}
\tablenotetext{9}{Membership probability computed following Equation \ref{eq:memb}.}
 
 \tablecomments{This table is available in its entirety in a machine-readable form in the online ApJ article. A portion is shown here for guidance regarding its form and content.  Stars not classified as members but with successful radial velocity measurements can be made available upon request to the authors.}
 
 \label{tab:stardata}
 \end{deluxetable*}

 \begin{deluxetable*}{p{0.6in}ccccccccc}
 \tablecolumns{10}
 
 \tablecaption{Summary of M31 dSph properties.}
 \tablehead{
   \colhead{Name\tablenotemark{1}} &
   \colhead{RA\tablenotemark{2}} &
   \colhead{Dec\tablenotemark{3}} &
   \colhead{$M_V$\tablenotemark{4}} &
   \colhead{$d_{\rm LOS}$\tablenotemark{5}} &
   \colhead{$r_{\rm M31}$\tablenotemark{6}} &
   \colhead{$N_{\rm memb}$\tablenotemark{7}} &
   \colhead{$\vsys$\tablenotemark{8}} &
   \colhead{$\sigma_{\rm LOS}$\tablenotemark{9}} &
   \colhead{Sources\tablenotemark{10}}
 }
\startdata

And I & 00:45:39.800 & +38:02:28.00 & $-11.8^{+1.0}_{-1.0}$ & $744.7^{+24.4}_{-23.6}$ & $58.4^{+35.4}_{-34.3}$ & 51 & $-376.3 \pm 2.2$\tablenotemark{*} & $10.2 \pm 1.9$\tablenotemark{*} & a,b,c \\
And III & 00:35:33.800 & +36:29:52.00 & $-10.2^{+0.3}_{-0.3}$ & $748.2^{+24.5}_{-23.7}$ & $75.2^{+35.5}_{-34.4}$ & 62 & $-344.3 \pm 1.7$\tablenotemark{*} & $9.3 \pm 1.4$\tablenotemark{*} & a,c,d \\
And V & 01:10:17.100 & +47:37:41.00 & $-9.6^{+0.3}_{-0.3}$ & $820.4^{+15.3}_{-15.0}$ & $118.2^{+29.9}_{-29.0}$ & 85 & $-397.3 \pm 1.5$ & $10.5 \pm 1.1$ & a,e,f \\
And VII & 23:26:31.700 & +50:40:33.00 & $-13.3^{+0.3}_{-0.3}$ & $762.1^{+25.0}_{-24.2}$ & $218.3^{+35.8}_{-34.7}$ & 136 & $-307.2 \pm 1.3$ & $13.0 \pm 1.0$ & a,c,g \\
And IX & 00:52:53.000 & +43:11:45.00 & $-8.1^{+0.4}_{-0.1}$ & $765.6^{+25.1}_{-24.3}$ & $40.5^{+35.9}_{-34.7}$ & 32 & $-209.4 \pm 2.5$ & $10.9 \pm 2.0$ & a,c,h \\
And X & 01:06:33.700 & +44:48:16.00 & $-7.4^{+0.1}_{-0.1}$ & $701.5^{+33.1}_{-31.6}$ & $109.4^{+41.9}_{-40.2}$ & 27 & $-164.1 \pm 1.7$ & $6.4 \pm 1.4$ & a,i \\
And XI & 00:46:20.000 & +33:48:05.00 & $-6.9^{+0.5}_{-0.1}$ & $871.0^{+84.0}_{-76.6}$ & $139.1^{+87.9}_{-80.6}$ & 2 & $-461.8 \pm 3.7$\tablenotemark{**} & \nodata & a,j \\
And XII & 00:47:27.000 & +34:22:29.00 & $-6.4^{+0.1}_{-0.5}$ & $831.8^{+47.3}_{-44.7}$ & $109.2^{+53.8}_{-51.2}$ & 2 & $-525.3 \pm 3.4$\tablenotemark{**} & \nodata & a,j,k \\
And XIII & 00:51:51.000 & +33:00:16.00 & $-6.7^{+0.4}_{-0.1}$ & $871.0^{+84.0}_{-76.6}$ & $150.0^{+87.9}_{-80.6}$ & 12 & $-185.4 \pm 2.4$ & $5.8 \pm 2.0$ & a,j \\
And XIV & 00:51:35.000 & +29:41:49.00 & $-8.5^{+0.1}_{-0.1}$ & $734.5^{+120.6}_{-103.6}$ & $162.3^{+123.3}_{-106.5}$ & 48 & $-480.6 \pm 1.2$ & $5.3 \pm 1.0$ & a,l \\
And XV & 01:14:18.700 & +38:07:03.00 & $-9.8^{+0.4}_{-0.4}$ & $770.0^{+70.0}_{-70.0}$ & $93.6^{+74.6}_{-74.3}$ & 29 & $-323.0 \pm 1.4$ & $4.0 \pm 1.4$ & a,m,n \\
And XVI & 00:59:29.800 & +32:22:36.00 & $-9.2^{+0.5}_{-0.5}$ & $525.0^{+50.0}_{-50.0}$ & $279.4^{+56.2}_{-55.8}$ & 7 & $-367.3 \pm 2.8$ & $3.8 \pm 2.9$ & a,m,n \\
And XVIII & 00:02:14.500 & +45:05:20.00 & $-9.7^{+0.1}_{-0.1}$ & $1355.2^{+83.6}_{-78.8}$ & $590.9^{+87.5}_{-82.6}$ & 22 & $-332.1 \pm 2.7$ & $9.7 \pm 2.3$ & a,o \\
And XXI & 23:54:47.700 & +42:28:15.00 & $-9.3^{+0.1}_{-0.1}$ & $859.0^{+51.0}_{-51.0}$ & $149.2^{+57.1}_{-56.7}$ & 6 & $-361.4 \pm 5.8$ & $7.2 \pm 5.5$ & a,p \\
And XXII & 01:27:40.000 & +28:05:25.00 & $-6.2^{+0.1}_{-0.1}$ & $794.0^{+239.0}_{-239.0}$ & $220.6^{+240.4}_{-240.3}$ & 7 & $-126.8 \pm 3.1$\tablenotemark{$\dagger$} & $3.54^{+4.16}_{-2.49}$\tablenotemark{$\dagger$} & a,p

\enddata

\tablenotetext{1}{Name of the dSph.}
\tablenotetext{2}{Right Ascension of the dSph.}
\tablenotetext{3}{Declination of the dSph.}
\tablenotetext{4}{V-band absolute magnitude of the dSph.}
\tablenotetext{5}{Heliocentric line-of-sight distance to the dSph in kpc.}
\tablenotetext{6}{Distance from M31 to the dSph in kpc.}
\tablenotetext{7}{Number of successful spectra with $P_{\rm memb} >0.1$, i.e., dSph member stars.}
\tablenotetext{8}{Systemic velocity of the dSph estimated according to \S \ref{ssec:modeling} in \kps.}
\tablenotetext{9}{Total velocity dispersion of the dSph estimated according to \S \ref{ssec:modeling} in \kps.}
\tablenotetext{10}{Sources for location, distance, size, and luminosity of the dSph.}

\tablerefs{ 
a. \citet{brasseur11},
b. \citet{Paturel00},
c. \citet{Mcconnachie05},
d. \citet{Karachentseva98},
e. \citet{Armandroff98},
f. \citet{Mancone08},
g. \citet{Karachentseva01},
h. \citet{zucker04},
i. \citet{zucker07},
j. \citet{martin06},
k. \citet{Chapman07},
l. \citet{maj07and14},
m. \citet{ibata07},
n. \citet{Letarte09},
o. \citet{mc08},
p. \citet{martin09}
}
  \tablenotetext{*}{Kinematics for these satellites may be moderately affected by contamination due to M31 substructure -- see subsection for details.}
  \tablenotetext{**}{These measurements depend on an uncertain detection of a cold spike -- see subsection for details.}
   \tablenotetext{$\dagger$}{The parameter estimates for this object is based on a different method than the other dSphs -- see subsection for details.}

 \label{tab:dsphsumm}
 \end{deluxetable*}
 
 With our reduction and analysis procedure outlined in \S \ref{sec:specreduct} and  \ref{sec:membership}, we present the full catalog of stars we identify as likely dSph members in Table \ref{tab:stardata}.  Non-member stars are available to the reader on request.  We also present a summary of the properties of each dSph studied here in Table \ref{tab:dsphsumm}, including our results for systemic velocities and total velocity dispersions.  Dwarfs with no reported  $\sigma_{\rm LOS}$ have too few likely members to reliably compute a velocity dispersion.

 In the sub-sections that follow, we describe our results and a brief discussion for each satellite in the sample.  These each include a Figure showing CMDs, spatial distributions, velocity-radius relations, and velocity histograms for each dSph.   The sub-sections also describe any deviation from the fiducial parameters described in the previous section for membership selection. Where relevant, we include a discussion of previous kinematics for each dSph and compare our results.

\afterpage{\clearpage}

\subsection{And I}
\label{ssec:andi}

\begin{figure*}[tbp!]

\begin{center}
 \bf \large And I
 \end{center}

 \epsscale{.5}
 \plotone{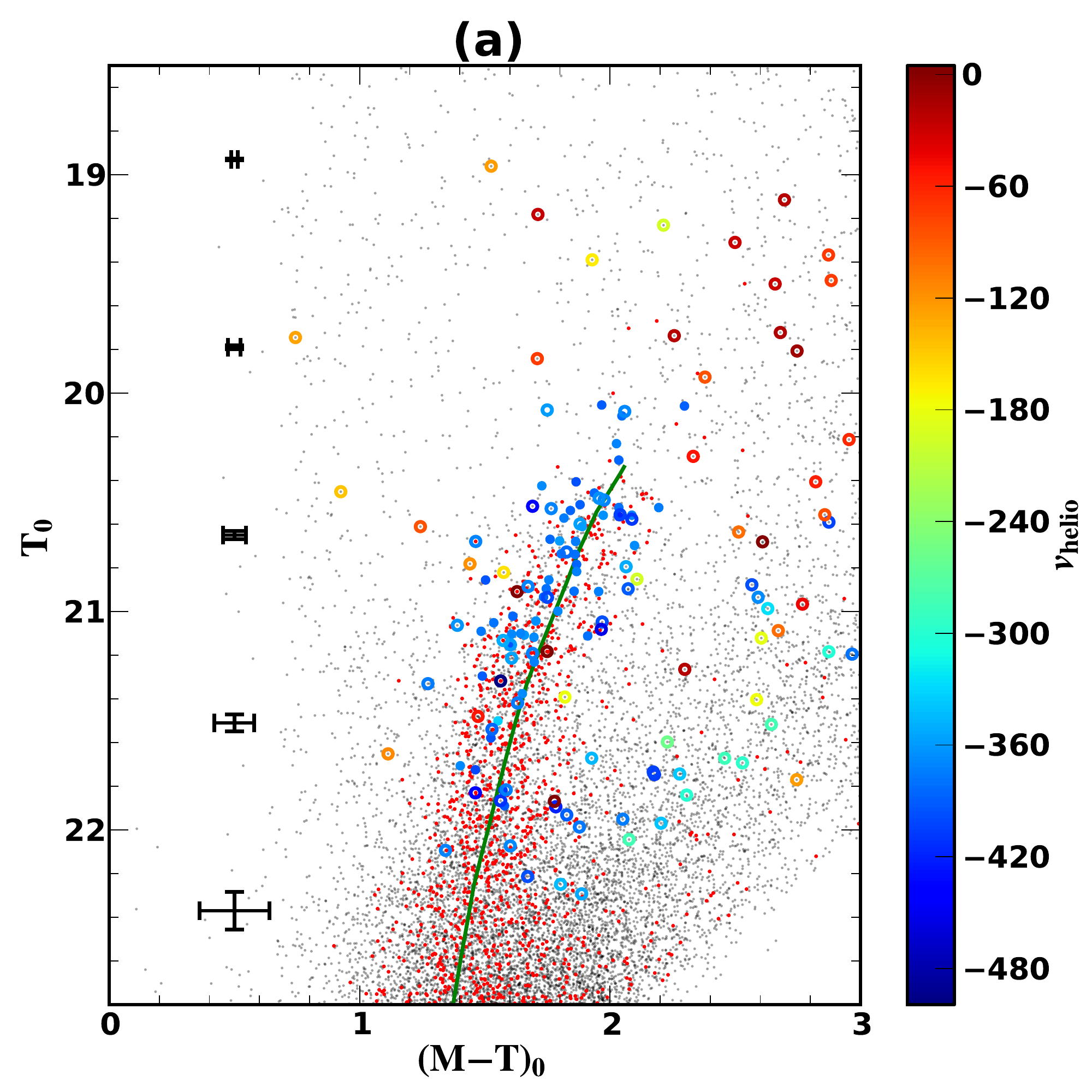}
 \plotone{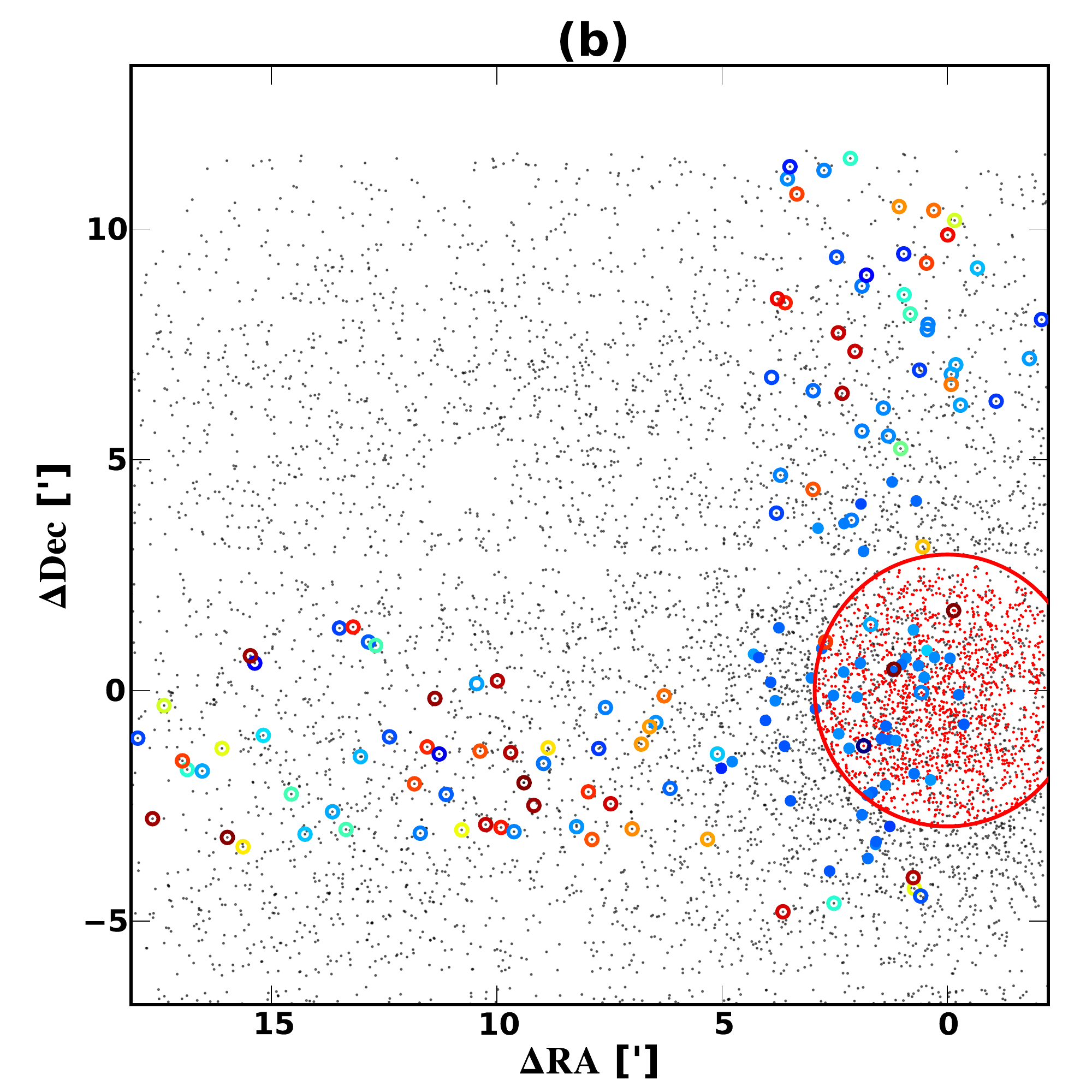}
 \plotone{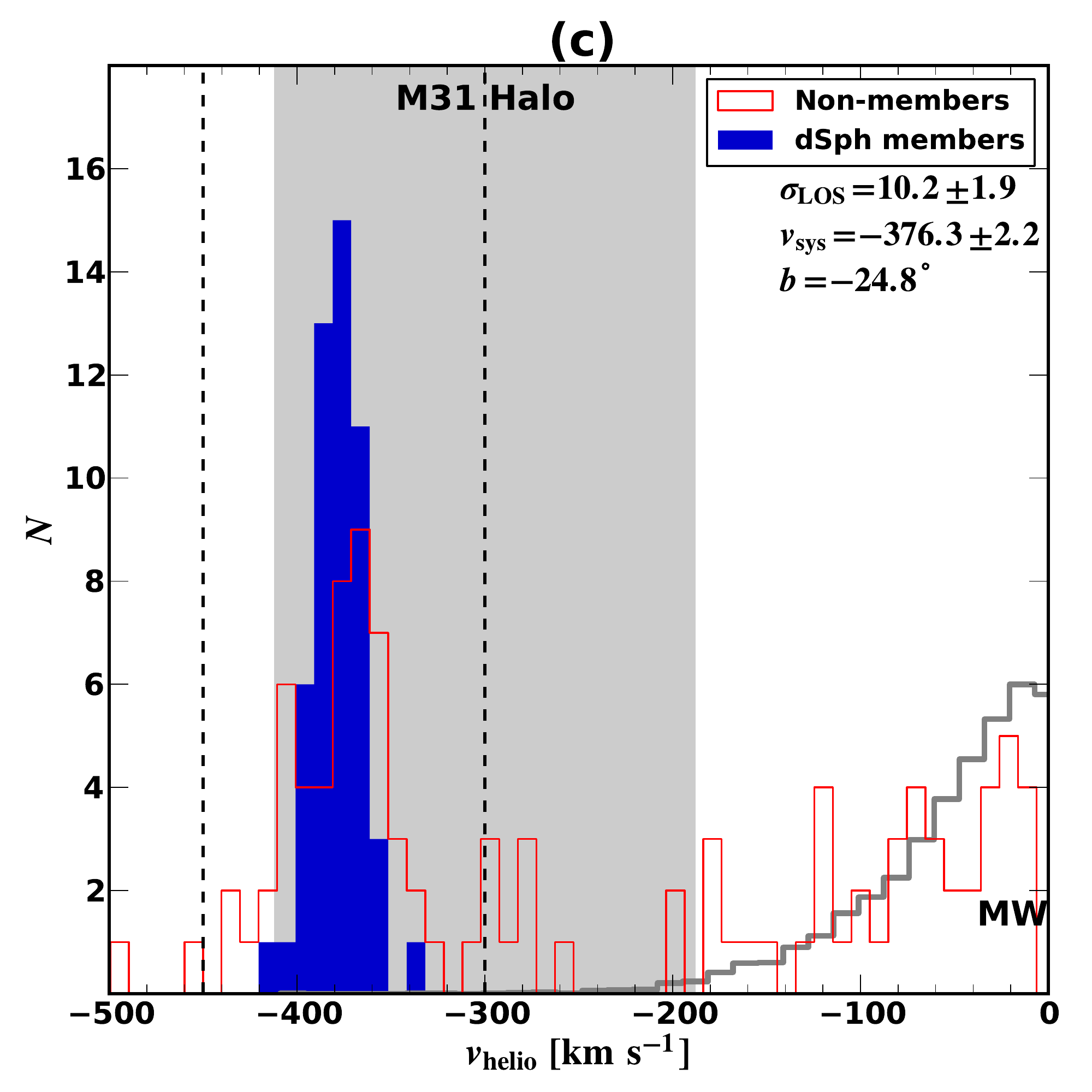}
 \plotone{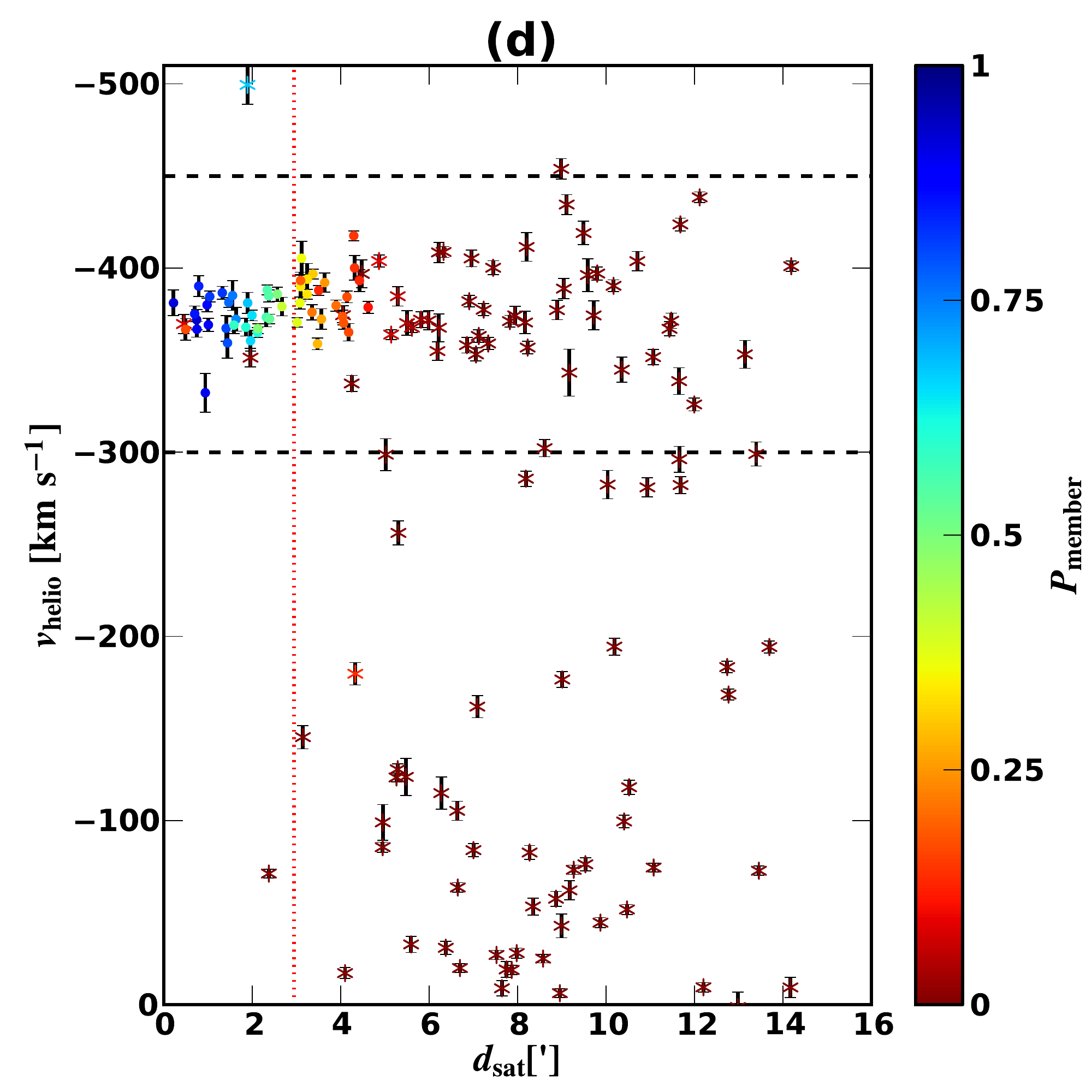}
 \caption{SPLASH results for And I. Panel (a) shows the (Galactic extinction 
          corrected) color-magnitude diagram for And I, and panel (b) shows the spatial 
          distribution of the same objects.   $R_{\rm eff}$ for the galaxy is indicated in (b) as the large red circle.  
          In both panels, circles are objects with spectra and measured 
          radial velocities.  Filled circles are those identified as members according
          to the criteria discussed in \S \ref{sec:membership}, open circles are non-members, and the color code gives
          the heliocentric radial velocity.  Small points are stellar-like objects detected photometrically, 
          but without spectra. The red points are those that lie inside $R_{\rm eff}$ and satisfy the
   DDO51 pre-selection criteria indicating  indicating a high likelihood of being a red giant, while the black points are the remainder. 
    The (green) line passing through 
          these points in panel (a) is the adopted isochrone.
          Panel (c) shows the heliocentric radial velocities of all spectra from the And I mask with sufficient signal-to-noise to 
          measure a velocity as the open (red) histogram. The filled (blue) histogram is for members.  Also shown is 
          a gray shaded region indicating the center and $1\sigma$ width of the M31 halo velocity distributions, as
          characterized by \citet{chapman06}. Versions of this figure for other dSph farther than 
          $100$ kpc (projected) do not show this halo distribution, as the typical number of halo stars per mask
          is very low at those distances.  Also shown as a (gray) histogram near $v_{\rm helio}=0$ is an arbitrarily normalized 
          distribution of MW foreground stars derived from the \citet{besancon} model in the direction of And I, based on a CMD
          selection box approximately matched to panel (a).    Panel (d) shows $\vh$ of all stars with successfully
          recovered velocities as a function of their distance from the center of the satellite, with error bars derived following the 
          procedure described by \S \ref{sec:specreduct}.  Circles are classified as members and star-shaped symbols are non-members,
          with the color code signifying the membership probability. The dotted (red) vertical line is $R_{\rm eff}$ for And I. 
          The dashed (black) vertical lines in both (c) and (d) indicate the minimum and maximum velocities for inclusion as a member. }
 \label{fig:and1plots}
\end{figure*}

And I, originally discovered by \citet{vdb72},  is one of the brightest of M31's dSph companions ($M_V=-11.8$, $L_V=4.5 \times 10^6 L_\odot$, similar to the MW dSph Leo I).  We present our data for this galaxy in Figure \ref{fig:and1plots}.  A cold spike is immediately apparent near $-400$ \kps{} in the velocity histogram.  However, the tails reveal an unusual aspect of And I that complicate the kinematics: it overlaps on the sky with M31's Giant Southern Stream  \citep[GSS,][]{ibata01gss}.  The GSS is M31's largest tidal feature, and while it is $\sim 100$ kpc behind And I \citep{McConnachie2003GSS3D}, it overlaps both on the sky and kinematically with And I, as described in \citet{gilbert09gss} and \citet{kalirai10}. In addition, while the GSS's overall metallicity and age distribution is distinct from that of And I, the tails of each distribution overlap each other, as well as the general halo population \citep{gilbert09gss}.  Thus, the GSS is a potentially major source of contamination, as stars at the boundaries between the distributions cannot be clearly assigned to one group over the other.  Additionally, And I is one of the closest dSphs to M31 and the overall morphology of the galaxy hints at tidal disruption \citep[e.g.,][]{mcc06}. Thus the choice of the $\eta$ parameter in Equation \ref{eq:memb} has a significant effect on our derived kinematics, because GSS stars are likely entering into the sample when $\eta$ is large.

\begin{figure}[tbp!]
 \epsscale{1}
 \plotone{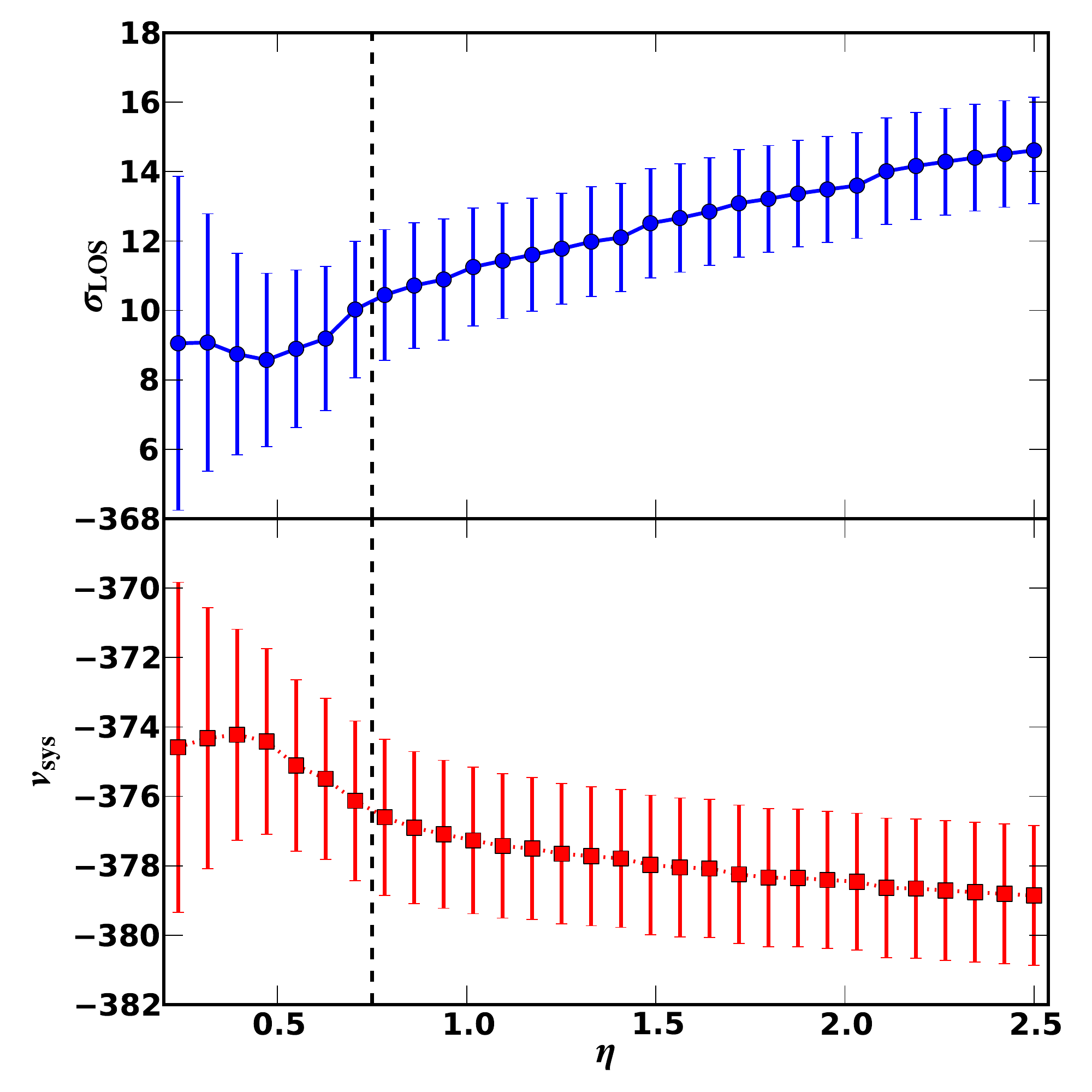}
 \caption{Dependence of kinematical parameter estimates on And I membership parameter $\eta$.  $\eta$ is defined as the free parameter of Equation \ref{eq:memb} that defines the scale relative to $\Reff$ outside which membership probabilities fall off.  The upper panel (blue) displays the effect of $\eta$ on the inferred velocity dispersion, while the lower (red) is for the systemic velocity.  In both cases the values and associated error bars are derived for each $\eta$ following the procedure described in \S \ref{ssec:modeling}. The vertical dashed (black) line is the $\eta$ value adopted for And I.}
 \label{fig:and1cog}
\end{figure}

The primary effect is that increasing $\eta$ serves to more strongly weight stars at larger radii, which have a higher probability of being GSS contaminant stars. We show how this influences our modeling in Figure \ref{fig:and1cog}, which plots how our parameter estimates for $\slos$ and $\vsys$ are affected by the choice of $\eta$. The systemic velocity varies slightly but within the error bars of any given measurement, while $\slos$ varies significantly for large $\eta$.  We note, however, that the variation is initially relatively flat, suggesting that at small radii the contamination is not affecting the kinematical parameter estimates (at least within the error bars).  Hence, we choose a value of $\eta=0.75$, at approximately the value where the upward trend begins.  This is significantly more conservative than our fiducial value ($\eta=1.5$), but serves to down-weight stars in the outskirts of the dSph.  We note that while more conservative in the sense of having fewer stars, our result using the analysis procedure presented here is close to (well within error bars of) the earlier analysis of the same dataset performed in \citet{kalirai10}.

Even with a choice of $\eta$ confining the selection to the core of the dSphs, And I is the only dSph in our sample for which iterative $3\sigma$ clipping filters any stars.  The spectra and CMD locations of these two stars do not show clear indications of being non-members - they show RGB-like spectral features, and they  both lie near the fiducial And I RGB in the CMD of Figure \ref{fig:and1plots}.  Additionally, the velocity distribution is consistent with Gaussian even before sigma clipping\footnote{Gaussianity was tested for And I using Shapiro--Wilk \citep{shapirotest}, Anderson--Darling  \citep{adtest}, and Kolmogorov--Smirnov tests, none of which could reject the null hypothesis at the $p<10\%$ level.}, as described further in \S \ref{ssec:modeling}.  Furthermore, the presence of one or two outliers at $\sim 3\sigma$ is expected for a dataset of the size of And I at the $1$ to $10 \%$ level, or for our SPLASH dataset as a whole at the $\sim 45 \%$ level.    Thus, while it is plausible that these stars are actual members, there is a chance that these stars are M31 contaminants or extra-tidal stars (if And I is tidally disrupting).  While we use the unfiltered value in our analyses below for consistency, we also compute the kinematic parameters \emph{without} these stars, finding $\sigma_{\rm LOS, clip}= 8.2  \pm 1.7$ \kps and  $v{\rm sys, clip}= -377.0 \pm 1.9$ \kps.  This dispersion is an $\approx 1\sigma$ offset from the unfiltered value, implying that if the stars are truly non-members, there is a similar level of offset in our mass estimates described below.


\subsection{And III}
\label{ssec:andiii}

\begin{figure*}[tbp!]

\begin{center}
 \bf \large And III
 \end{center}

 \epsscale{.5}
 \plotone{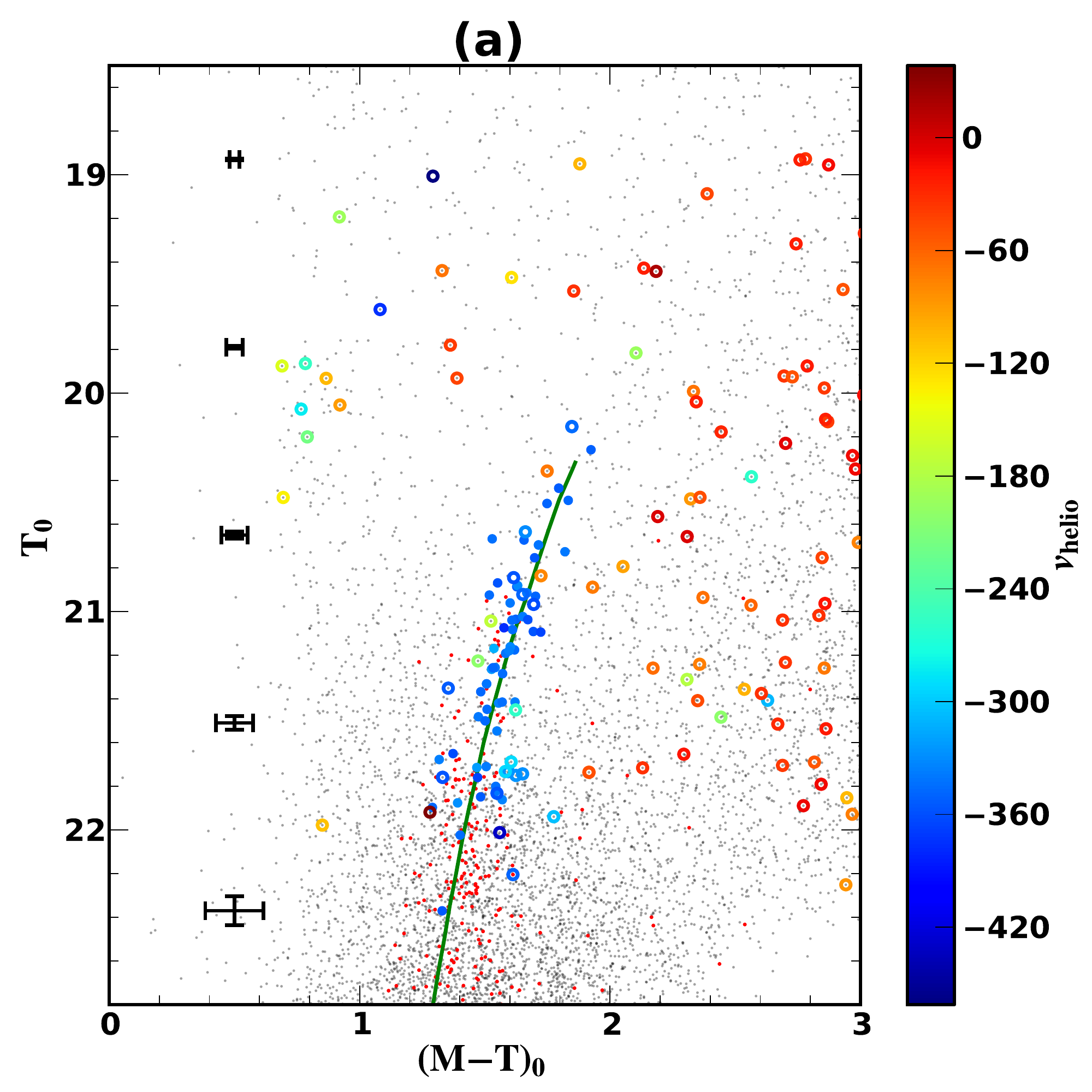}
 \plotone{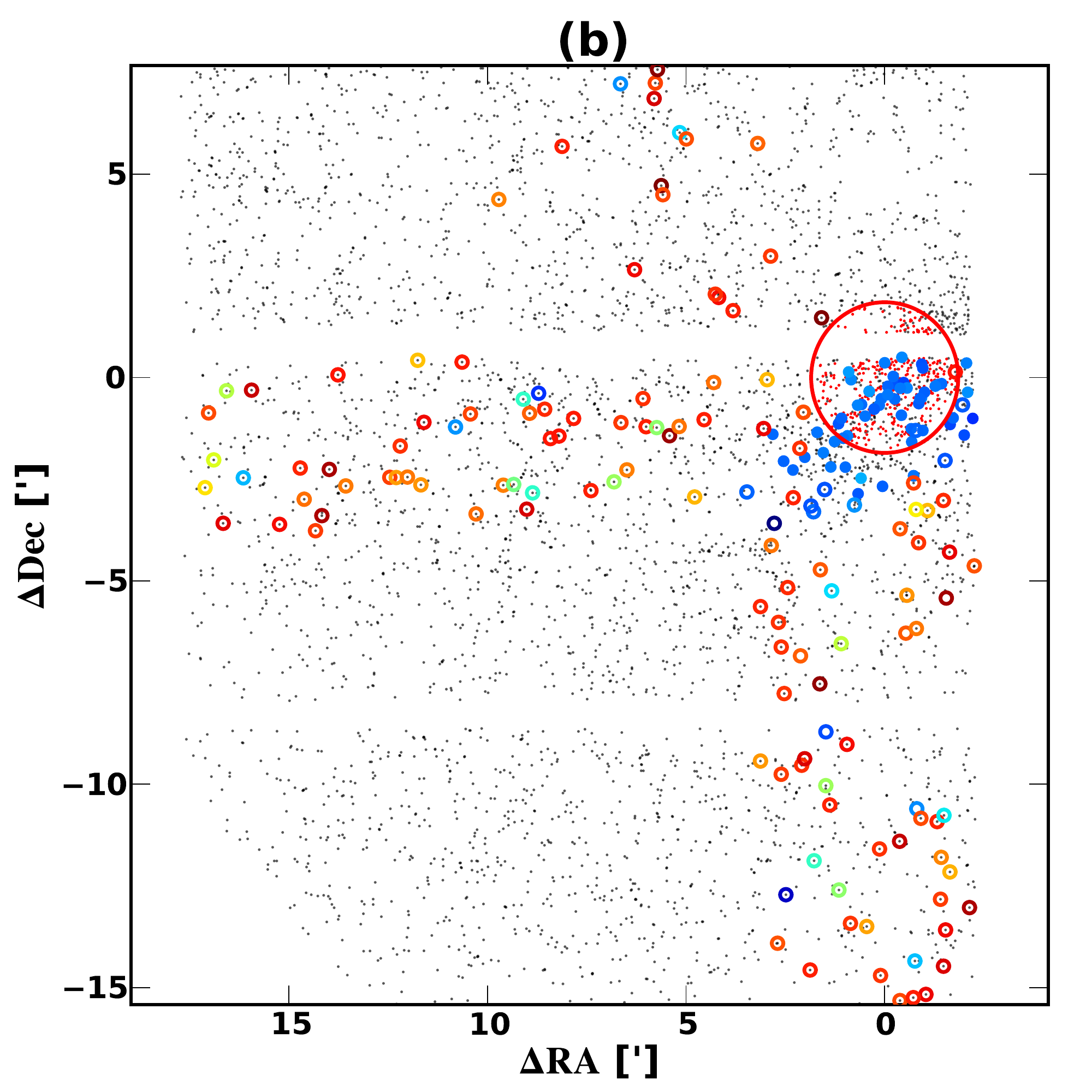}
 \plotone{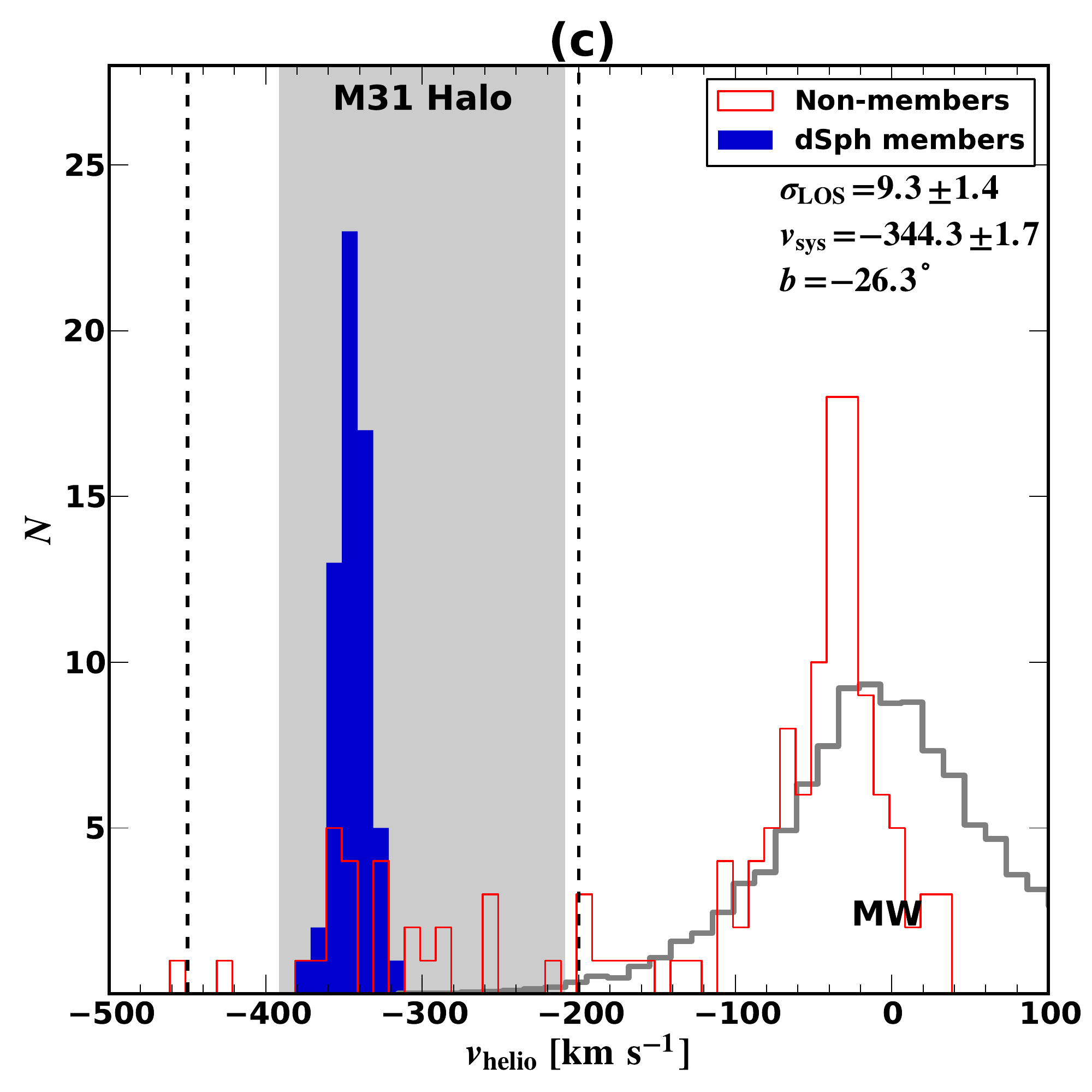}
 \plotone{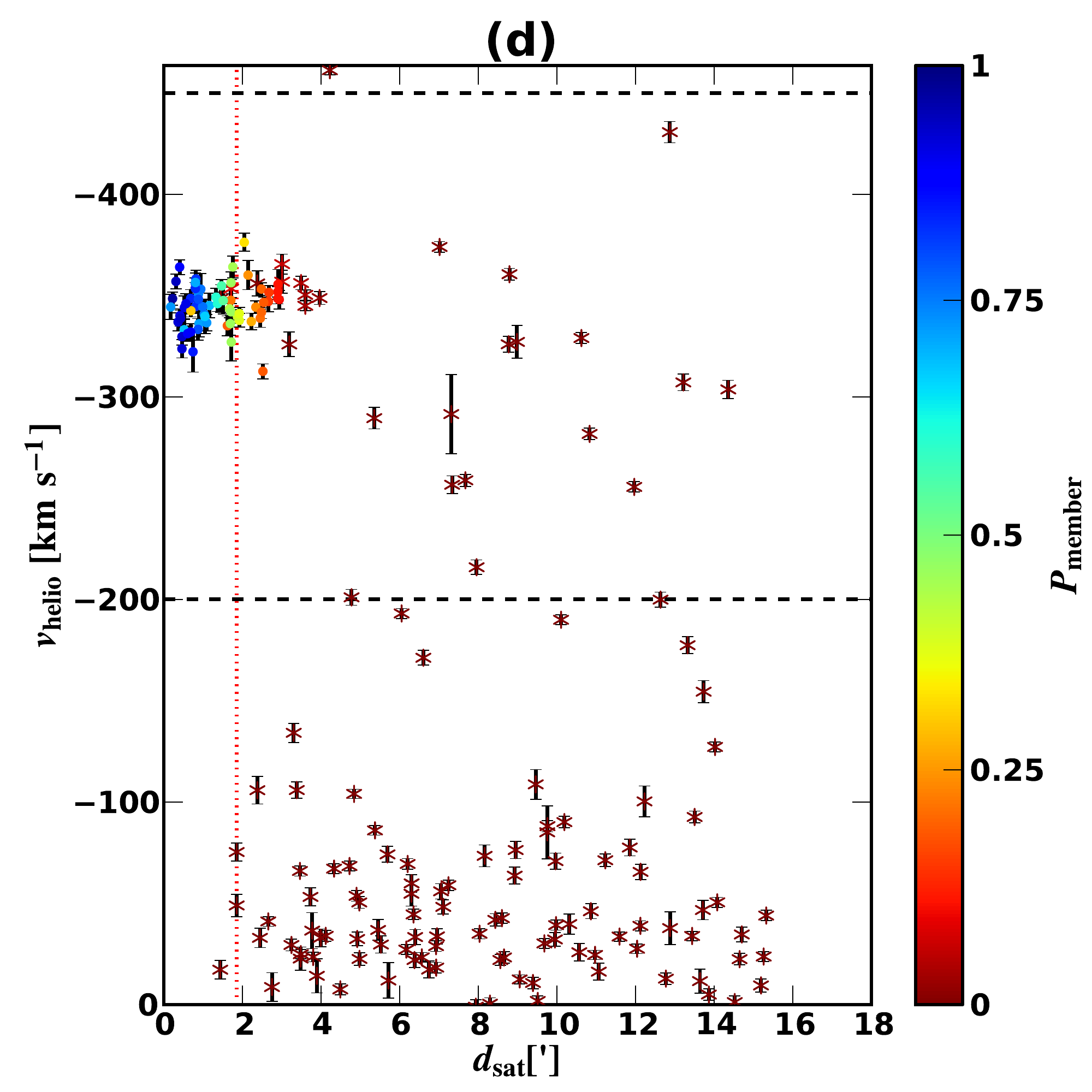}
 \caption{Same as Figure \ref{fig:and1plots}, but for And III.}
 \label{fig:and3plots}
\end{figure*}

\begin{figure}[tbp!]
 \epsscale{1}
 \plotone{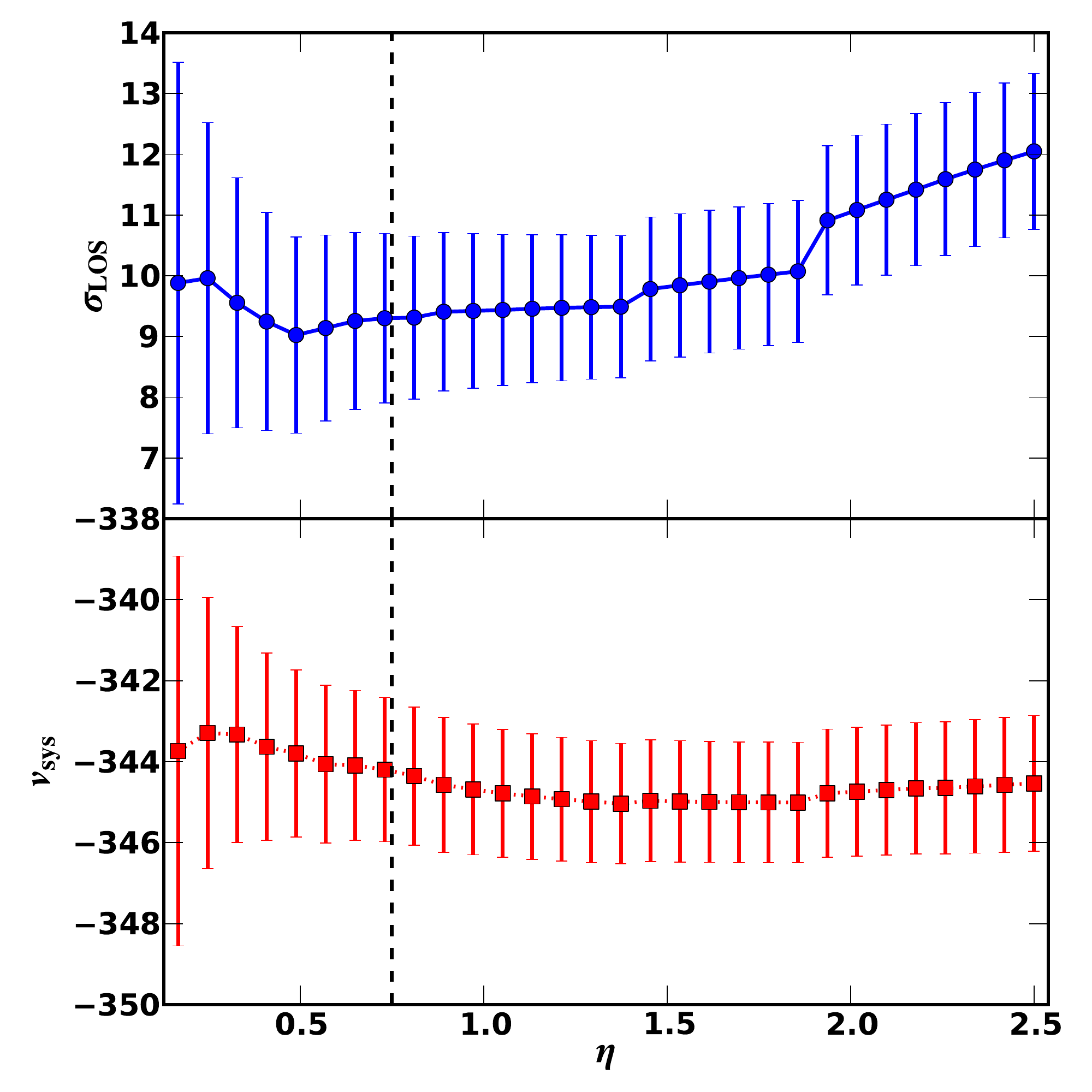}
 \caption{Same as Figure \ref{fig:and1cog}, but for And III.}
 \label{fig:and3cog}
\end{figure}

And III is another relatively bright M31 dSph ($M_V=-10.2$, $L_V=1.0 \times 10^6 L_\odot$), also discovered by \citet{vdb72}. We present our SPLASH results for this satellite in Figure \ref{fig:and3plots}.  As for And I, a cold spike is clearly present, in this case near $-350$ \kps.  While near the fringe of the GSS, there is still a hint of possible GSS structure in the form of a slight excess of stars with overlapping velocities far from the center of the satellite.  Additionally, there will be a contribution from the tails of the M31 halo distribution given that And III is relatively close to M31 on the sky ($5^{\circ}$, or $68$ kpc projected).    We plot the variation of $\slos$ and $\vsys$ with $\eta$ in Figure \ref{fig:and3cog}, revealing  only a weak dependence, as in And I.  However, there is a slight upturn in $\slos$ near our fiducial value.  Therefore, we adopt $\eta=0.75$, matching the choice for And I.

Our velocity dispersion for And III here is higher than
\citet{kalirai10}, which found $4.7 \pm 1.8$.  This is driven in part by the inclusion of a
third mask with additional members, and differences in our procedure for determining
per-star errors address the remainder of the difference. The primary cause, however, is likely
 the different methods used for determining membership between this work 
and \citet{kalirai10}.  In particular, the two most extreme velocity members (the highest and lowest velocity members from Figure \ref{fig:and3plots})
are included here, and are not in \citet{kalirai10}.  
As described in \Ss \ref{sec:membership} and \ref{ssec:modeling}, such outliers can influence  the kinematical parameter estimates,
but because of the overlap between the M31 halo/spheroid distribution and the dSph member populations, it is ambiguous whether or not these 
stars should actually be included as members.  For this data set, $3\sigma$ clipping does not 
eliminate them, and even with them included, the velocity distribution is consistent with Gaussian
(see \S \ref{sec:membership}).  
Hence, for consistency with the remainder of this paper we keep these stars as members 
and report the resulting kinematical parameters.

\subsection{And V}
\label{ssec:andv}

\begin{figure*}[tbp!]

\begin{center}
 \bf \large And V
 \end{center}

 \epsscale{.5}
 \plotone{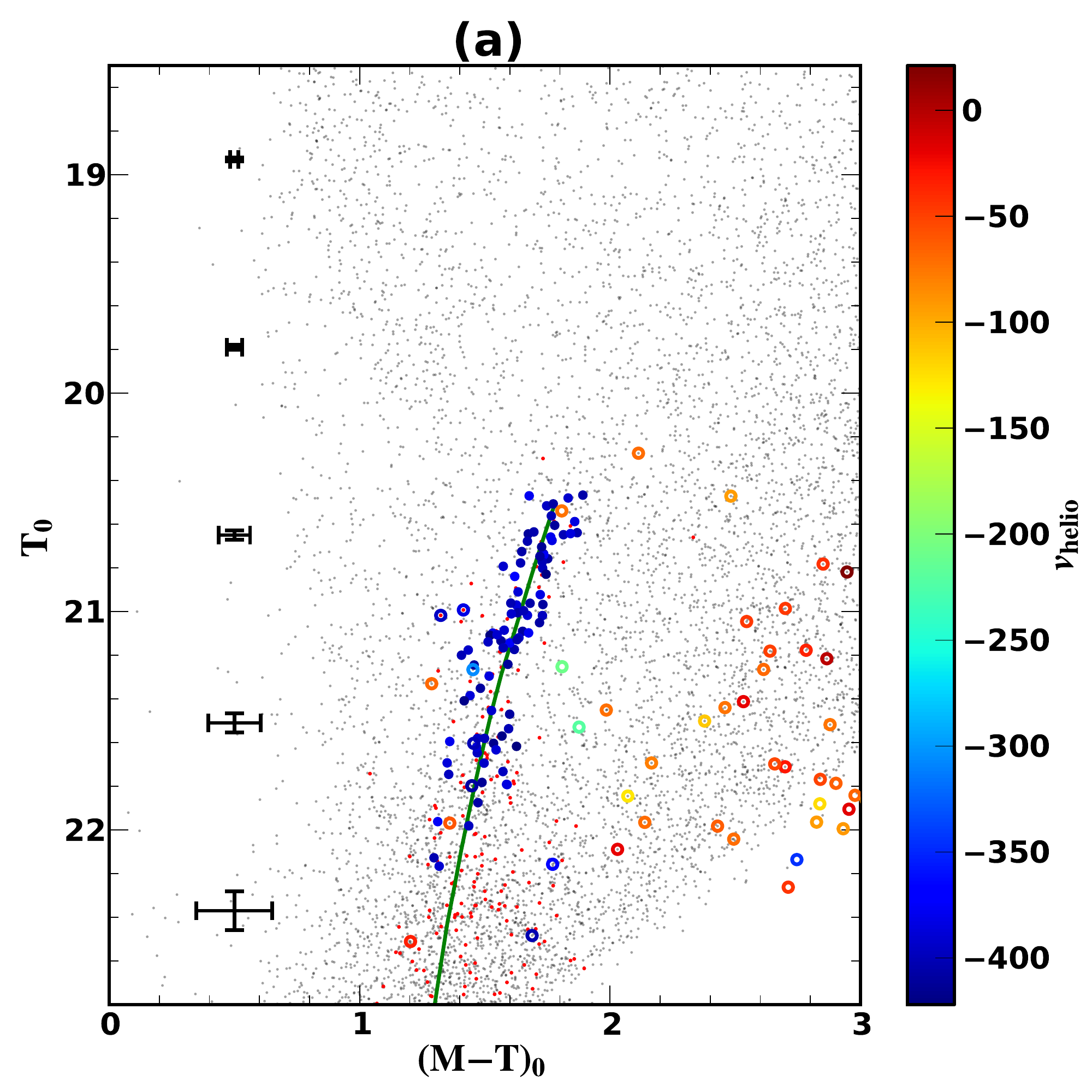}
 \plotone{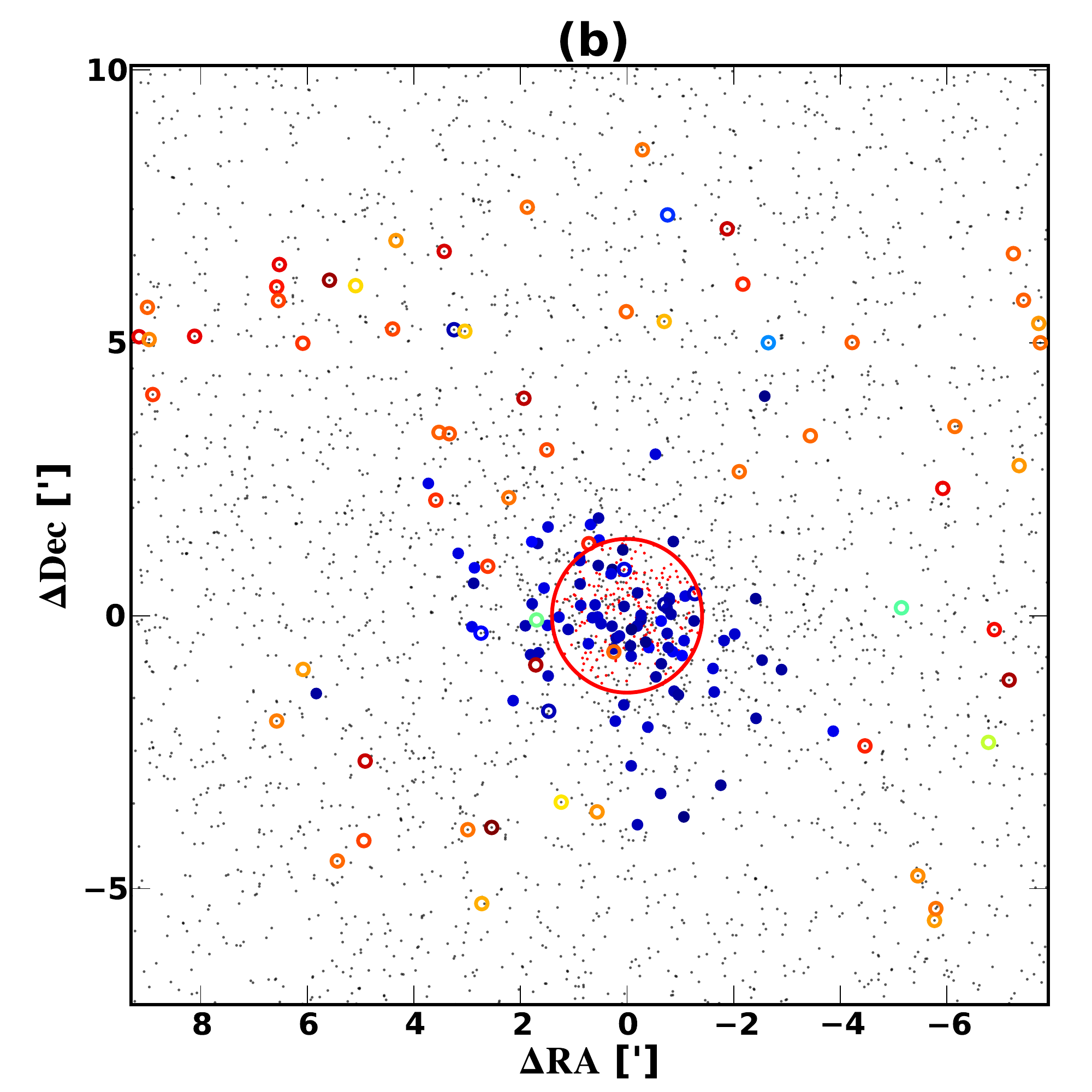}
 \plotone{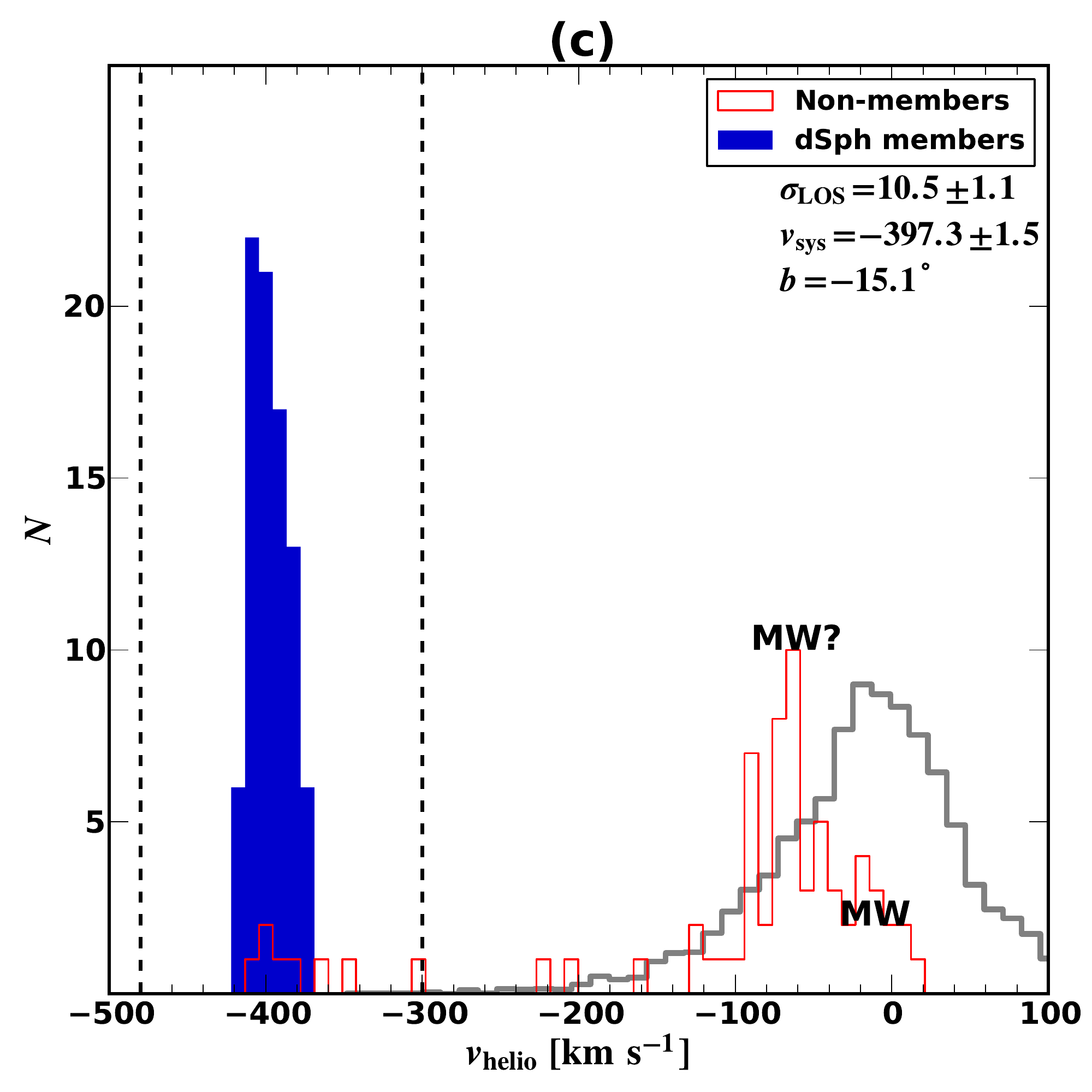}
 \plotone{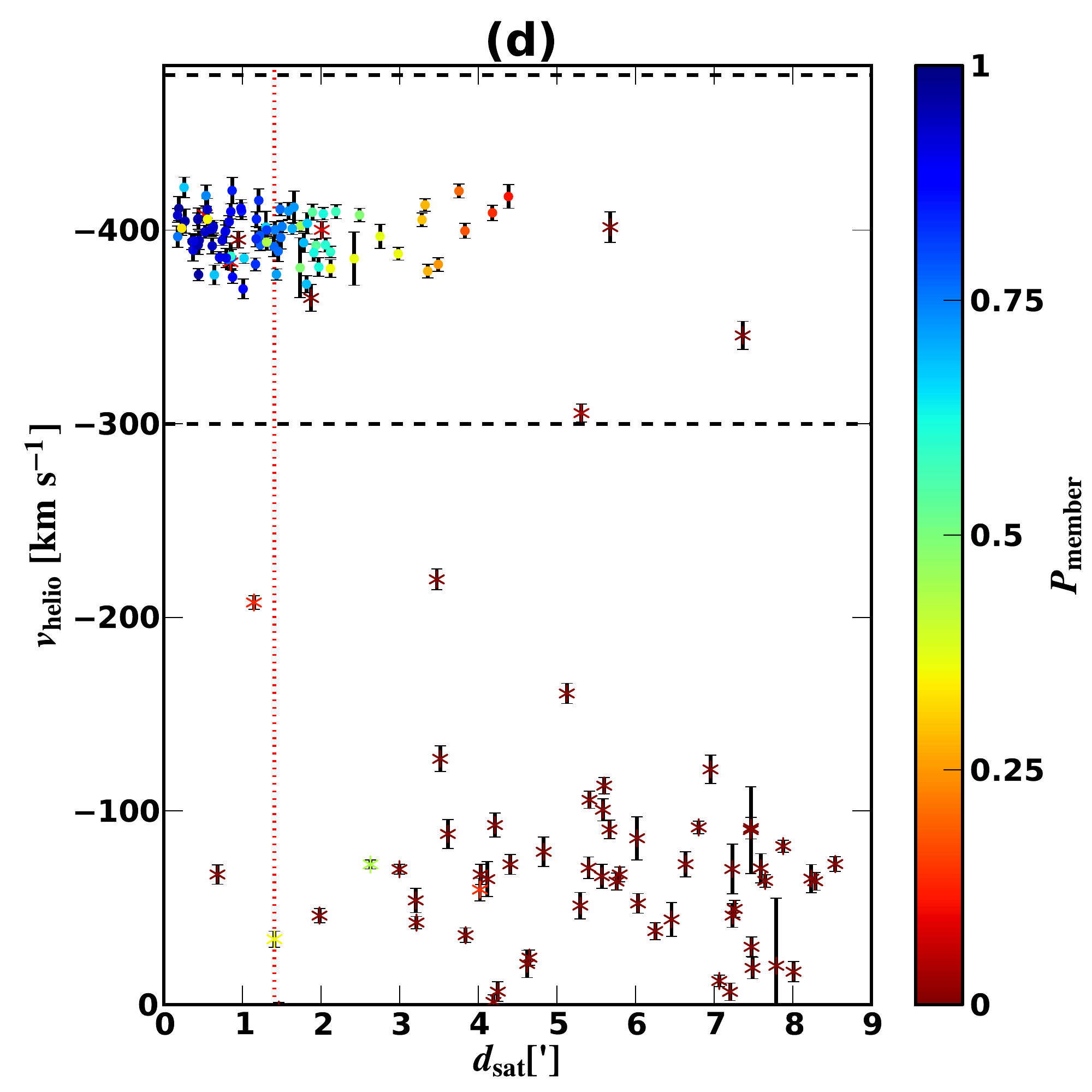}
 \caption{Same as Figure \ref{fig:and1plots}, but for And V.}
 \label{fig:and5plots}
\end{figure*}

We present our results for And V (discovered by \citealt{Armandroff98}) in Figure \ref{fig:and5plots}.  And V ($M_V=-9.6$, $L_V=5.9 \times 10^5 L_\odot$) shows a pronounced and very clean cold spike near $-400$ \kps, and our kinematic parameter estimates show no significant dependence on $\eta$.   Because the velocity peak is well away from any contaminants in velocity space, and this dSphs is some distance on the sky from M31 ($8.0^{\circ}$, or $110$ kpc projected), the membership and parameter estimates are quite secure.  Thus, for this dSph, we use our fiducial membership parameters.  Additionally, our derived systemic velocity is consistent within error bars of the value adopted in \citet{evans00} from the data set of \citet{guhathakurta00}. 

We also note the presence of another possible cold spike in this field at $\sim -70$ \kps.   At such low heliocentric velocity, this feature is likely to be associated with the MW, particularly given the low galactic latitude of And V ($b=-15.1^{\circ}$). One possible candidate may be the Monoceros ring feature \citep{newberg02mon}, as this overdensity has been noted in fields near M31 \citep{ibata03}.  The cold feature in Figure \ref{fig:and5plots} has a radial velocity consistent with extrapolation of the  velocity--Galactic longitude relation for the Monoceros ring \citep[][Figure 2]{crane03}, but it is not clear that the Mon should have a high density in this field.  Alternatively, the cold spike may be related to the Triangulum-Andromeda (Tri-And) feature (described in more detail in \S \ref{ssec:andix}), but it is not at the expected radial velocity \citep[$v_{\rm helio} \sim -130$ \kps{} for this field, ][]{rp04triand}.  A detailed analysis of the source of this feature is beyond the scope of this paper.

\subsection{And VII}

\begin{figure*}[tbp!]

\begin{center}
 \bf \large And VII
 \end{center}

 \epsscale{.5}
 \plotone{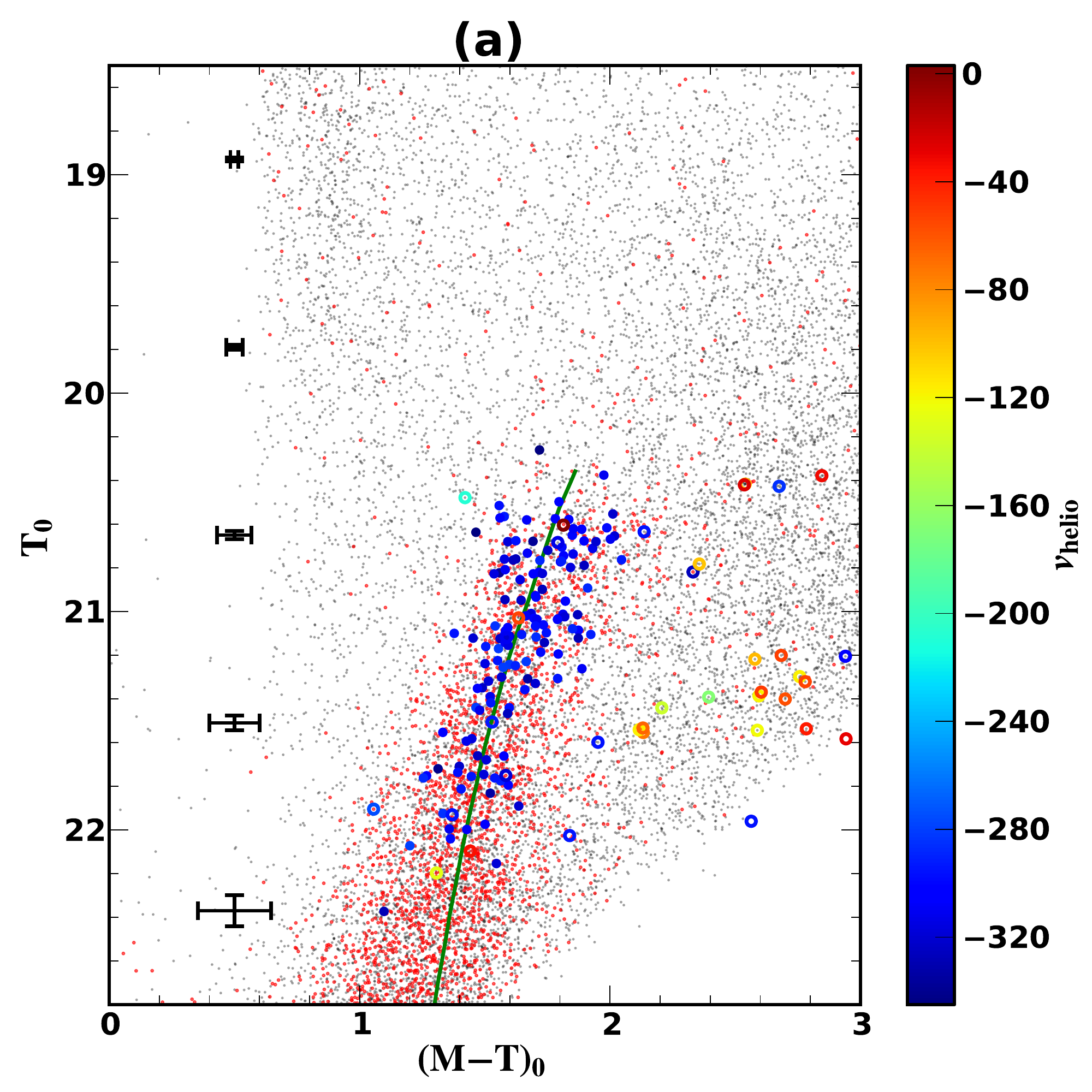}
 \plotone{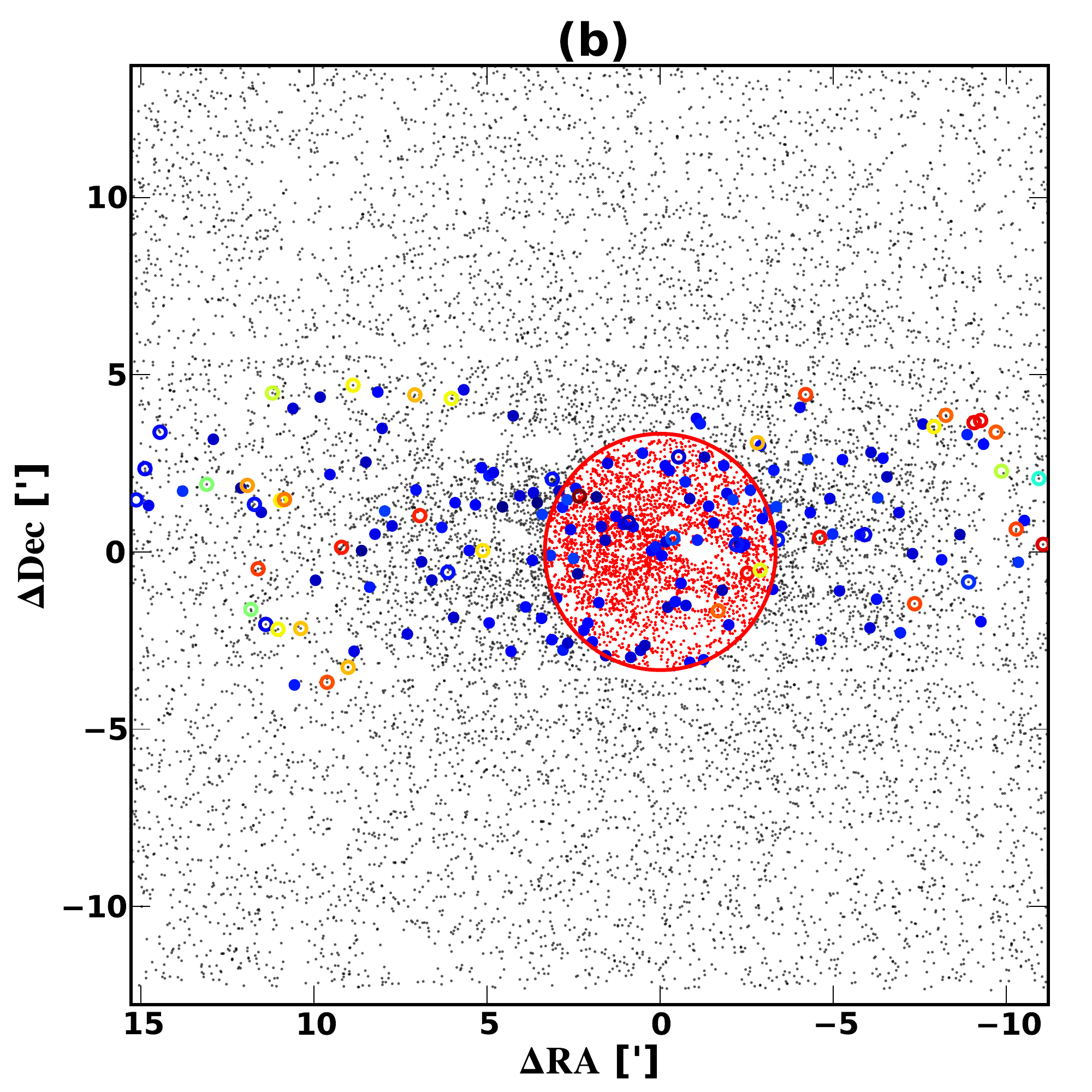}
 \plotone{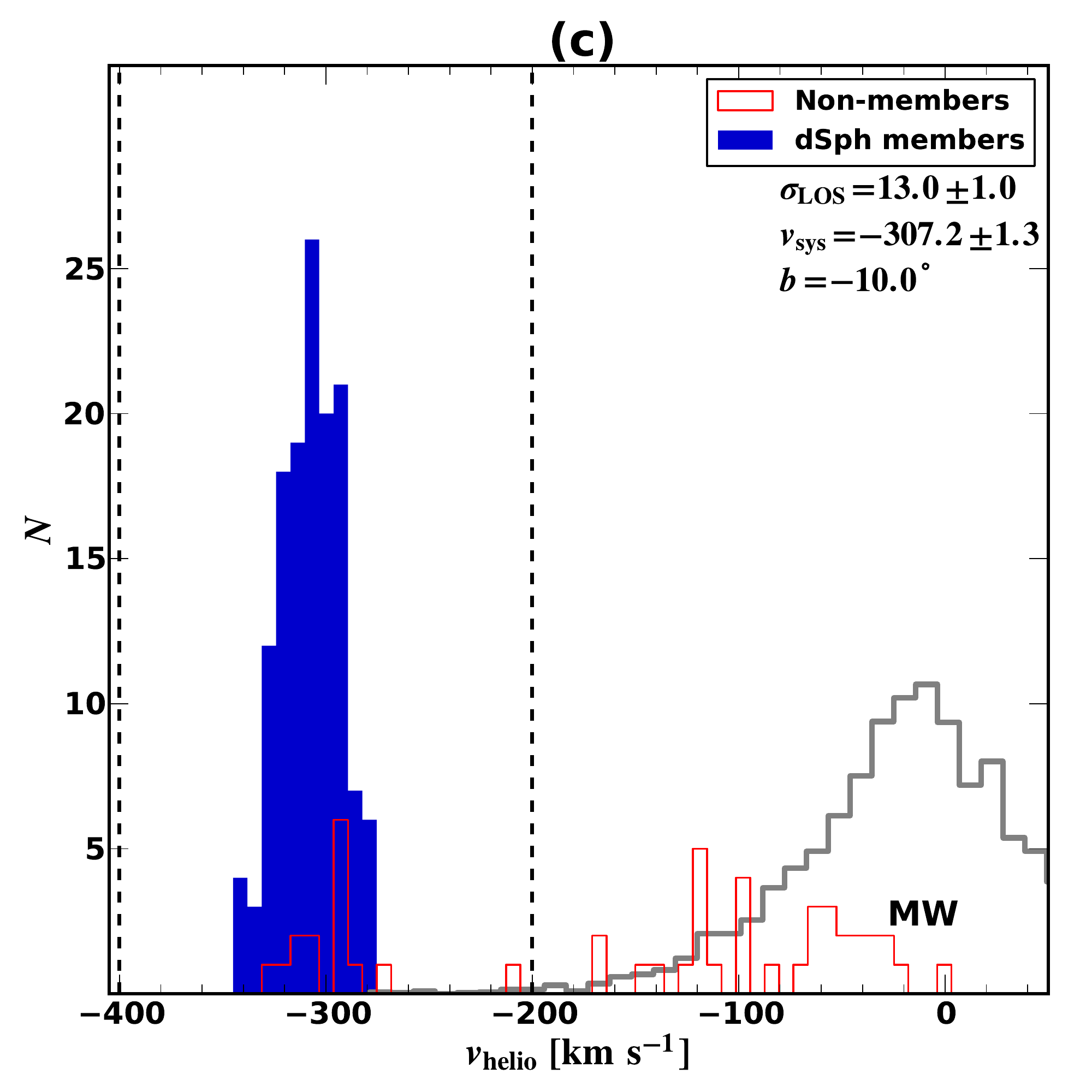}
 \plotone{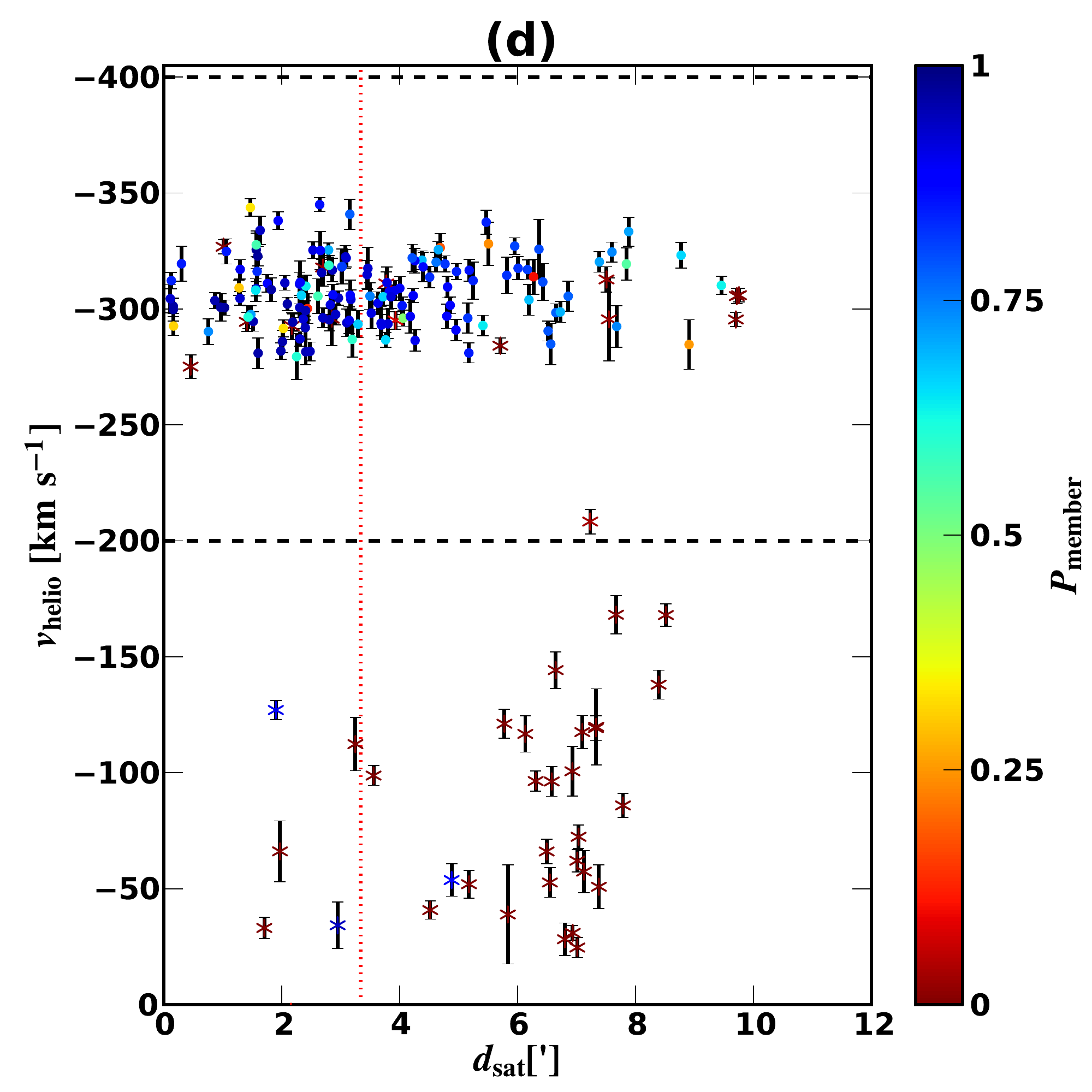}
 \caption{Same as Figure \ref{fig:and1plots}, but for And VII.}
 \label{fig:and7plots}
\end{figure*}

\begin{figure}[tbp!]
 \epsscale{1}
 \plotone{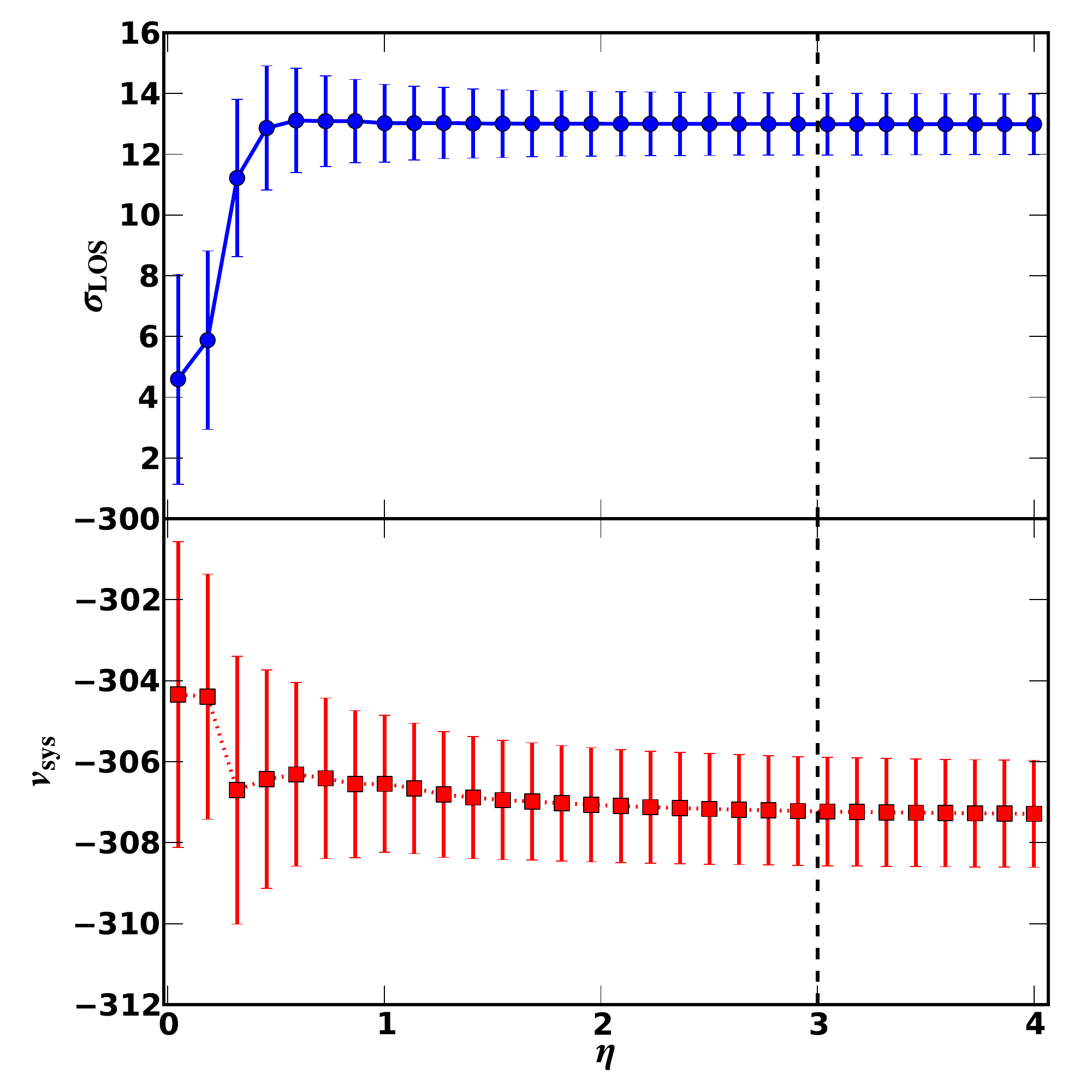}
 \caption{Same as Figure \ref{fig:and1cog}, but for And VII.  The $x$-axis has been expanded relative to Figure \ref{fig:and1cog}.}
 \label{fig:and7cog}
\end{figure}

And VII, discovered by \citet{kk99}, is the brightest of M31's dSph companions, at $M_V=-13.3$, $L_V=1.8 \times 10^7 L_\odot$ (although M31's dE satellites are only 1-2 mags brighter).  We present our results in Figure \ref{fig:and7plots}.  A cold spike is immediately apparent near $-300$ \kps.  While this velocity is close to that of the peak of the M31 halo, And VII has a very large projected distance from M31 ($16.2^{\circ}$, or $220$ kpc projected), and hence M31 halo contamination is likely negligible for this galaxy.  Additionally, as is apparent from Figure \ref{fig:and7cog}, our recovered kinematical parameters are nearly independent of $\eta$, motivating our choice of $\eta=3$.  Because this dSph is so bright, the RGB is very well populated, and the tails of the metallicity distribution are well populated, we increase our isochrone color width $\sigma_c=0.25$.  Hence, most of our spectra for these masks are classified as likely members, even well beyond the half-light radius, and these stars cluster tightly in the cold peak.  This further underscores the reliability of our DDO51 pre-selection's ability to select giants over dwarfs when there are enough giants available.   Additionally, our recovered $\vsys$ is almost identical to that of \citet{evans00} and \citet{guhathakurta00}.

\subsection{And IX}
\label{ssec:andix}

\begin{figure*}[tbp!]

\begin{center}
 \bf \large And IX
 \end{center}

 \epsscale{.5}
 \plotone{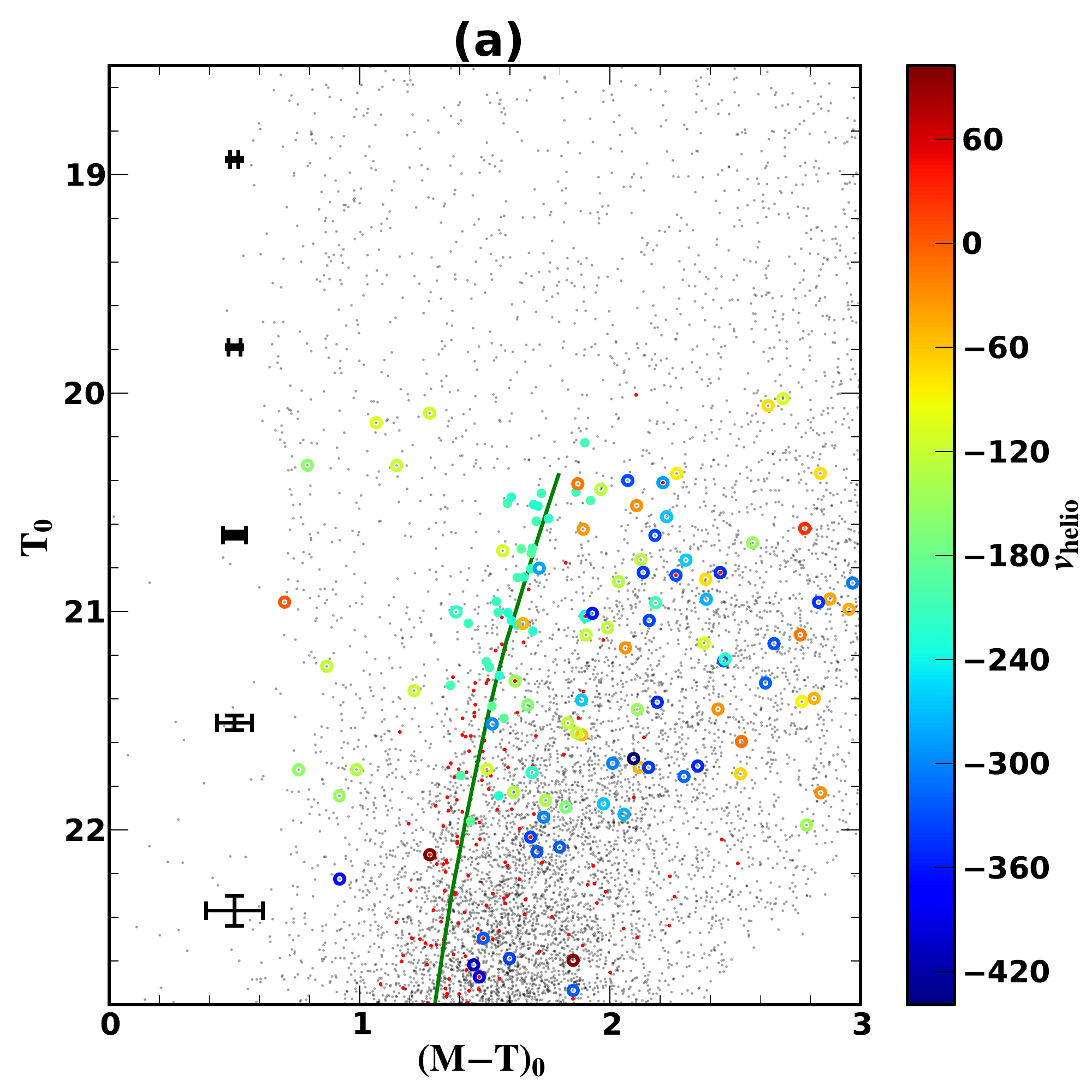}
 \plotone{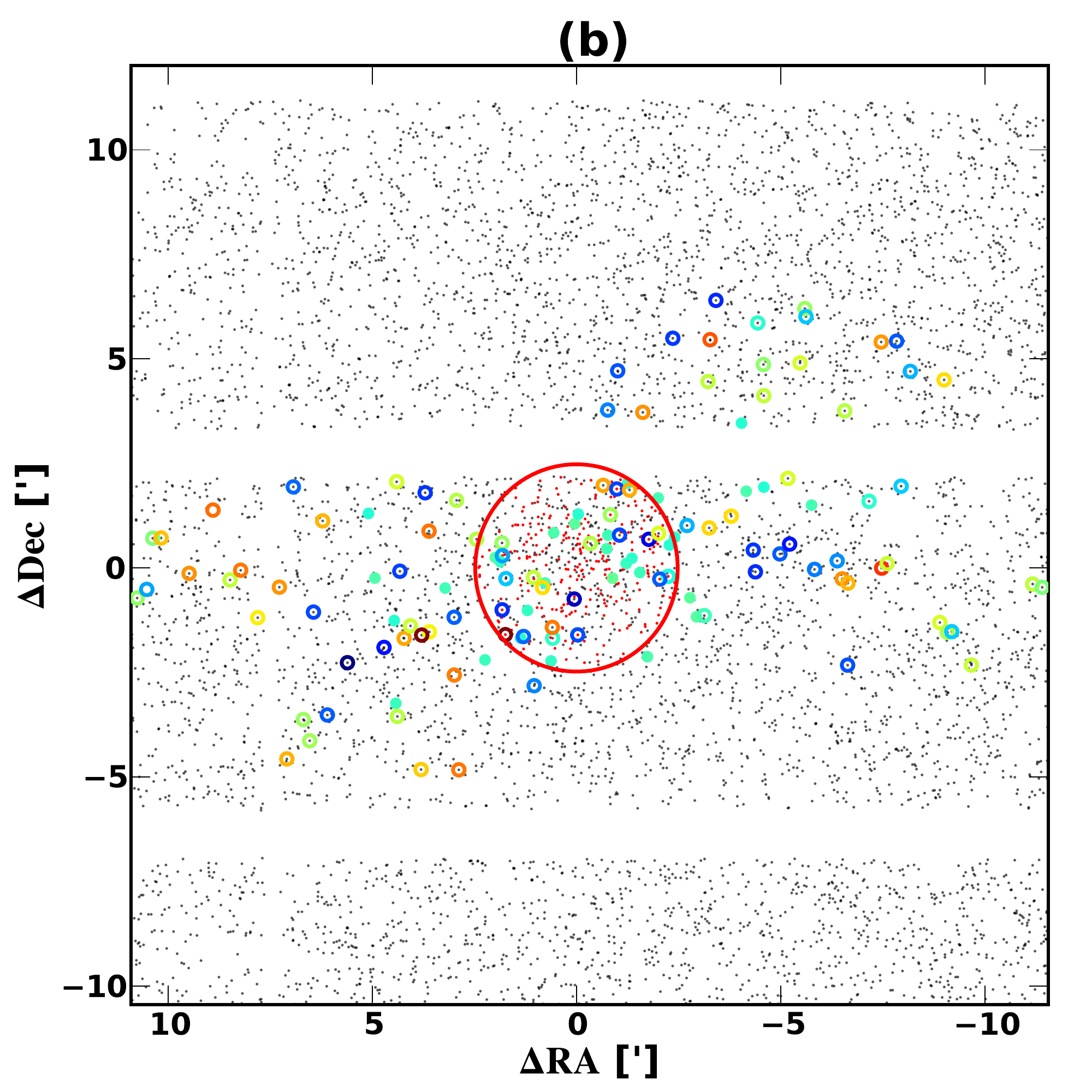}
 \plotone{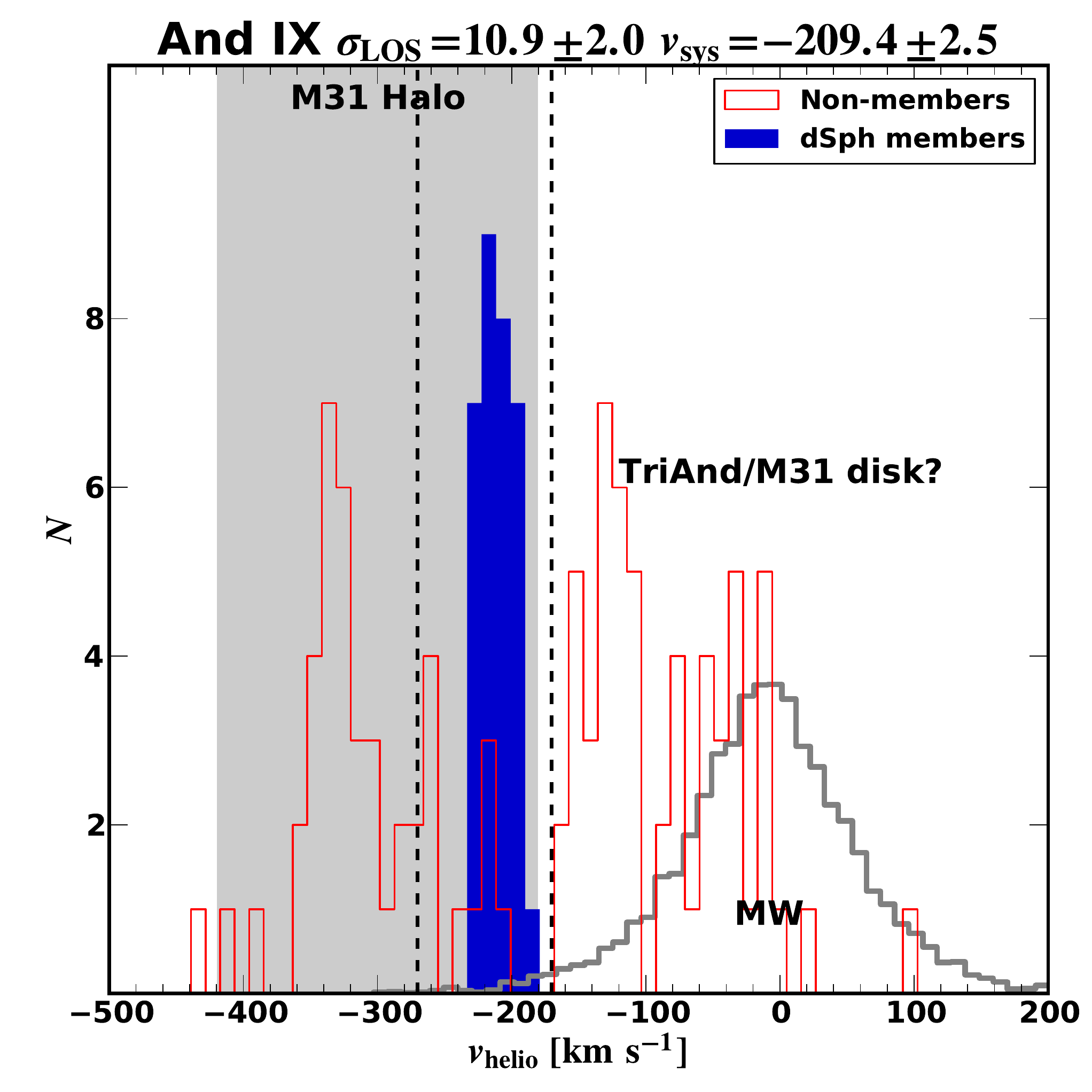}
 \plotone{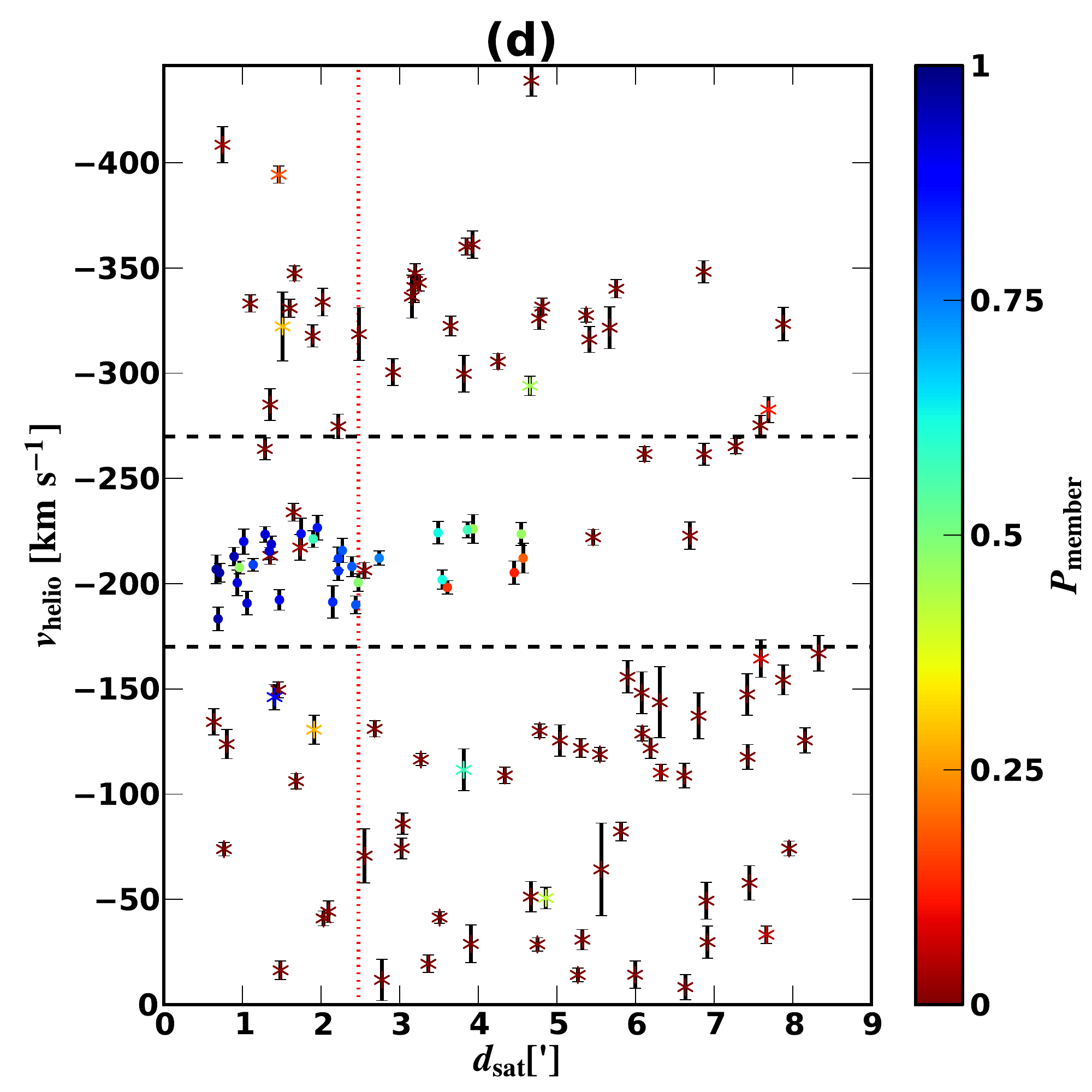}
 \caption{Same as Figure \ref{fig:and1plots}, but for And IX.}
 \label{fig:and9plots}
\end{figure*}

And IX, discovered by \citet{zucker04},  is one of the fainter M31 dSph we consider here ($M_V=-8.1$, $L_V=1.5 \times 10^5 L_\odot$). We present our observations  in Figure \ref{fig:and9plots}.  A cold spike is present near $-200$ \kps, but the velocity histogram also shows  spikes near $-330$ and $-130$ \kps.  The former is consistent with M31 contamination, strong in this field because And IX is quite close to M31 ($2.7^{\circ}$, or $37$ kpc projected).  This first peak is most likely a mix of smooth halo stars with perhaps a contribution from features such as the NE shelf \citep{fardal07}.    The spike near $-130$ \kps, however, is far from the expected peak for the M31 halo, while the MW distribution for red stars in this field peaks at -20 \kps \citep{besancon}.  A possible explanation for this peak is the Tri-And  overdensity \citep{rp04triand}.  And IX lies near a high density region of the TriAnd feature, and given the inhomogeneity of the TriAnd feature, a significant TriAnd population in this field is quite plausible.  Additionally, this peak is fully consistent with the systemic velocity and width of TriAnd \citep[][Figure 4; And IX is at $l=123$ and $\vh=-130$ implies $v_{\rm gsr}=46$ \kps for And IX]{rp04triand}.  An alternative explanation for this feature may be an extension of the M31 disk population, given And IX's proximity to M31.  The radial velocity of the peak broadly matches the expectations from \citet{ibata05} for this field, but some of the stars in the feature are in parts of the CMD that are not consistent with M31 expectations.  Thus, it is likely that this peak is a mixture of TriAnd and M31 disk stars.

Both of the aforementioned peaks are nearly uniformly spread across the field, however, while  stars near $-200$ \kps are spatially concentrated at the photometric center of the satellite.  
Thus, we take the peak at $-200$ \kps as the peak for And IX, consistent with the results of   \citet{chapman05and9}. While our membership  criteria eliminate most stars in the other two peaks, a small number of outliers nevertheless pass our membership cuts (while no starlist is published  in \citealt{chapman05and9}, this likely includes the outlier mentioned based on the distance given in that work).  Thus, for this satellite, we impose more stringent final velocity restrictions (vertical dashed lines in Figure \ref{fig:and9plots}) to filter out these outliers, as they are consistent with falling within the distributions of properties for the other populations.  We note that kinematic results we obtain with this procedure differ at the $1-2 \sigma$ level from   \citet{chapman05and9} and \citet{collins10}, although our sample has roughly twice the number of member stars.

\subsection{And X}

\begin{figure*}[tbp!]

\begin{center}
 \bf \large And X
 \end{center}

 \epsscale{.5}
 \plotone{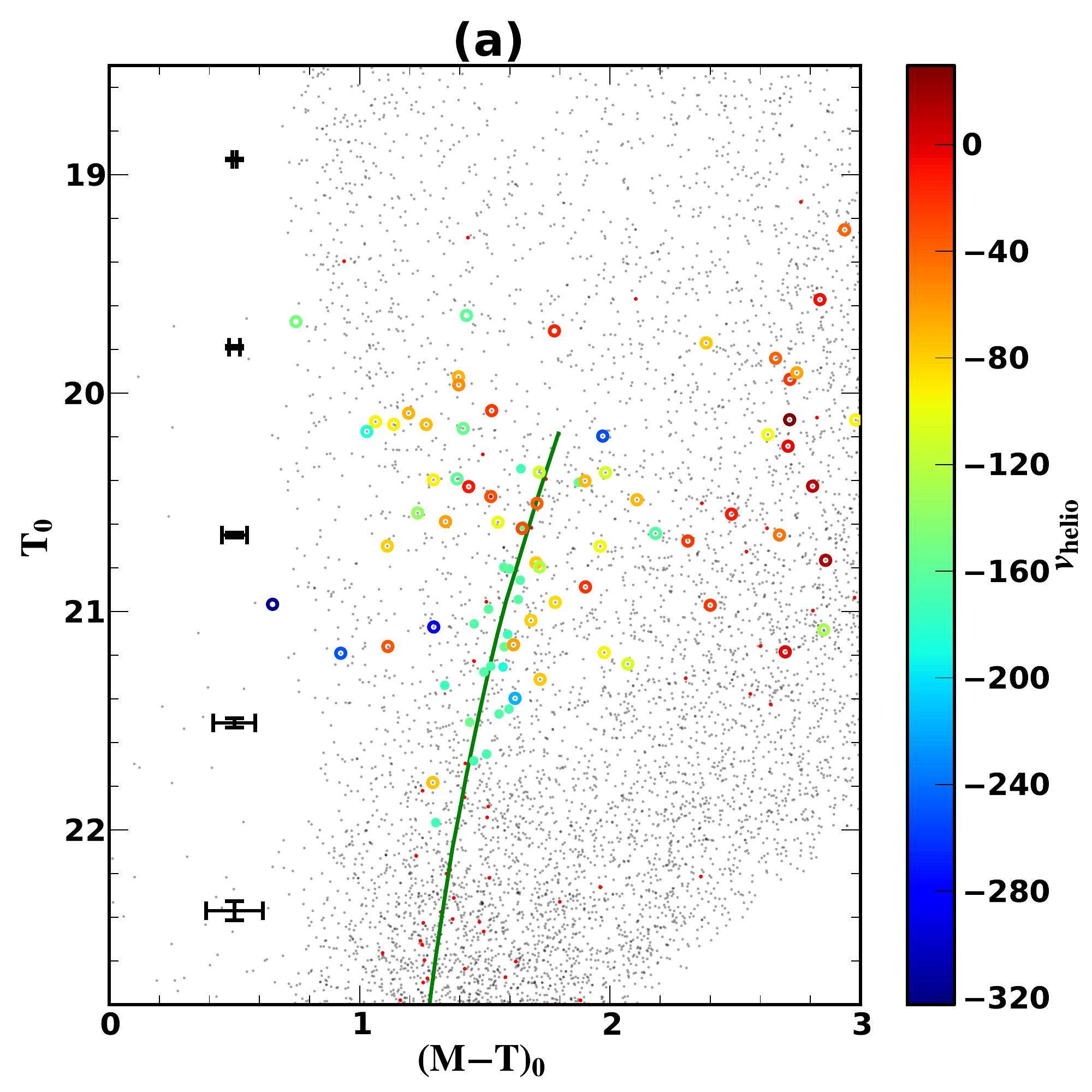}
 \plotone{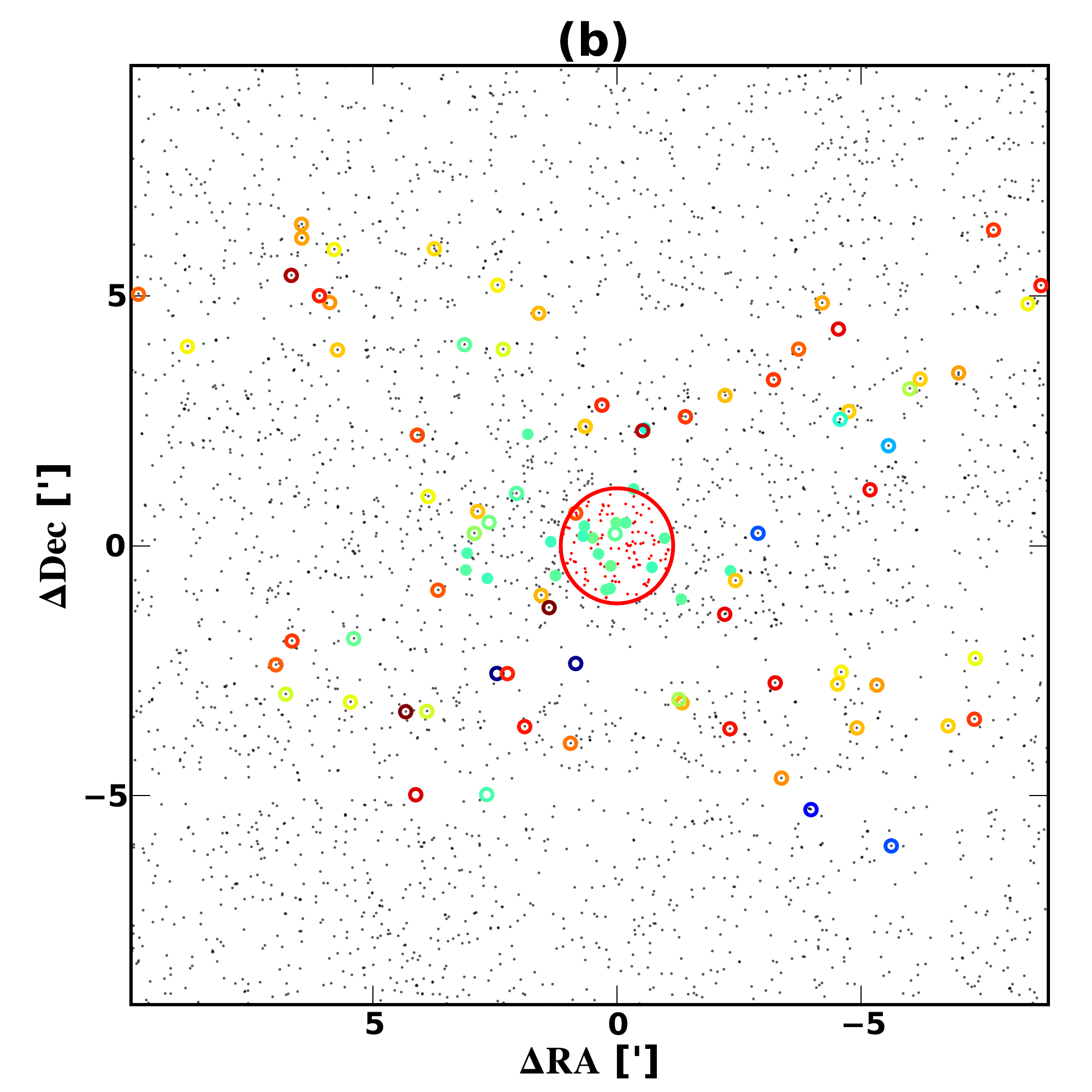}
 \plotone{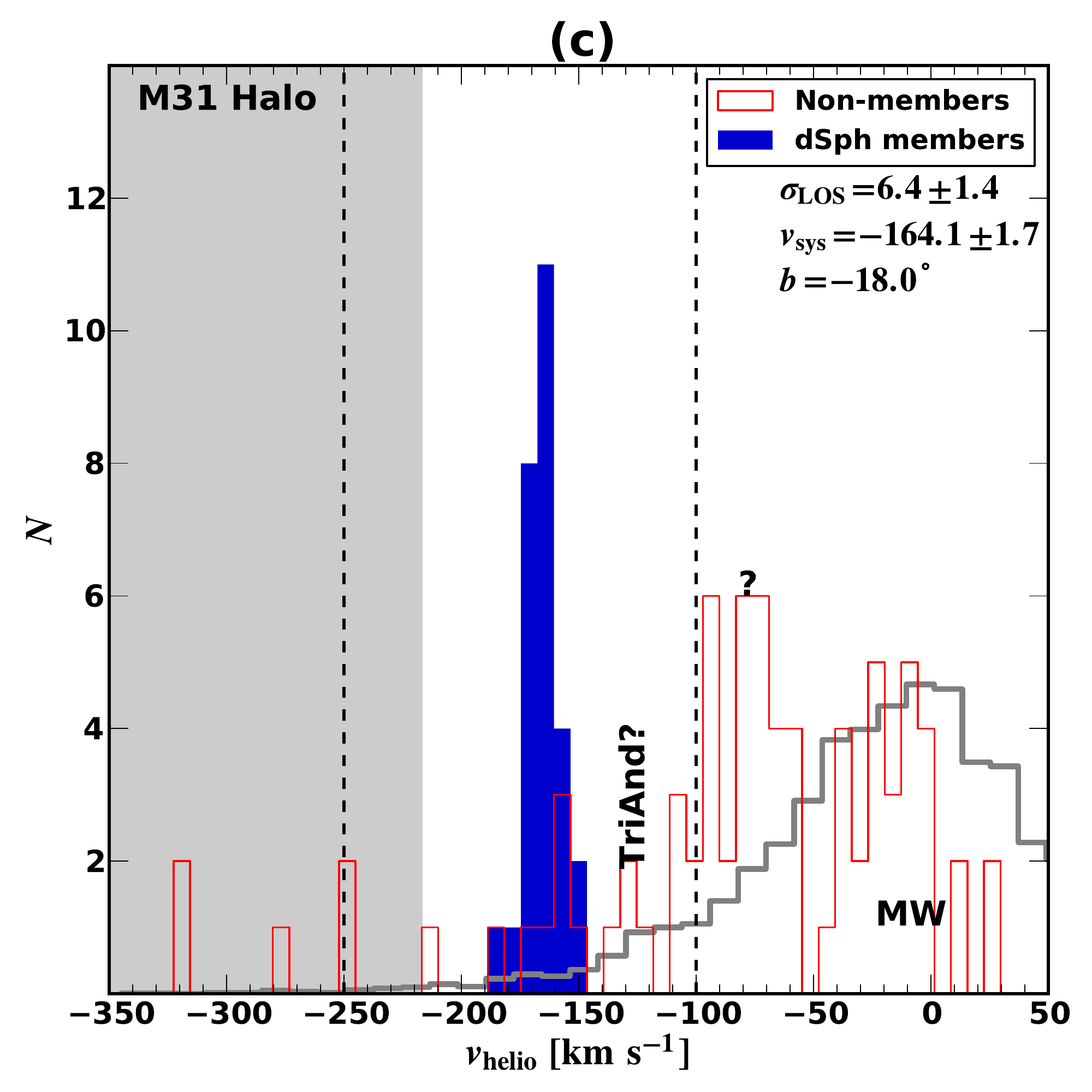}
 \plotone{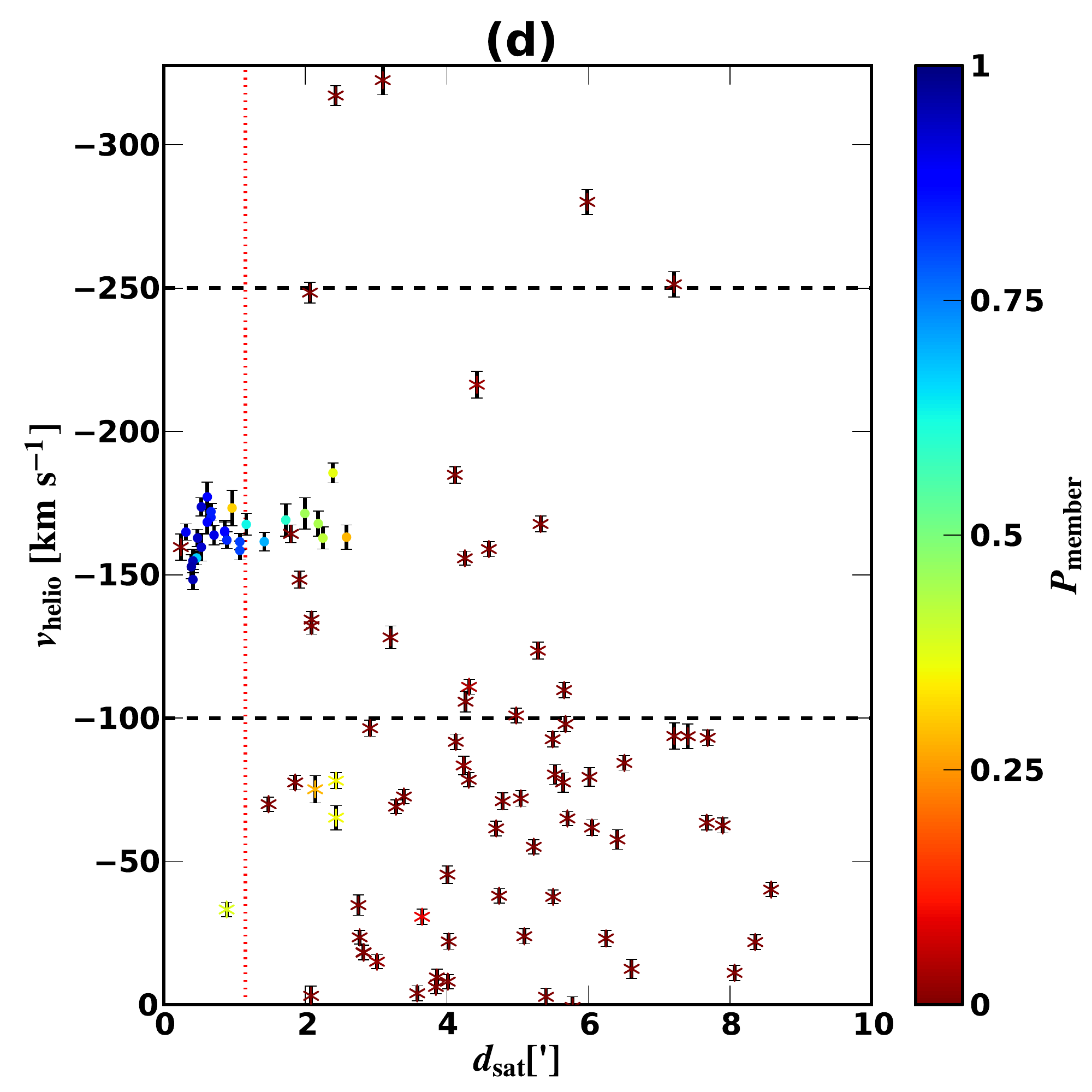}
 \caption{Same as Figure \ref{fig:and1plots}, but for And X.}
 \label{fig:and10plots}
\end{figure*}

Our results for And X (discovered by \citealt{zucker07}, $M_V=-7.4$, $L_V=7.5 \times 10^4 L_\odot$) are shown in Figure \ref{fig:and10plots}. A clear cold spike is apparent near $-165$ \kps.  This field is significantly farther from the center of M31 relative to And IX ($5.6^{\circ}$, or $77$ kpc projected), and hence the M31 halo contamination is much lower.  We thus use our fiducial parameters, as changing $\eta$ does not affect our kinematical results.  We note that the dispersion we find here is $1.4\sigma$ from the result of \citet{kalirai09andx} using the same observations, due to the fact that our method does not formally reject all of the velocity outliers mentioned in \citet{kalirai09andx}.  If we reject outlier stars in the same fashion, we reproduce a very similar sample as that paper and nearly identical kinematic results.

We also note the presence of another peak in the histogram well offset from the MW expectation at  $\sim -80$ \kps, and a possible smaller peak at $\sim -130$ \kps.  This field is in a relatively high density portion of the TriAnd feature, and the kinematics of \citet{rp04triand} suggest this latter peak is due to this feature.  However, the broader $\sim -80$ \kps{} peak is well offset from the expected velocity for this feature, suggesting either TriAnd has a wider velocity dispersion than previously though, or this peak is a different (most likely MW) substructure.

\subsection{And XI}

\begin{figure*}[tbp!]

\begin{center}
 \bf \large And XI
 \end{center}

 \epsscale{.5}
 \plotone{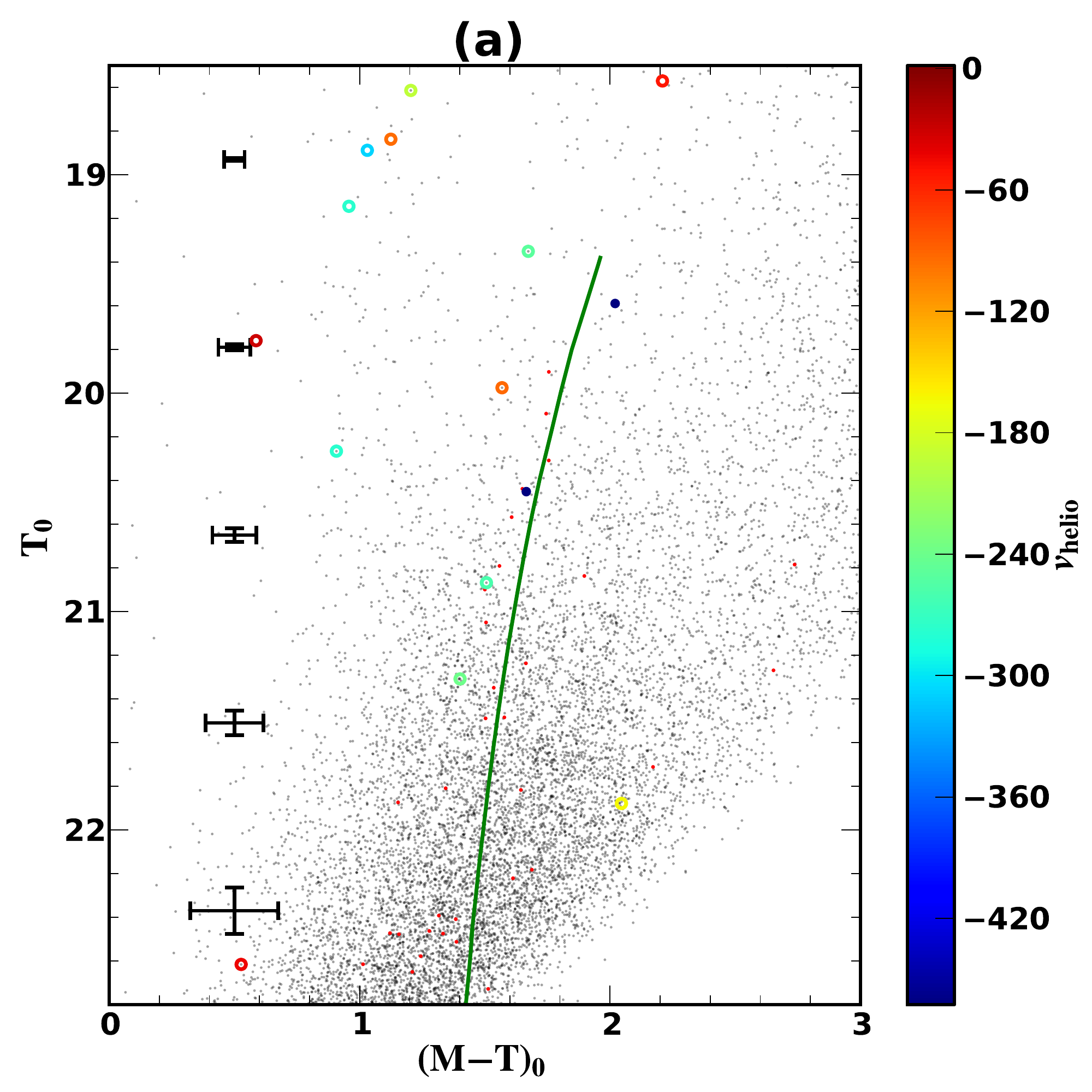}
 \plotone{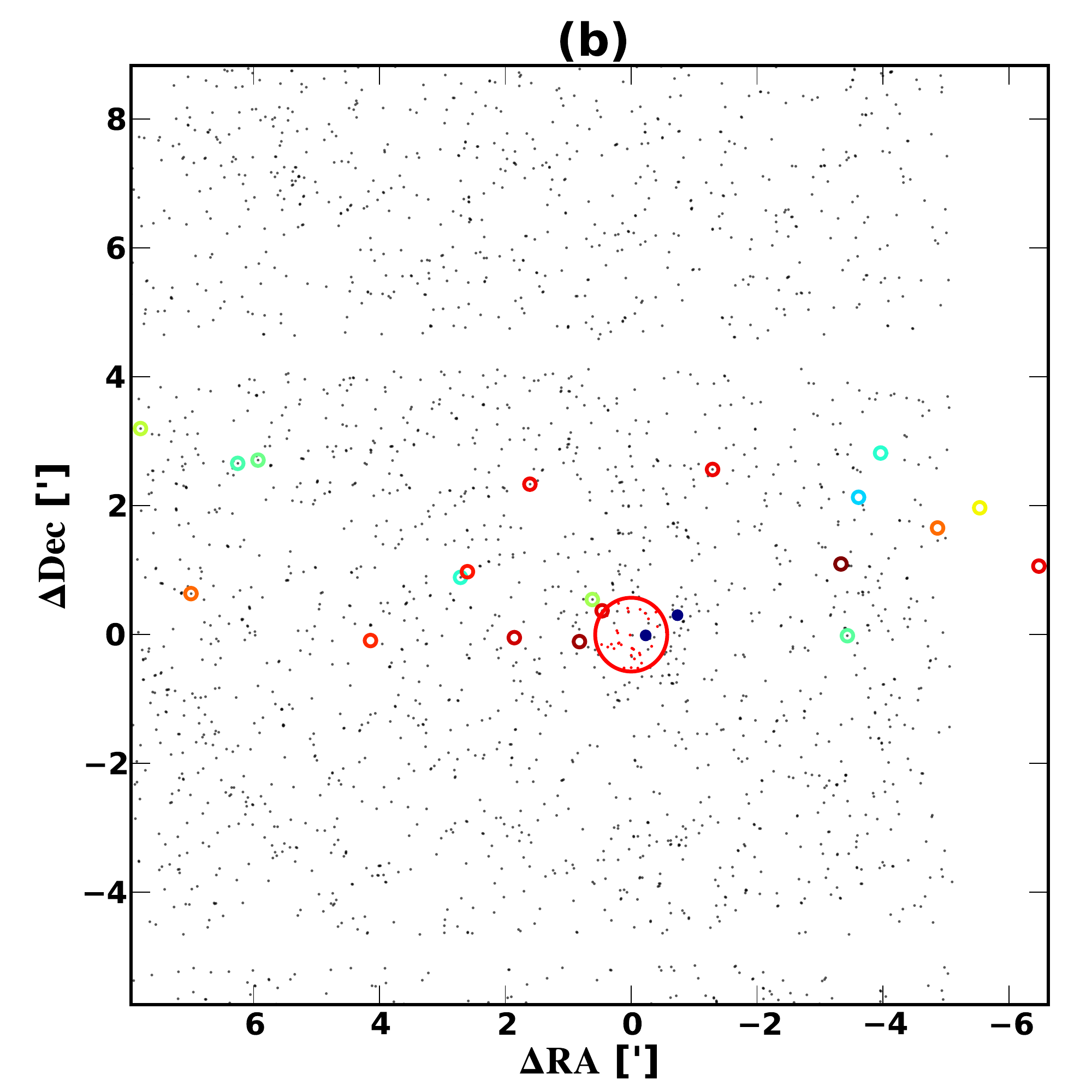}
 \plotone{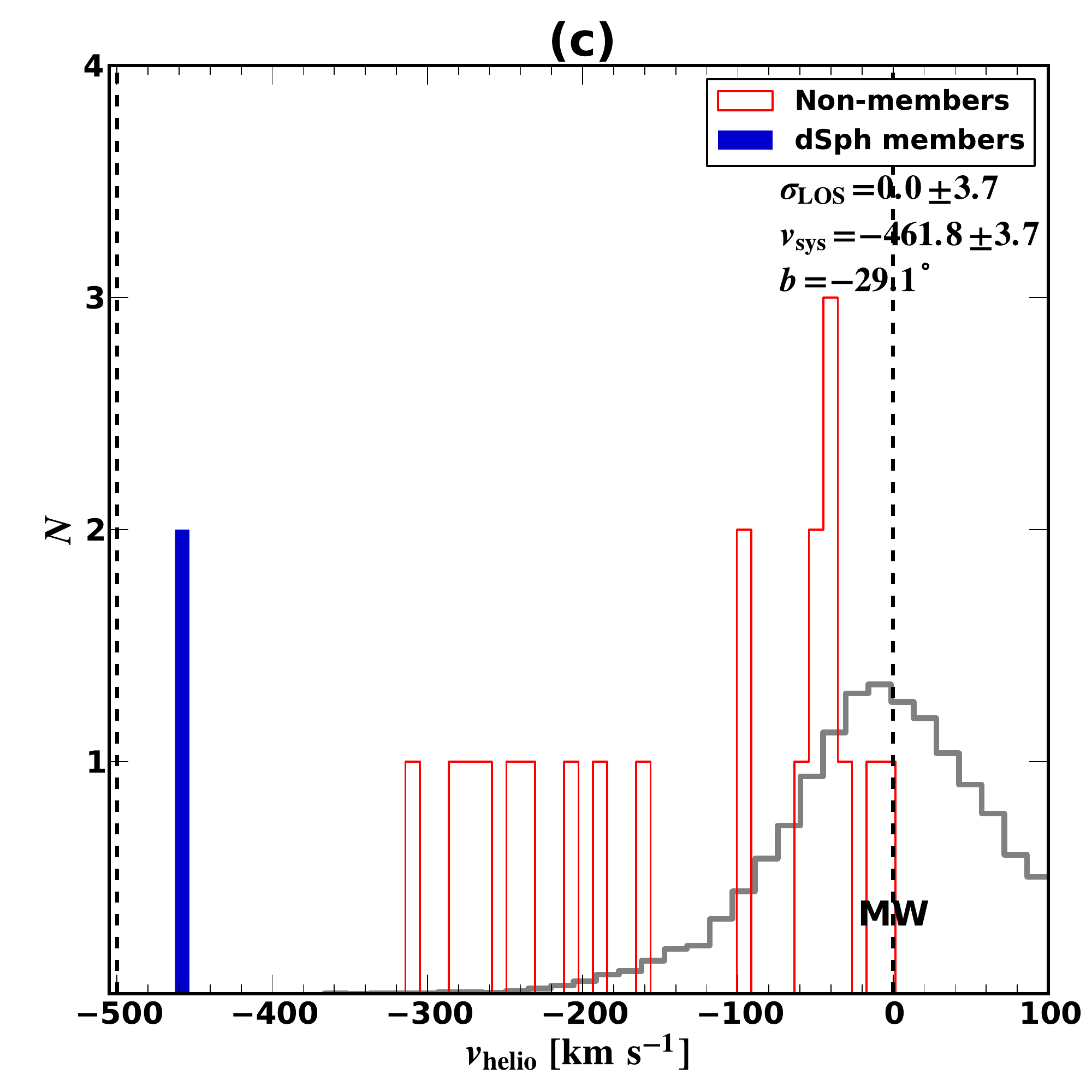}
 \plotone{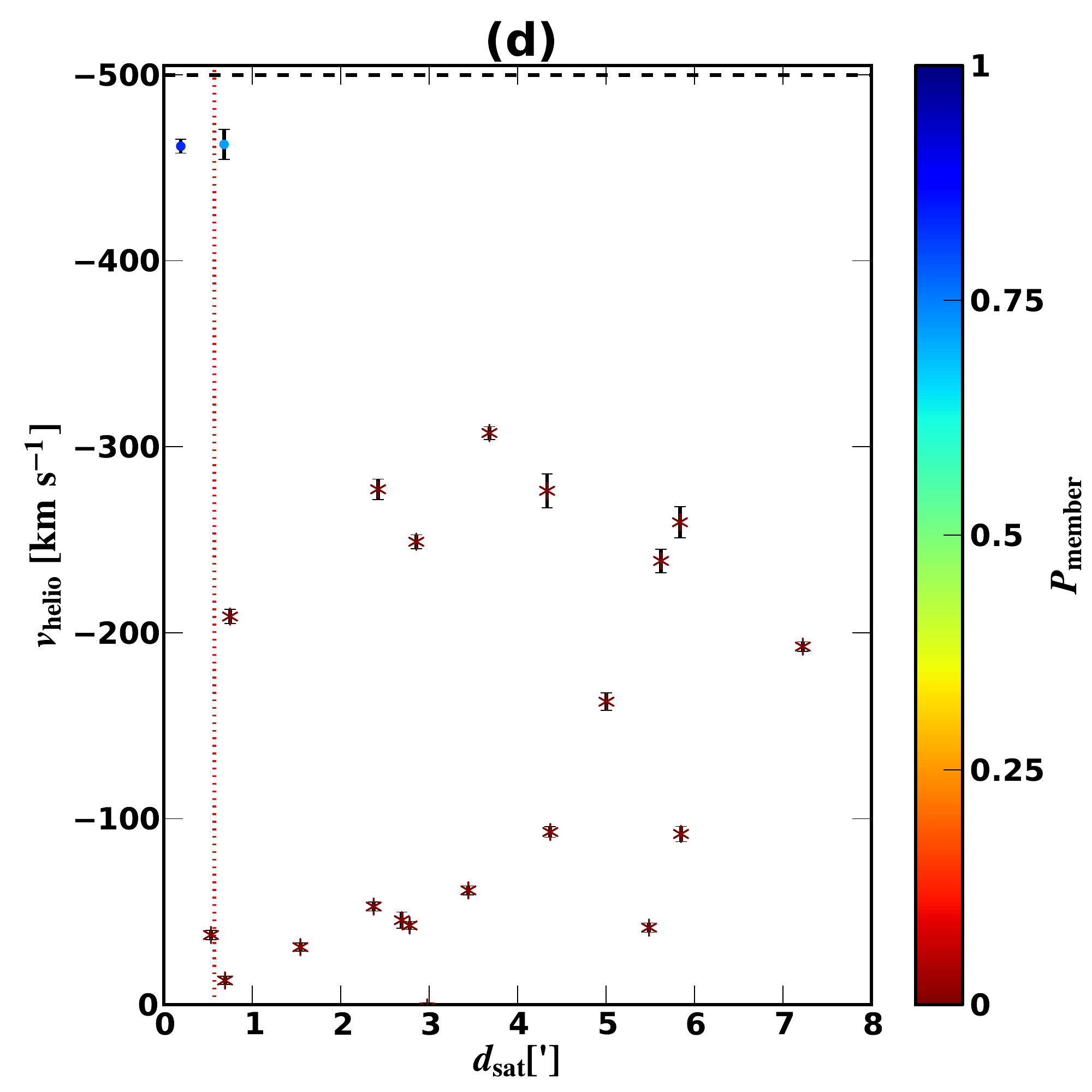}
 \caption{Same as Figure \ref{fig:and1plots}, but for And XI.}
 \label{fig:and11plots}
\end{figure*}

Our spectroscopy for And XI (discovered by \citealt{martin06}) is shown in Figure \ref{fig:and11plots}.  And XI is one of the faintest dSphs in our sample ($M_V=-6.9$, $L_V=4.9 \times 10^4 L_\odot$), and the field is both small and has few bright RGB candidates.  Hence, we do not see an obvious cold spike.  We do obtain spectra of two stars near the center of the dwarf that have very close radial velocities.  The large distance from M31 ($7.5^{\circ}$, or $102$ kpc projected) combined with their proximity to each other and the center of the dSph renders it plausible that these are dSph members rather than M31 halo stars.  Our reported $\vsys$ is derived assuming these two are members.  However, this result is in conflict with \citet{collins10}, which report five stars near $-420$ \kps.  Thus, while we report the mean velocity of these two stars, it is quite possible that these are simple M31 contaminants and we have detected no actual members. 


\subsection{And XII}

\begin{figure*}[tbp!]

\begin{center}
 \bf \large And XII
 \end{center}

 \epsscale{.5}
 \plotone{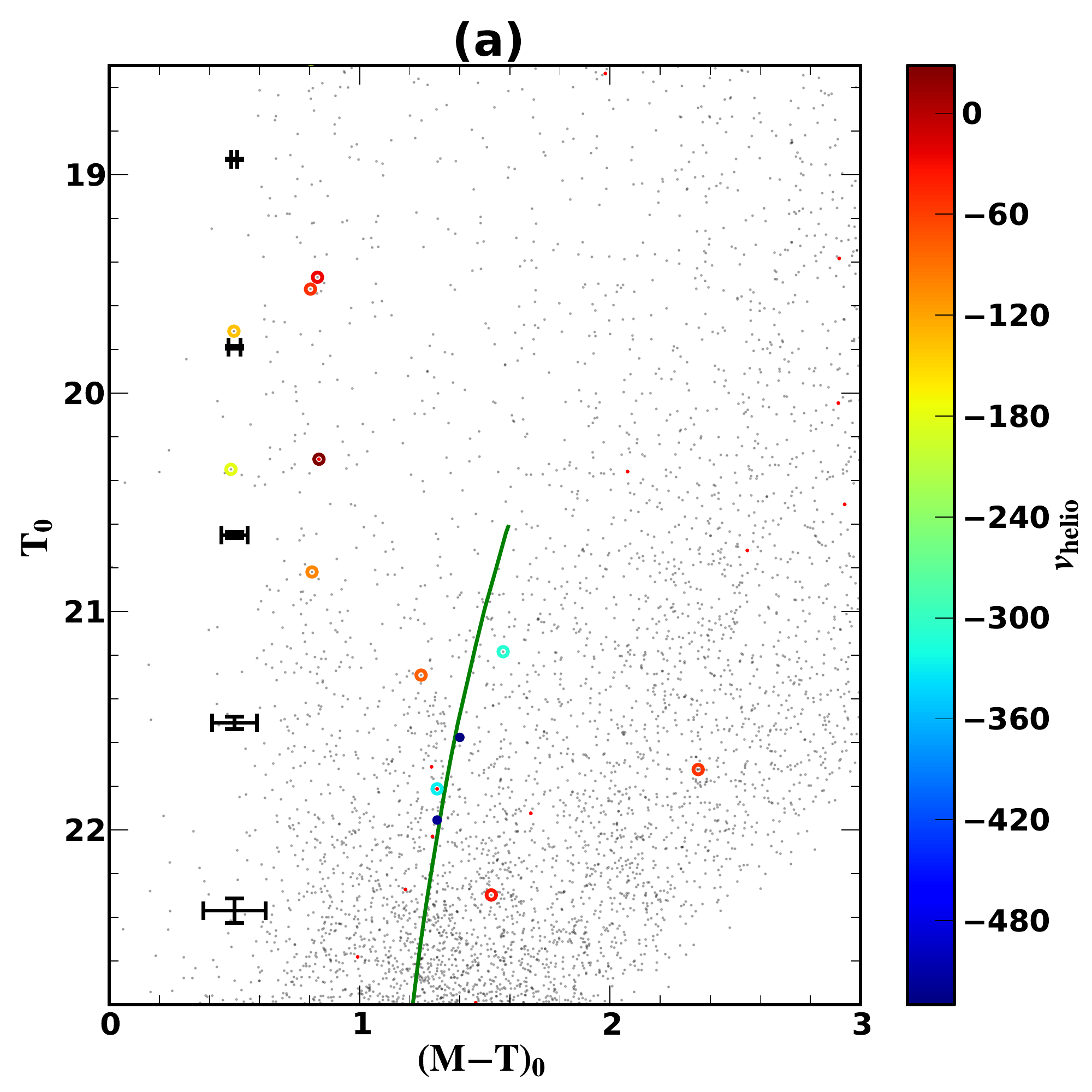}
 \plotone{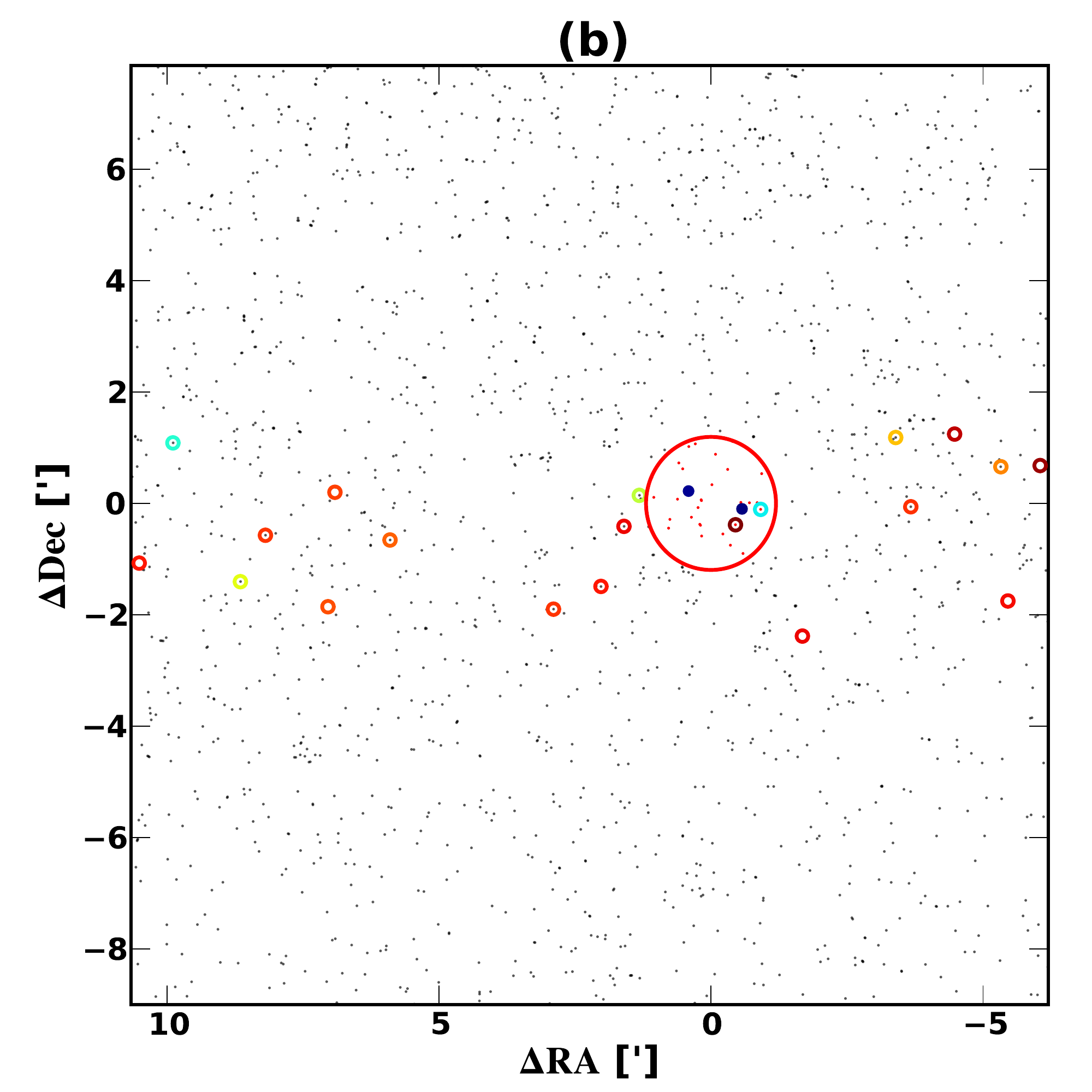}
 \plotone{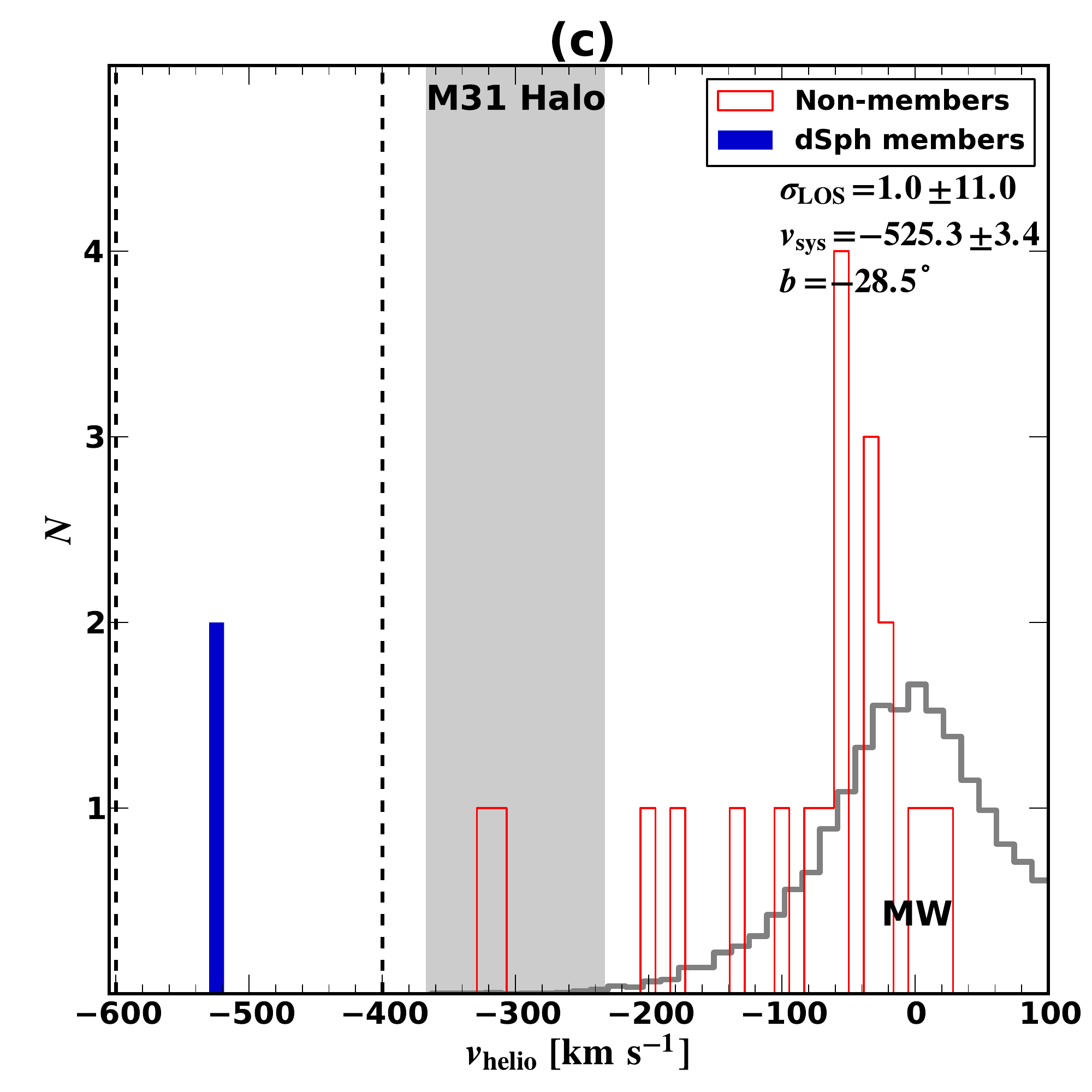}
 \plotone{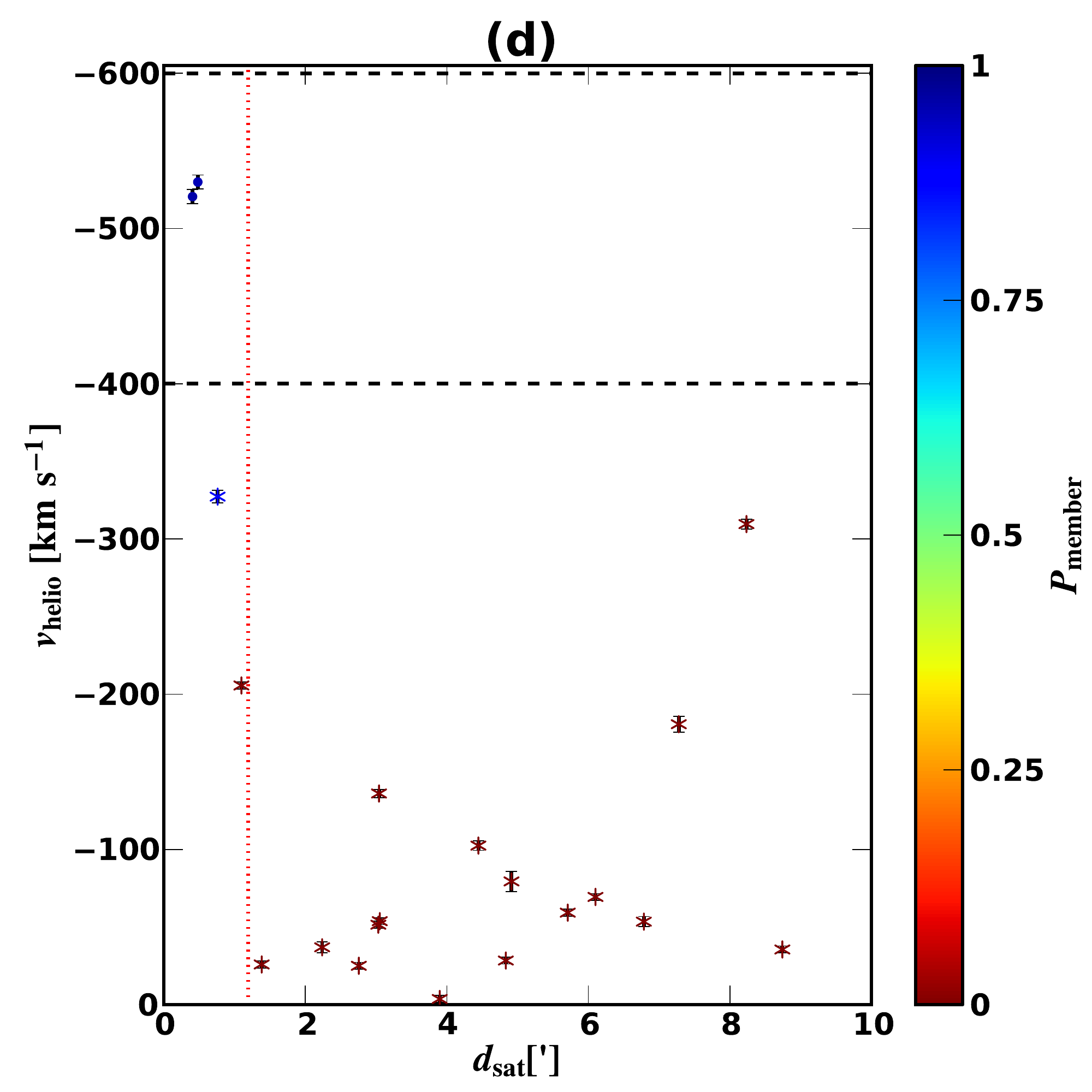}
 \caption{Same as Figure \ref{fig:and1plots}, but for And XII.}
 \label{fig:and12plots}
\end{figure*}

Our spectroscopy for And XII (discovered by \citealt{martin06}) is shown in Figure \ref{fig:and12plots}. As with And XI, this field is very spare due to the faintness of the satellite ($M_V=-6.4$, $L_V=3.1 \times 10^4 L_\odot$) and distance from M31 ($7.0^{\circ}$, or $95$ kpc projected).  We see no clear cold spike, but there are two stars that lie at much more negative velocities than expected for the M31 halo. As with And XI, we report the mean of these two as $\vsys$, and again, this is offset from \citet{collins10} by $\sim 30$ \kps, so these $\vsys$ come with the caveat that they may well be non-members.

\subsection{And XIII}

\begin{figure*}[tbp!]

\begin{center}
 \bf \large And XIII
 \end{center}

 \epsscale{.5}
 \plotone{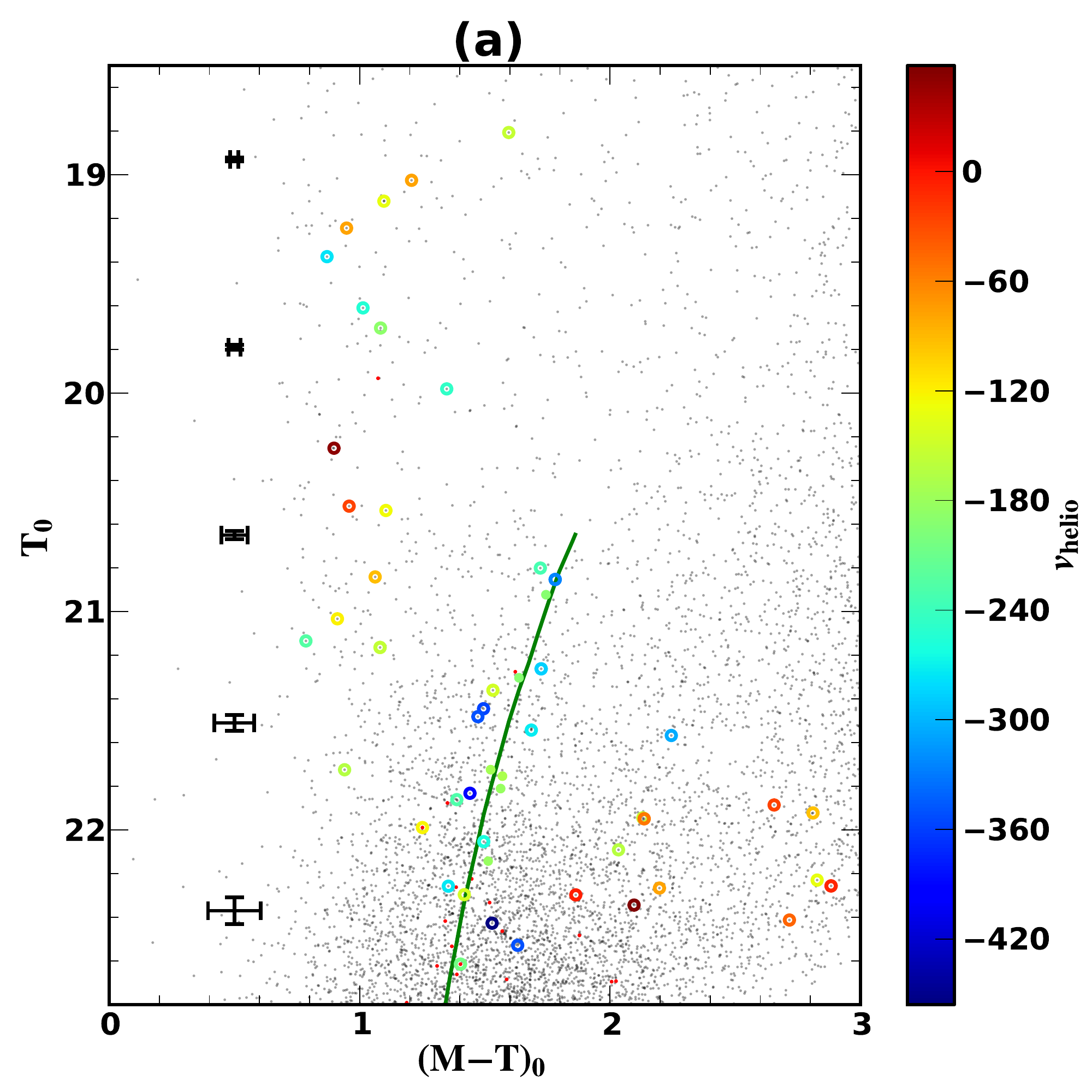}
 \plotone{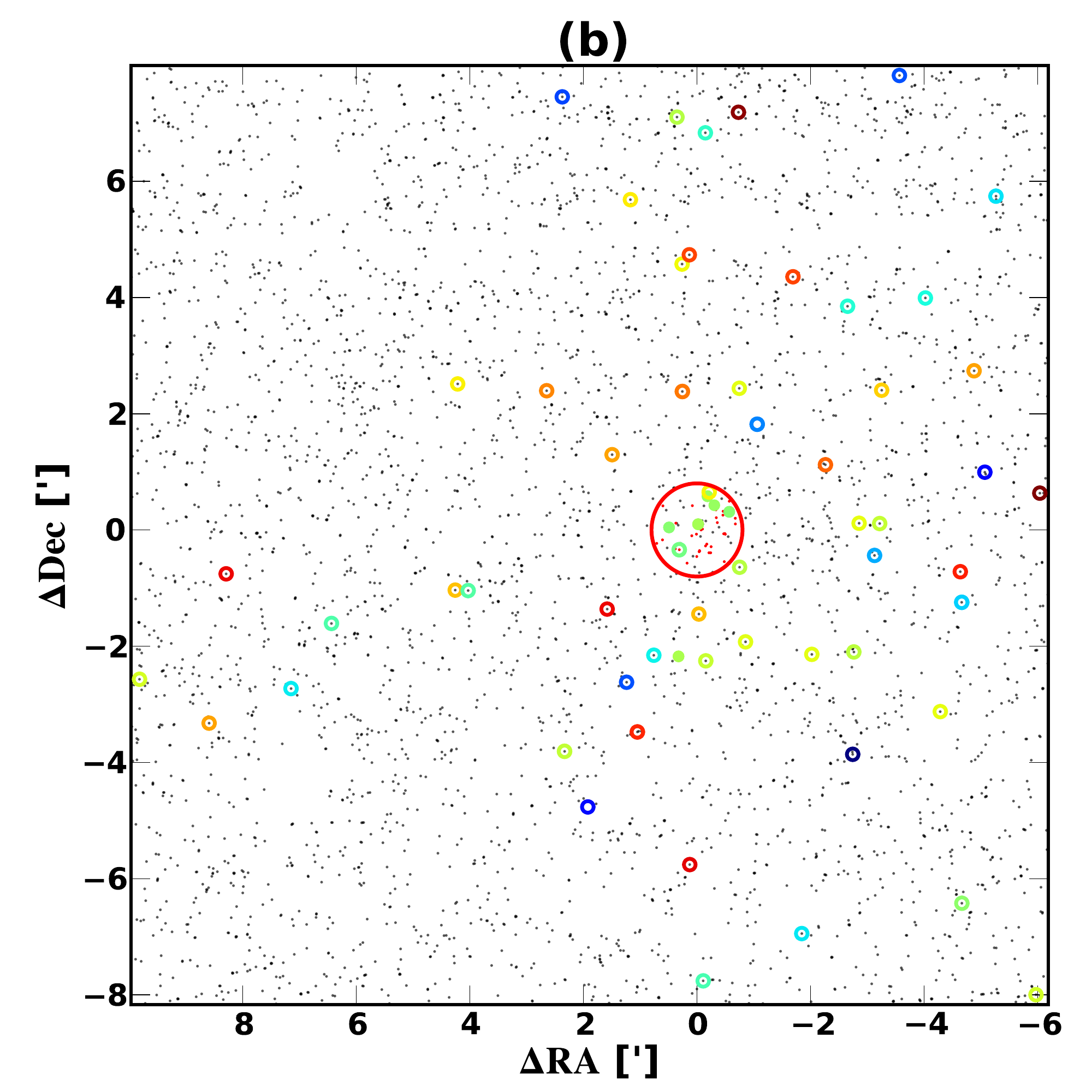}
 \plotone{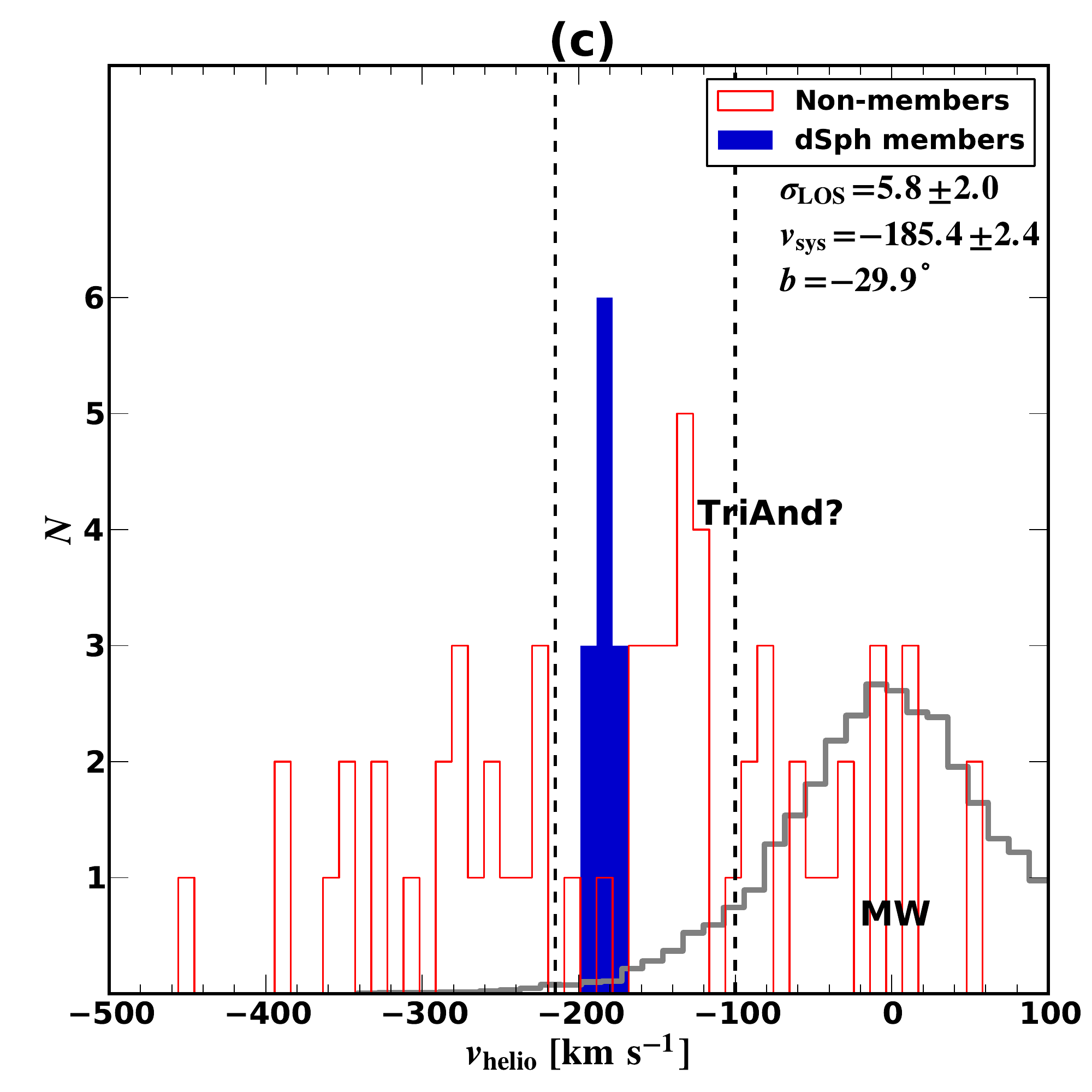}
 \plotone{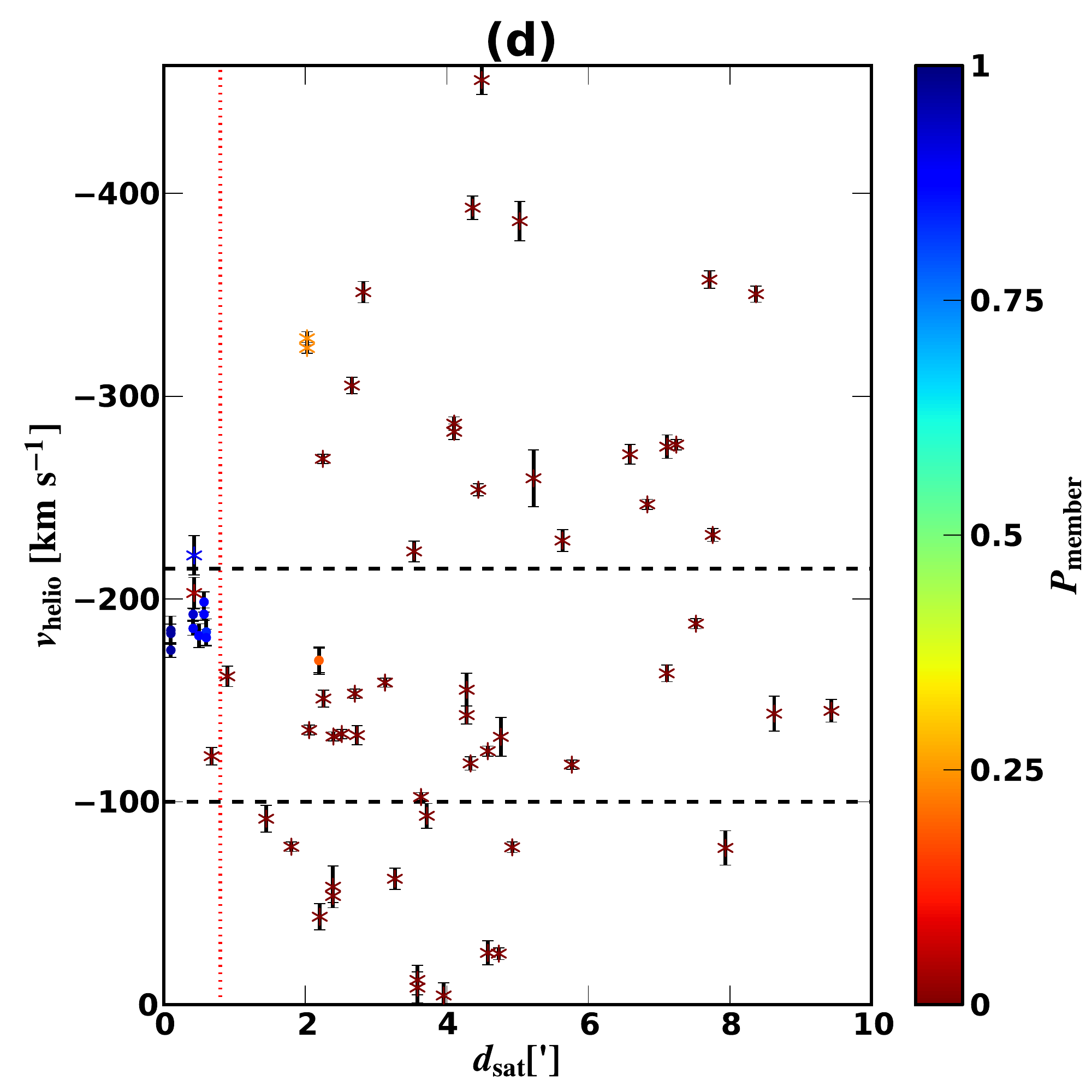}
 \caption{Same as Figure \ref{fig:and1plots}, but for And XIII.}
 \label{fig:and13plots}
\end{figure*}

Our spectroscopy for And XIII (discovered by \citealt{martin06}), another faint dSph ($M_V=-6.7$, $L_V=4.1 \times 10^4 L_\odot$) is shown in Figure \ref{fig:and13plots}. We adopt all fiducial parameters for this dwarf, and see a cold spike of 12 stars that pass membership cuts near $-200$ \kps.  The vertical dashed lines in the velocity histogram clearly indicate that we have explicitly cut out stars not near the cold spike, which may at first appear to artificially shrink the dispersion.  However, this actually only serves to eliminate two stars (all others failed the membership cuts with the fiducial parameter choices).  One of these stars is far removed from the dSph in all three diagrams, and hence is likely an M31 contaminant.  The other star removed by the cut has $v_{\rm helio}=-221$ \kps{} and is very close to the And XIII isochrone, as well as lying within the half-light radius.  However, it is also the faintest star in this spectroscopic sample, with a S/N $ = 1.9$.  Hence, only a single CaT line is detected in the spectrum, and may be biased by nearby sky lines.  We therefore reject that particular star and compute the kinematic parameters from the more well-defined cold spike.  The resulting parameters are roughly $1\sigma$ discrepant from the \citet{collins10} result, but we note that our sample is 4 times larger, and hence likely more robust.

This field also includes a relatively large number of non-members, despite its large distance from M31 ($8.5^{\circ}$, or $116$ kpc projected) and relatively large distance from the Galactic plane ($b=-29^{\circ}$).  Most of this structure is uniform across the field, suggesting that it is either foreground or large-scale halo structure.  In particular, there is a hint of a peak at $\sim-130$ \kps, which is consistent with the expected velocities for the TriAnd feature in this field.  The stars with radial velocities in the $-500$ to $-300$ \kps{} range are very unlikely to be of MW origin, however.  The distance from M31, assuming the halo model of \citet{guhathakurta05}, implies the number of M31 RGB stars present in this field is higher than expected, suggesting that the environs of And XIII include some sort of overdensity of M31 stars.  Whether this is due to substructure in M31's halo, tidal stripping of other satellites, or a connection to M33 (And XIII lies roughly along the M31-M33 axis), or simply statistical coincidence, are questions beyond the scope of this paper. 

\subsection{And XIV}

\begin{figure*}[tbp!]

\begin{center}
 \bf \large And XIV
 \end{center}

 \epsscale{.5}
 \plotone{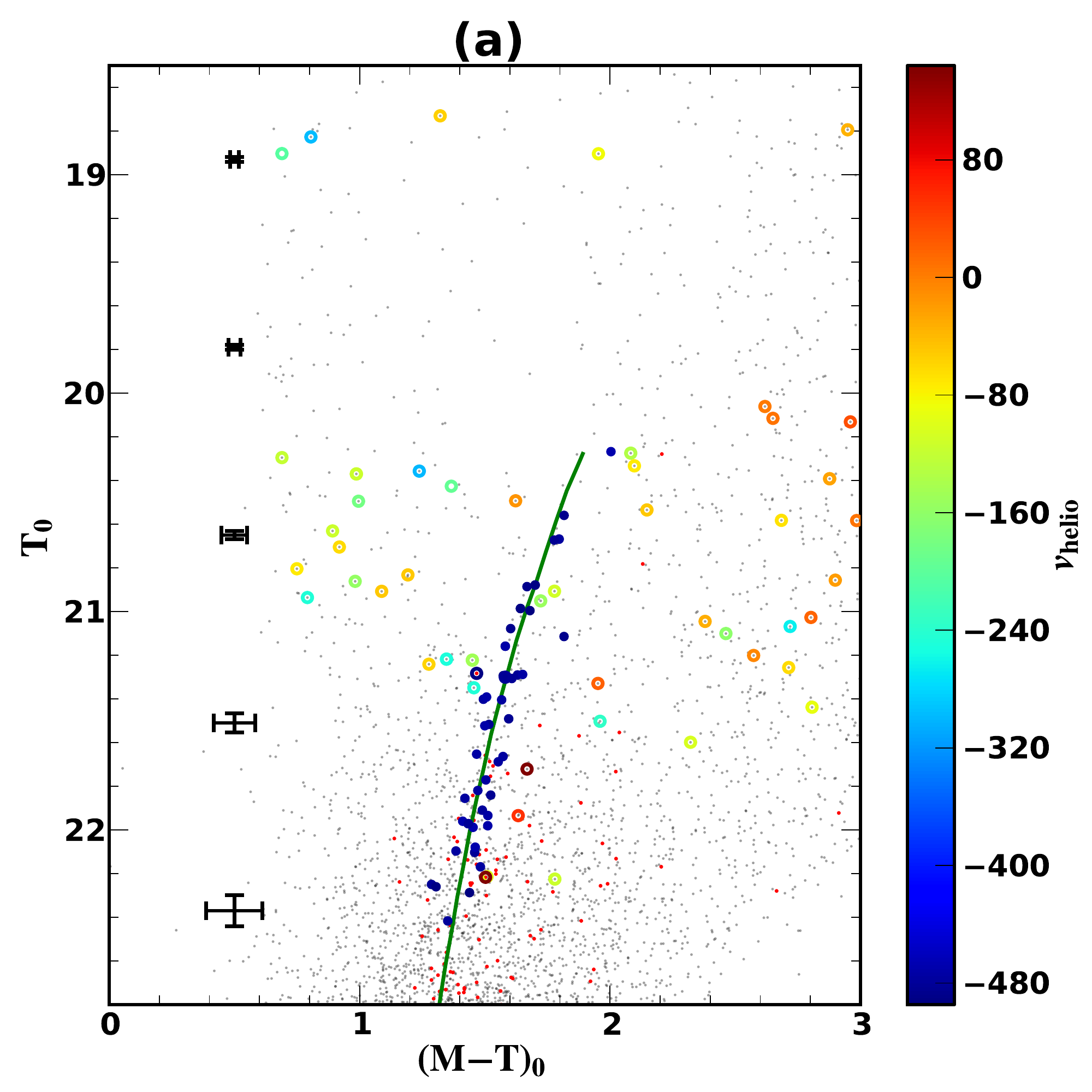}
 \plotone{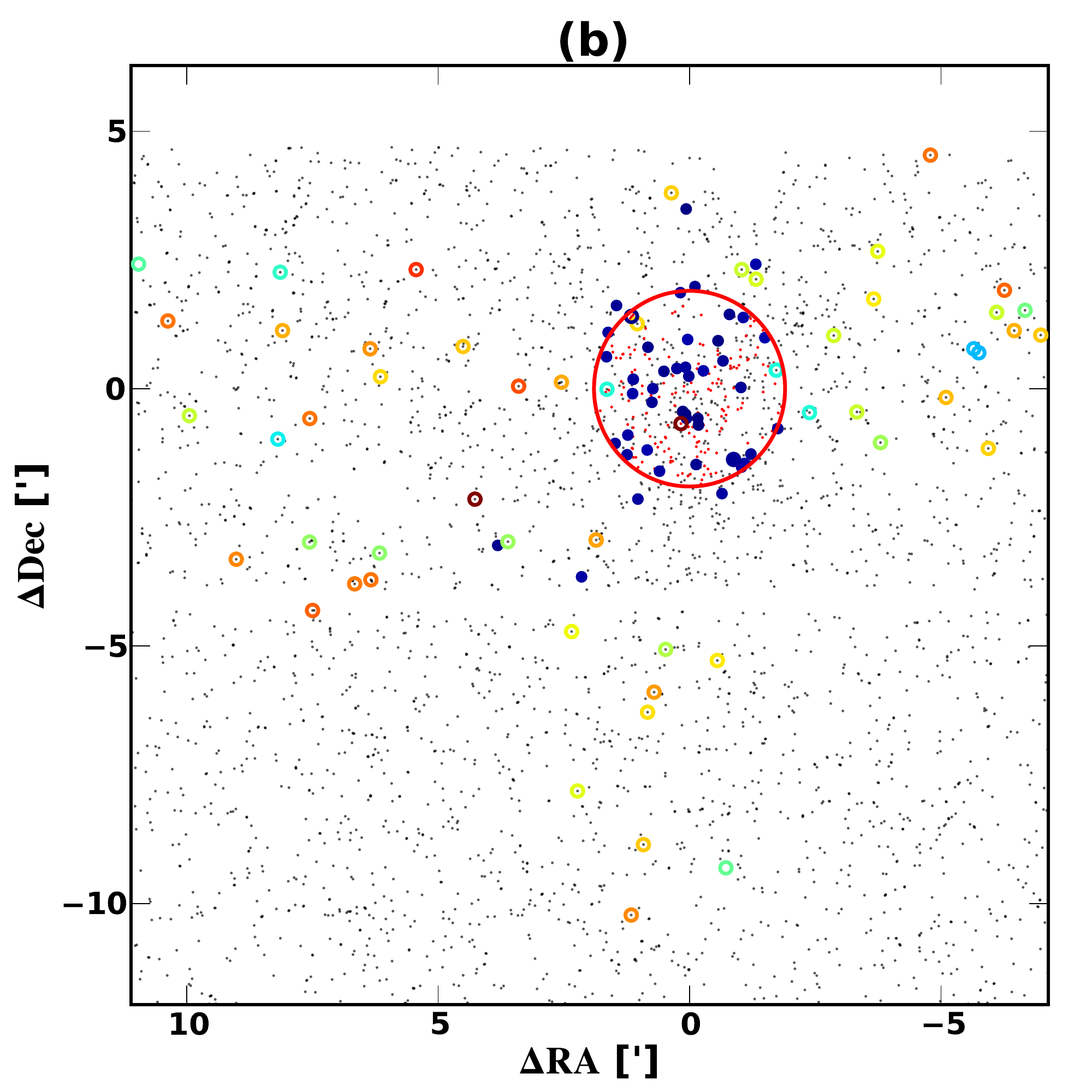}
 \plotone{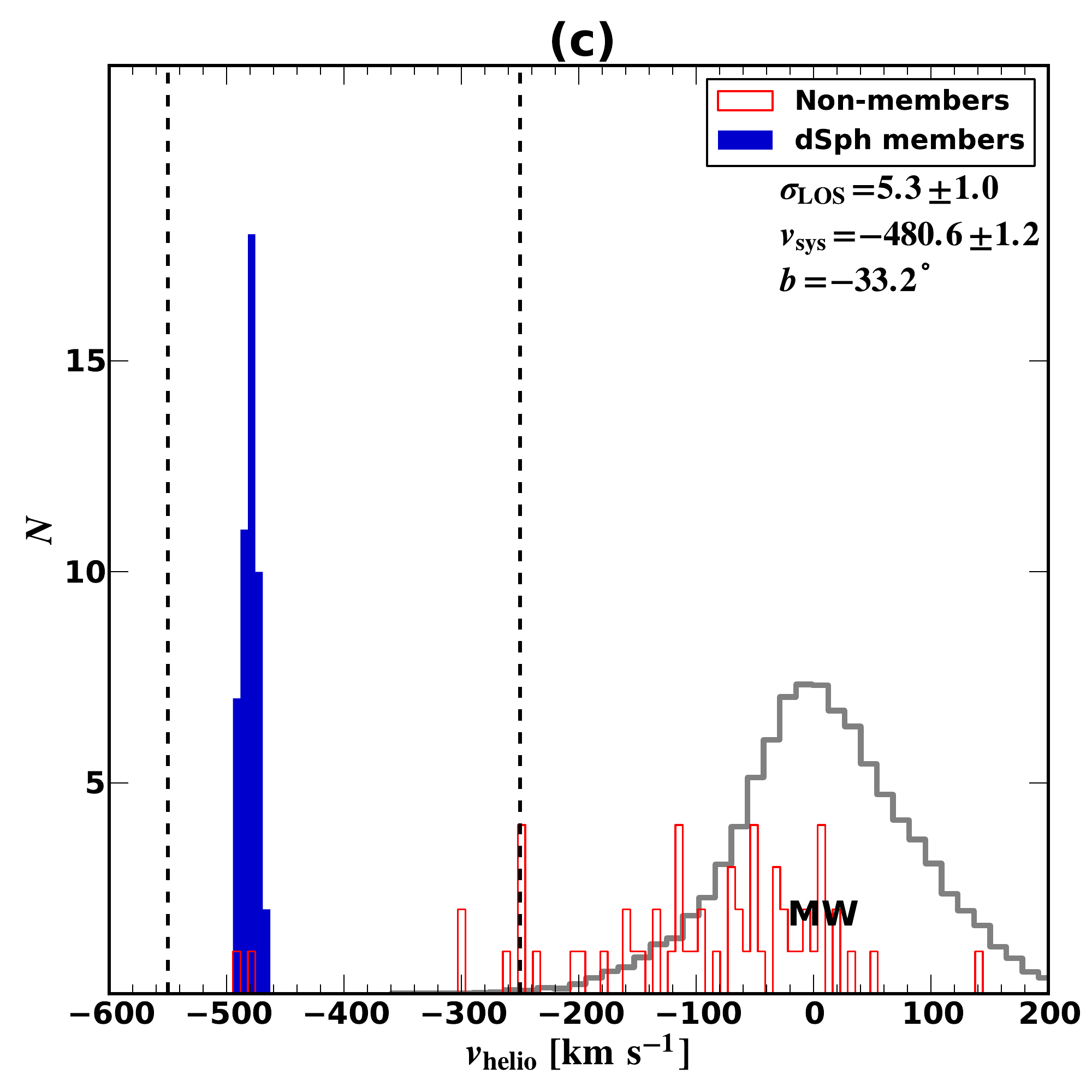}
 \plotone{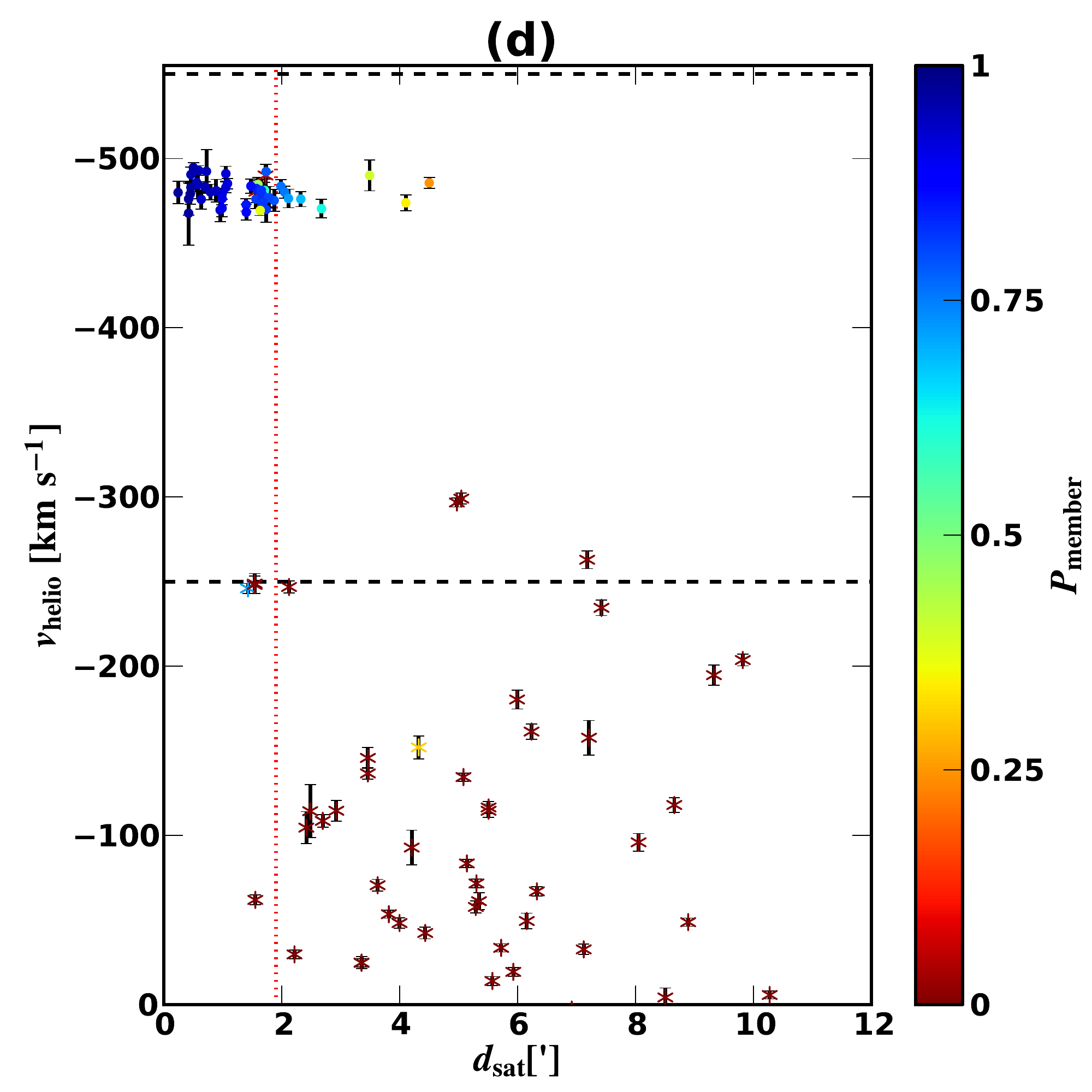}
 \caption{Same as Figure \ref{fig:and1plots}, but for And XIV.}
 \label{fig:and14plots}
\end{figure*}

And XIV (discovered by \citealt{maj07and14},   $M_V=-8.5$, $L_V=2.1 \times 10^5 L_\odot$), is unusual because of its  large velocity relative to M31.  This may imply that it is unbound or barely bound to M31, further supported by its large on-sky distance from M31 ($11.7^{\circ}$, or $160$ kpc projected).  Because of this isolation, it shows a clean cold peak, immediately apparent in Figure \ref{fig:and14plots} near $-480$ \kps. We note that our results here are fully consistent with \citet{maj07and14}, although this should come as no surprise given that two of our three masks are from the same data set.

\subsection{And XV}
\label{ssec:andxv}

\begin{figure*}[tbp!]

\begin{center}
 \bf \large And XV
 \end{center}

 \epsscale{.5}
 \plotone{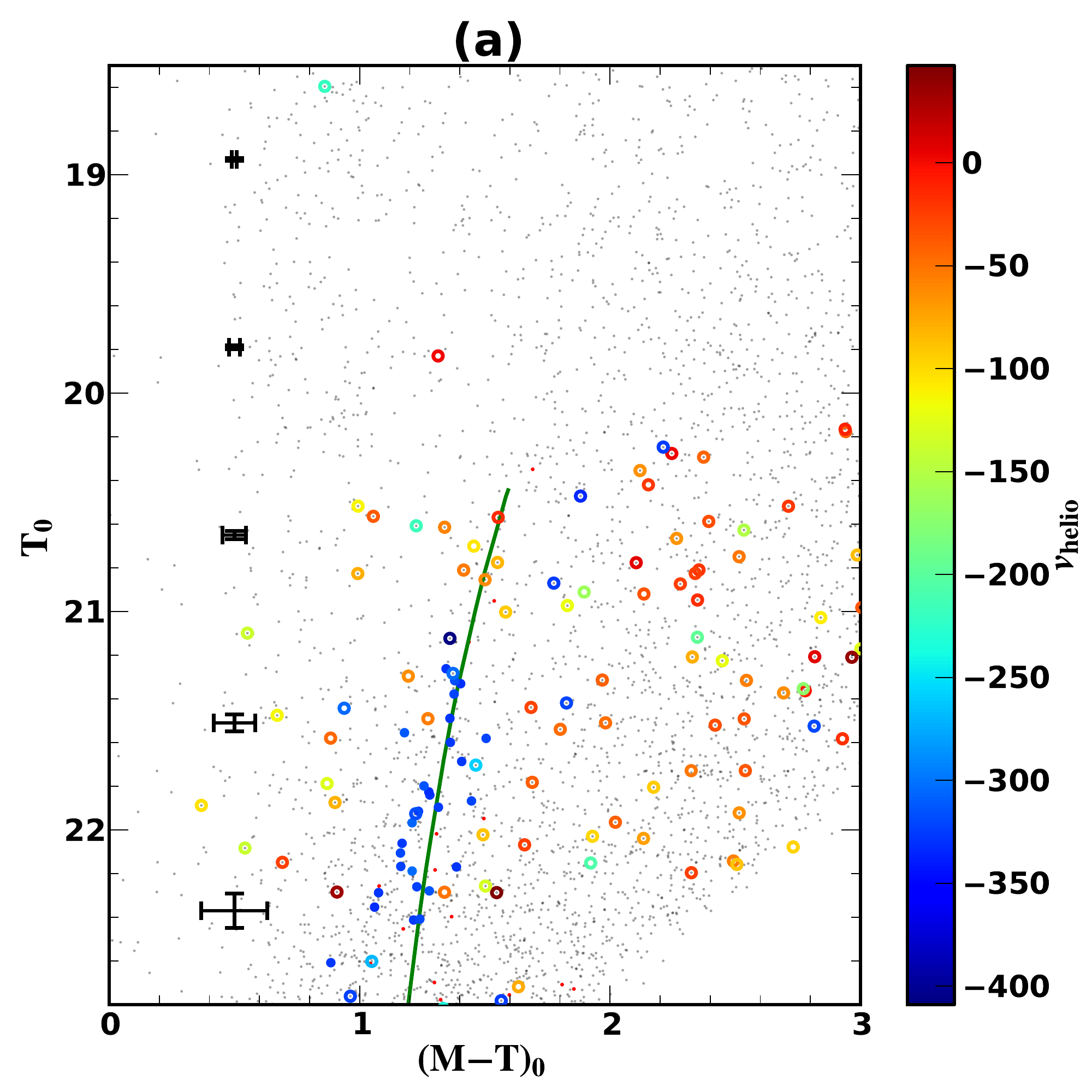}
 \plotone{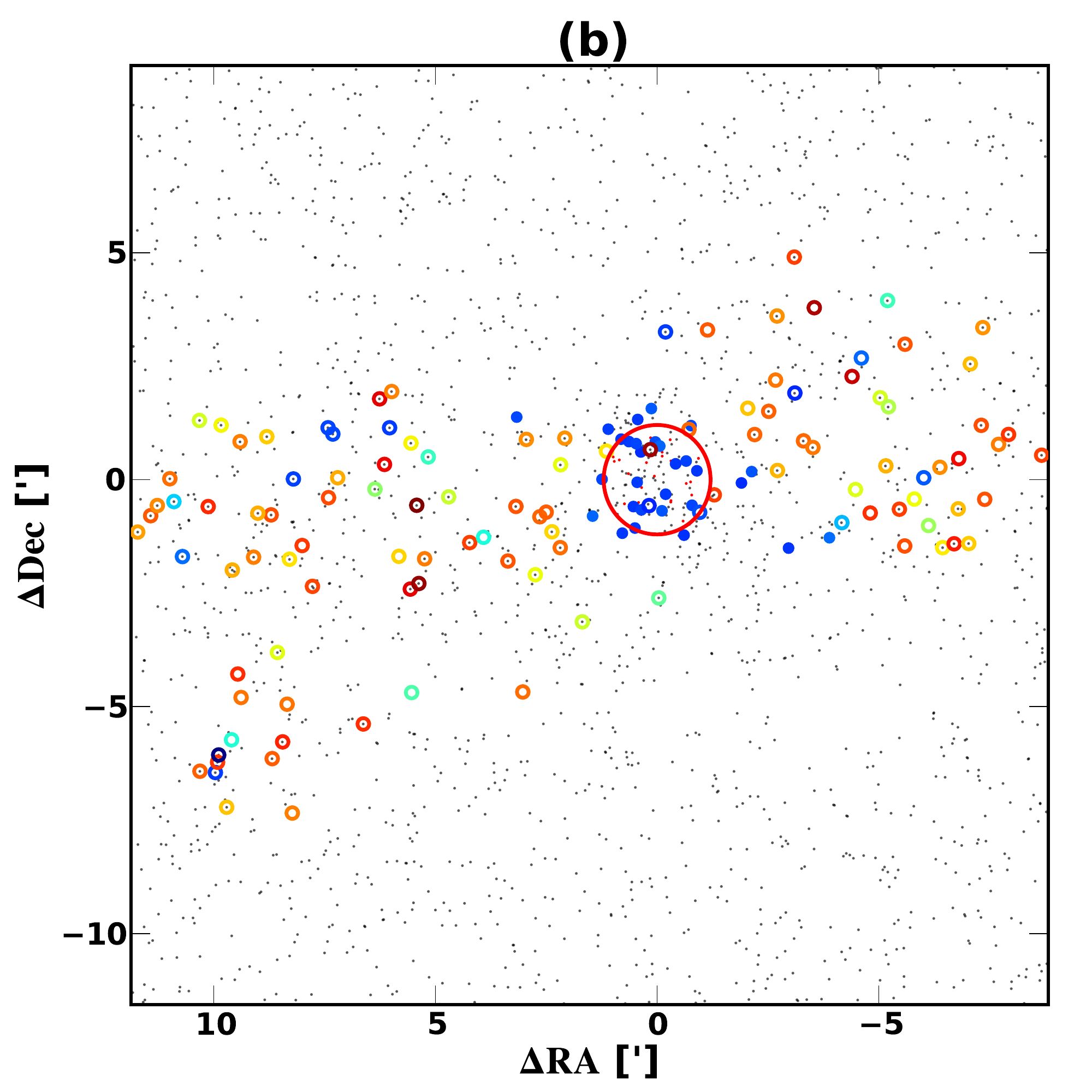}
 \plotone{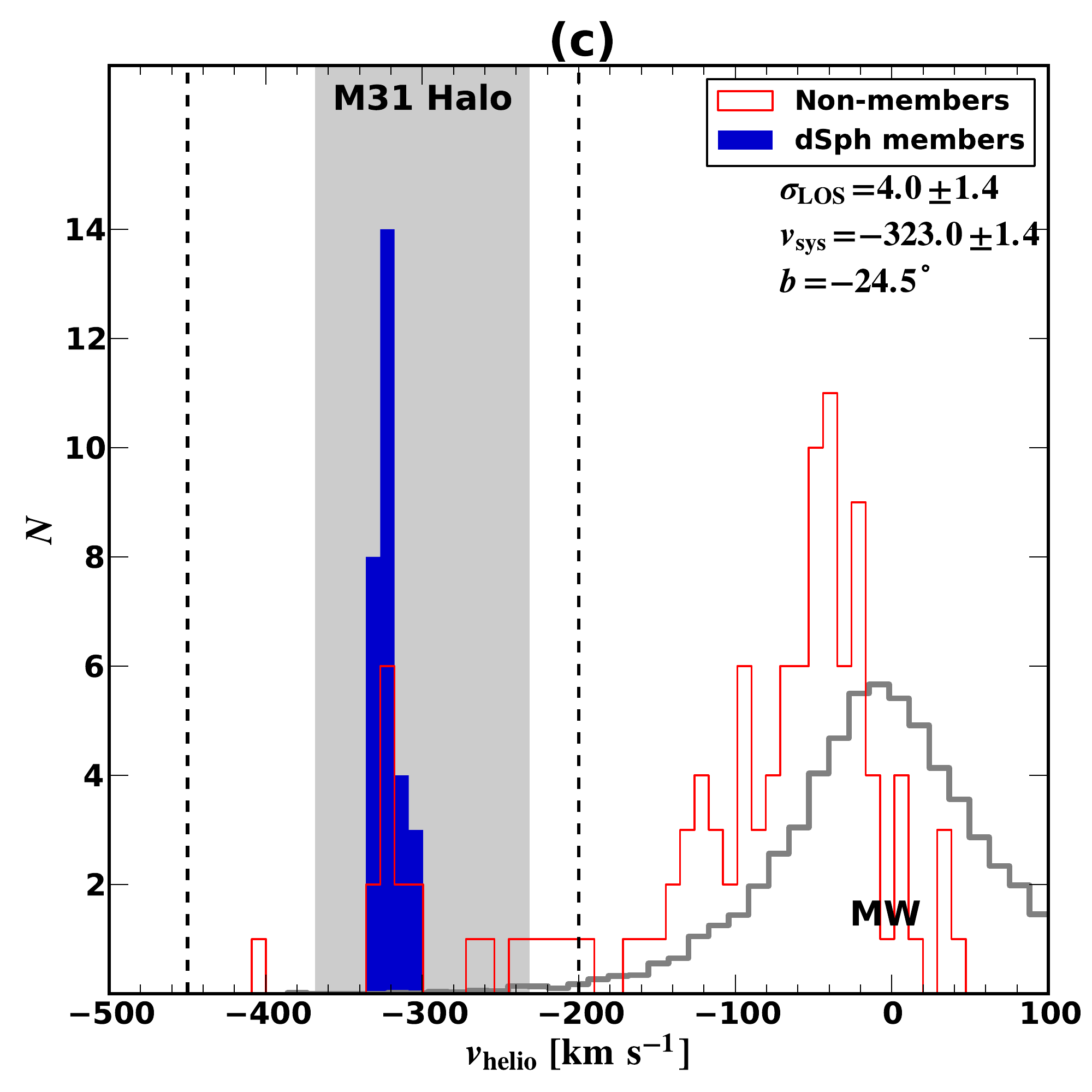}
 \plotone{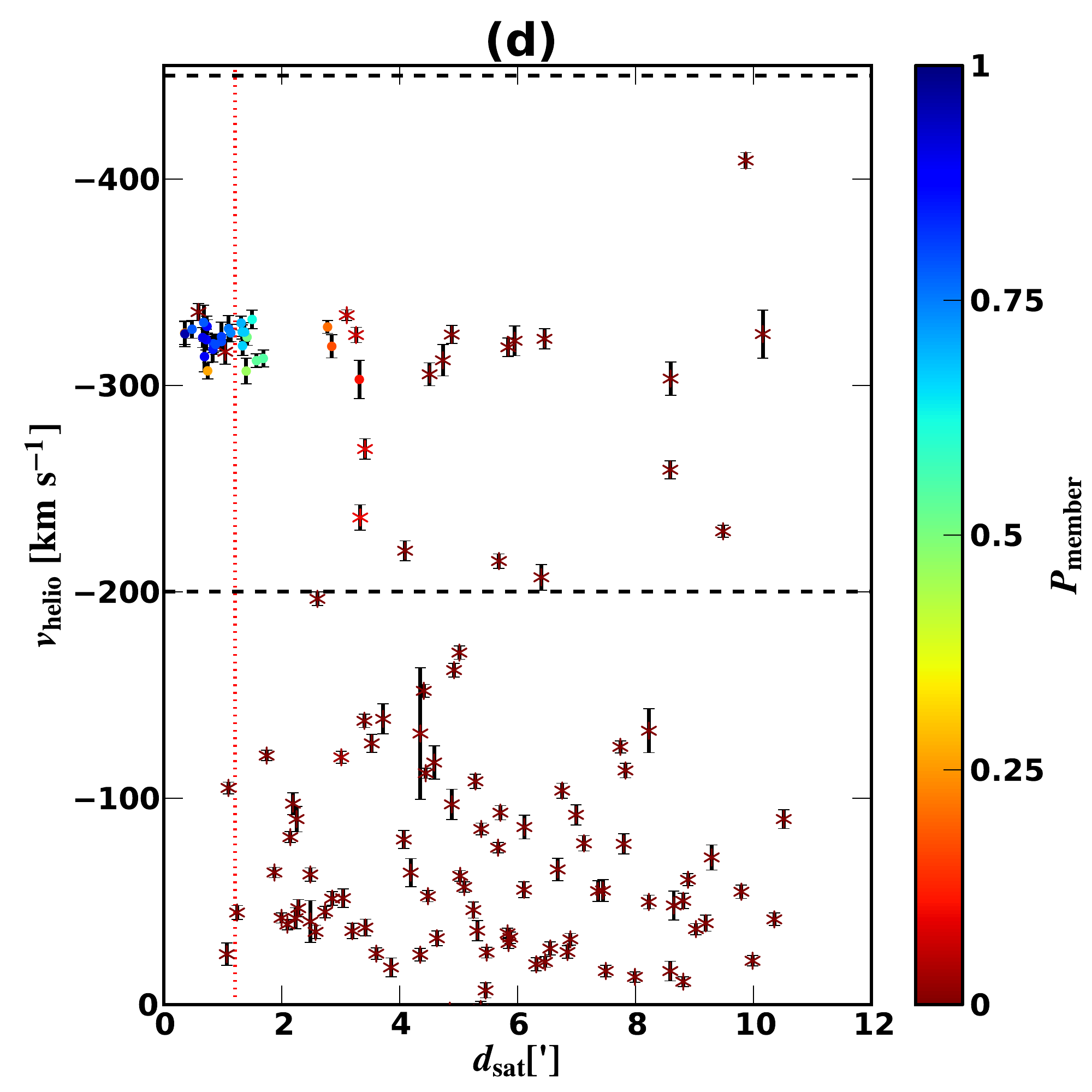}
 \caption{Same as Figure \ref{fig:and1plots}, but for And XV.}
 \label{fig:and15plots}
\end{figure*}

And XV (discovered by \citealt{ibata07}), is a reasonably bright ($M_V=-9.8$, $L_V=7.1 \times 10^5 L_\odot$), relatively isolated dSph ($6.8^{\circ}$, or $93$ kpc projected).  Hence, Figure \ref{fig:and15plots} has a readily apparent cold spike near $-300$ \kps.  However, caution is warranted for this dSph, as it is clear that a significant number of stars in the cold spike do not pass our membership criteria.  Examining the spatial diagram demonstrates clearly that there are stars with velocities consistent with the cold spike that are many $\Reff$ away from the galaxy center.  While the velocities of these stars are near that of the M31 halo distribution, some appear to be kinematically nearly as cold as the satellite itself, and lie along the dSph locus in the CMD.  This suggests that these stars have been tidally stripped from And XV, a conclusion further supported by the dSph's disturbed morphology \citep{ibata07}.  Other stars near that velocity are far from the dSph in the CMD, however.  These may be stars from the ``Stream B'' feature of the M31 halo that lies near And XV and has very similar kinematics \citep{chapman08}.

\begin{figure}[tbp!]
 \epsscale{1}
 \plotone{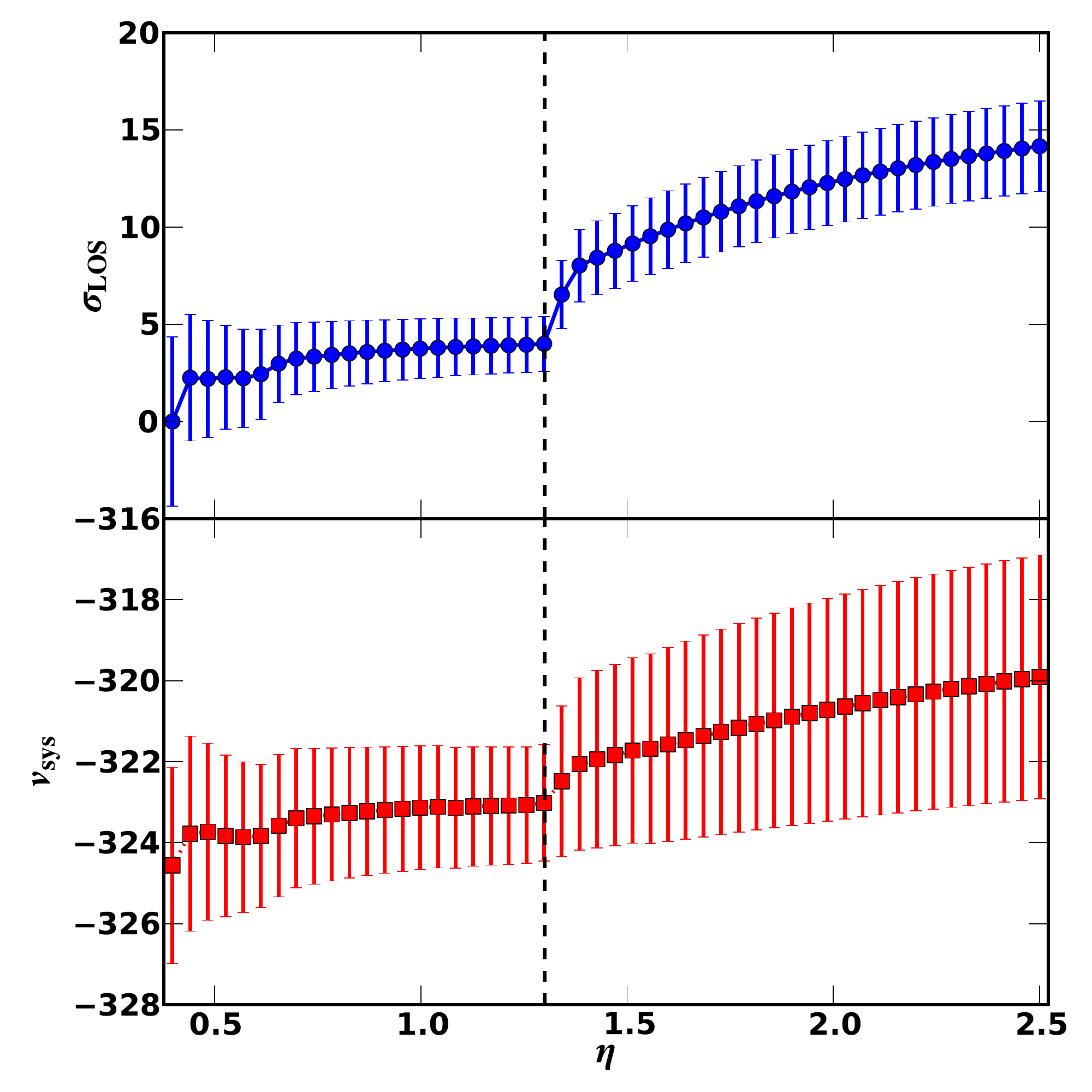}
 \caption{Similar as Figure \ref{fig:and1cog}, but for And XV.  The x-axis has been more densely sampled relative to Figure \ref{fig:and1cog} to better capture the variation near the chosen $\eta$ value.}
 \label{fig:and15cog}
\end{figure}

Given the presence of these contaminants, it is clear that the choice of $\eta$ will influence our kinematical results.  This is apparent from Figure \ref{fig:and15cog}, where the sudden jumps in $\slos$ and increasing trend towards larger $\eta$ indicates the effect of contaminants.  Fortunately,  for $\eta  \lesssim 1.3$, the kinematical parameter estimates are constant (within the error bars).  Hence, we use $\eta=1.3$ for this dSph.  Additionally, we have adopted a value of $\sigma_c=0.2$ to account for a somewhat wider  CMD for this object (although we note that we obtain similar kinematical parameters if we use our fiducial $\sigma_c=0.1$).  With these parameters, the velocities of the stars we identify as members are stable to iterative $3\sigma$ clipping and are consistent with a Gaussian distribution (see \S \ref{sec:membership}).   Our resulting $\slos$ and $\vsys$ are $1-2\sigma$ discrepant from \citet{Letarte09}, however.  This is likely due to a combination of sample size (our sample is 2--3x larger), star-by-star errors (ours are smaller due to use of a higher resolution grating), and our use of a sample that is more strongly weighted towards the center of the dSph.  Given the contamination issues outlined above, the stars closest to the center are likely of great importance.

\subsection{And XVI}

\begin{figure*}[tbp!]

\begin{center}
 \bf \large And XVI
 \end{center}

 \epsscale{.5}
 \plotone{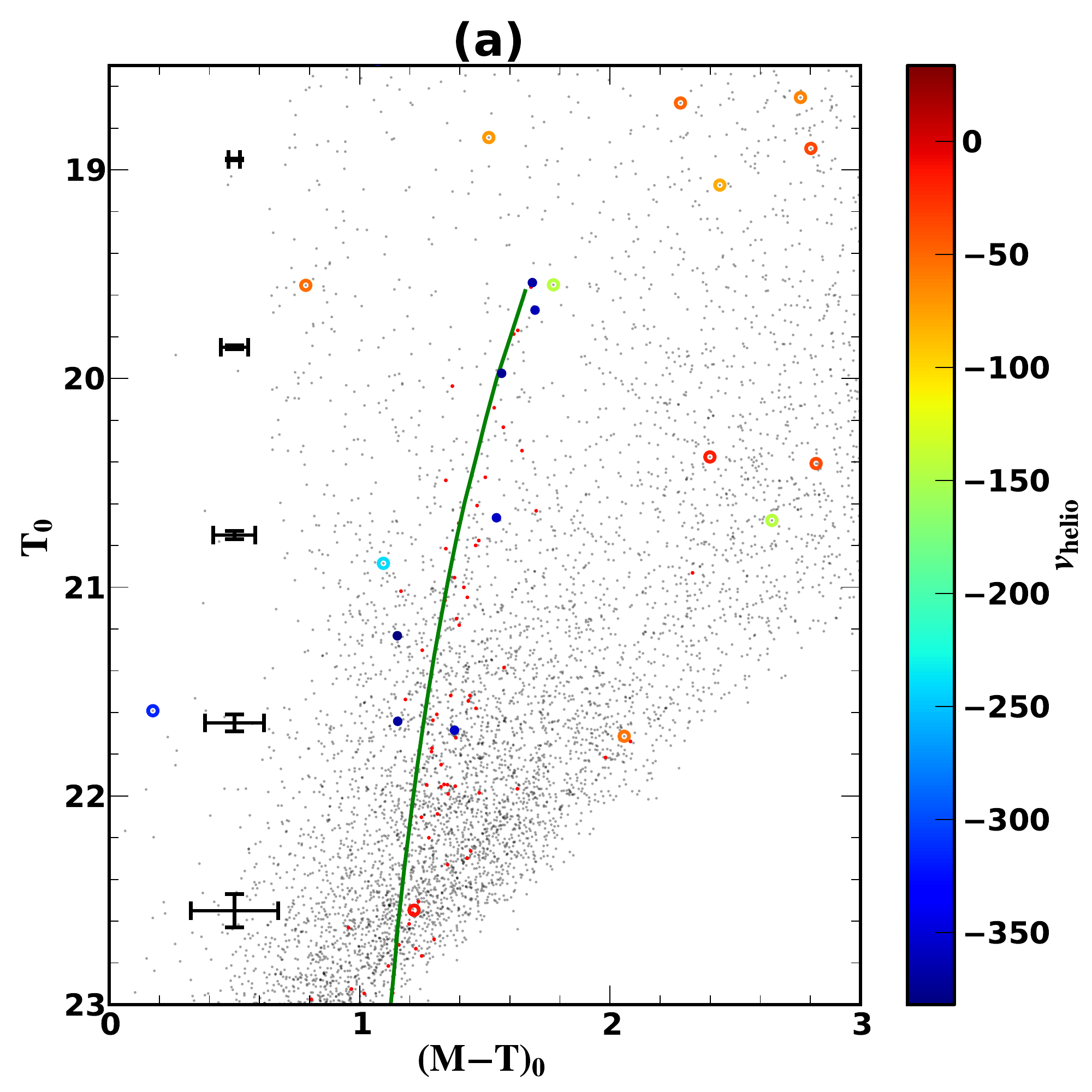}
 \plotone{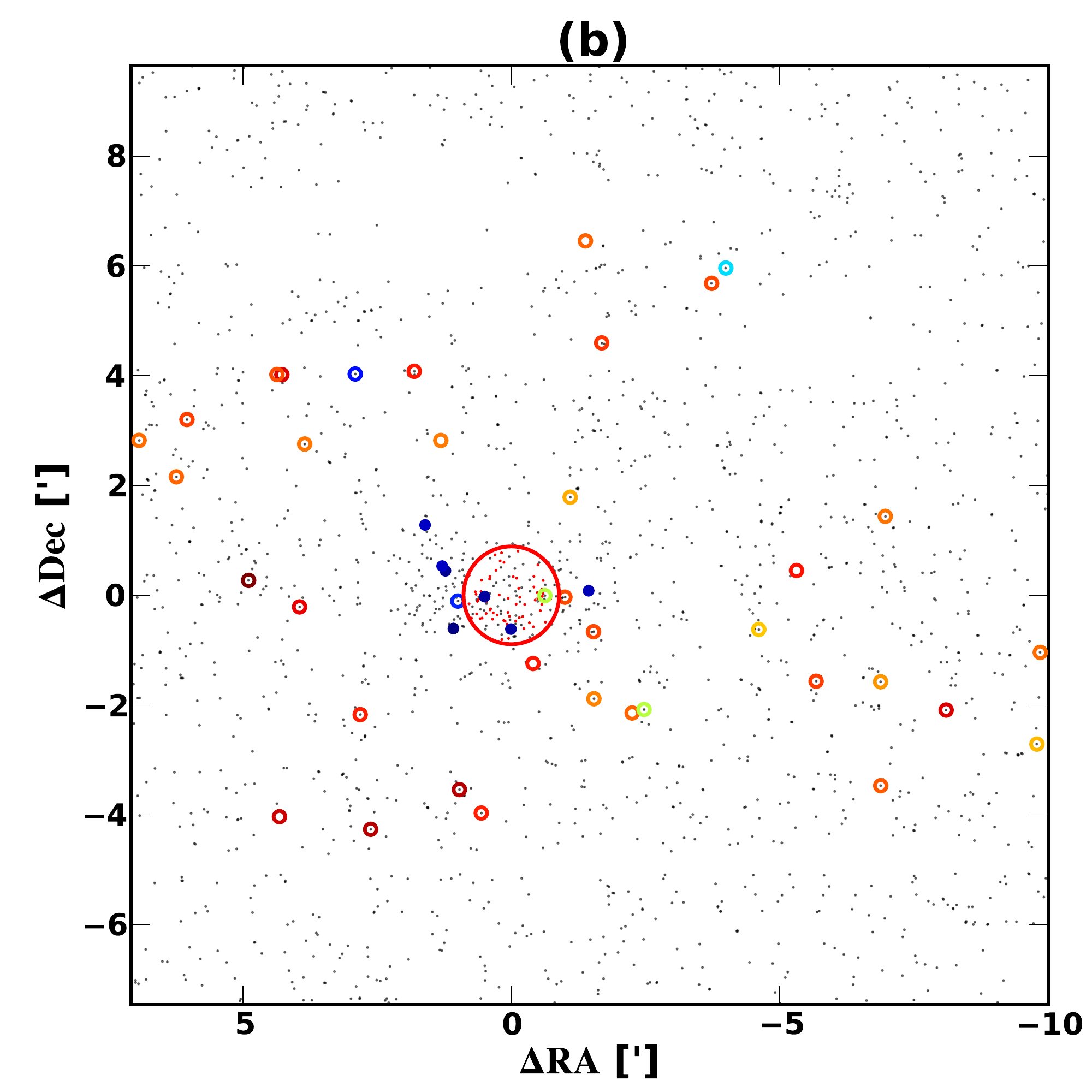}
 \plotone{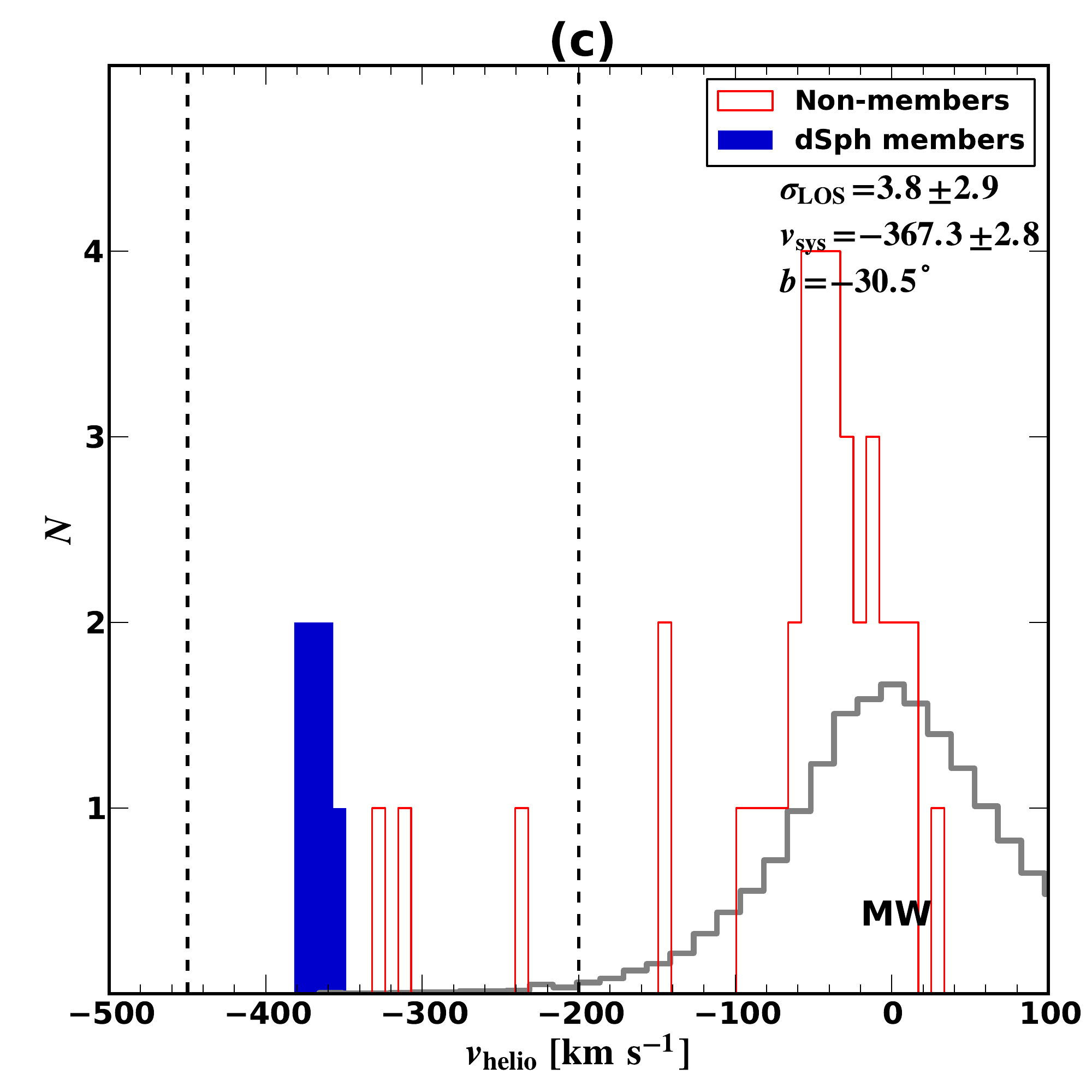}
 \plotone{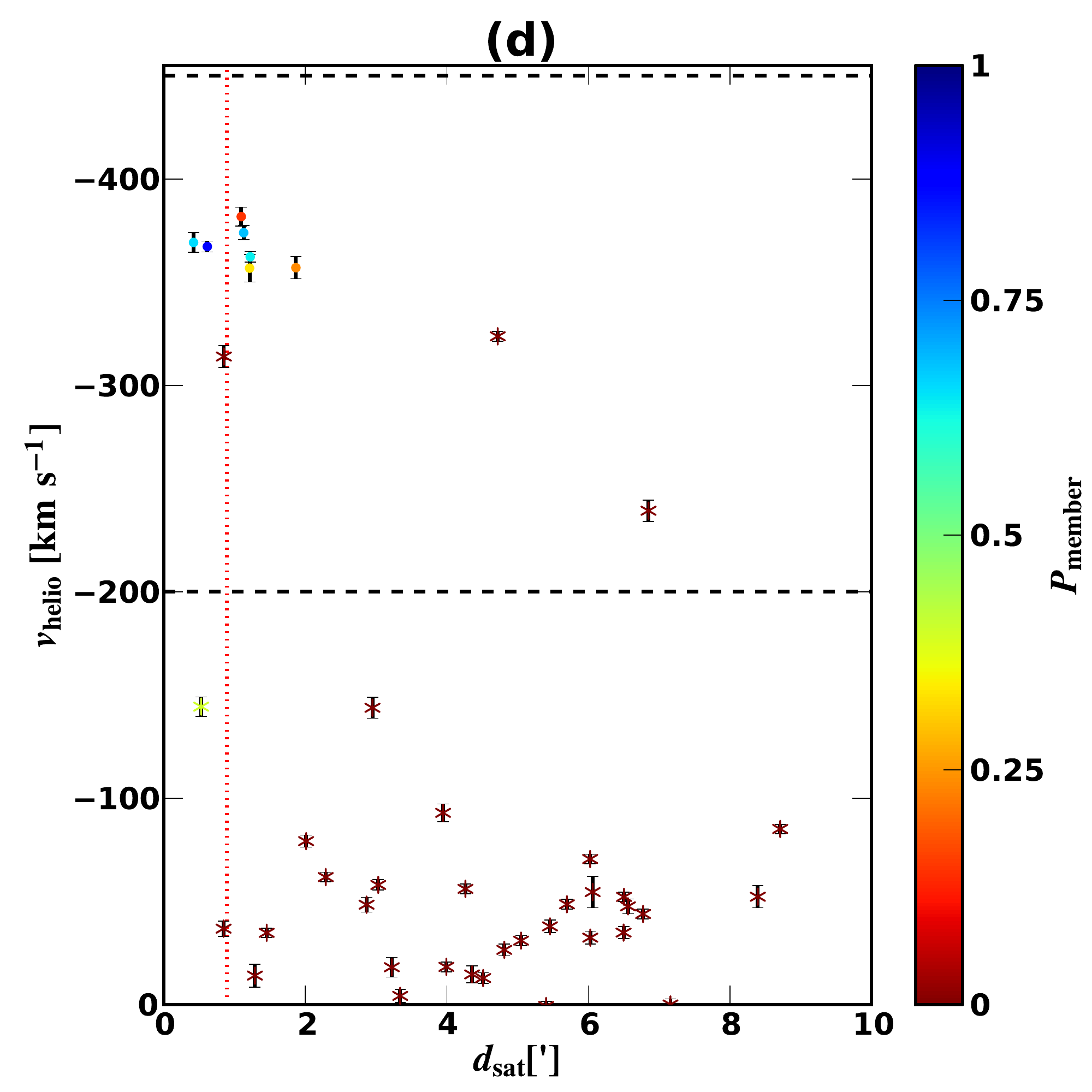}
 \caption{Same as Figure \ref{fig:and1plots}, but for And XVI.}
 \label{fig:and16plots}
\end{figure*}

Our And XVI (discovered by \citealt{ibata07}, $M_V=-9.2$, $L_V=4.1 \times 10^5 L_\odot$) field is rather sparse in successful targets, but Figure \ref{fig:and16plots} does show a centrally-concentrated cluster of stars near $-360$ \kps. Fortunately, And XVI's relatively large distance from M31 ($9.5^{\circ}$, or $130$ kpc projected) and substantial distance from the Galactic plane ($b=-30^{\circ}$) implies very little contamination.  Our fiducial parameters result in a marginally resolved velocity dispersion.  The small number of stars (7) imply that this measurement should be taken with caution, however, as excluding a few stars from the analysis typically results in a  dispersion formally consistent with zero.  We note that our $\vsys$ is discrepant from \citet{Letarte09} at the $\sim 3\sigma$ level, however, even when a dispersion is not resolved.


\subsection{And XVIII}
\label{ssec:and18}

\begin{figure*}[tbp!]

\begin{center}
 \bf \large And XVIII
 \end{center}

 \epsscale{.5}
 \plotone{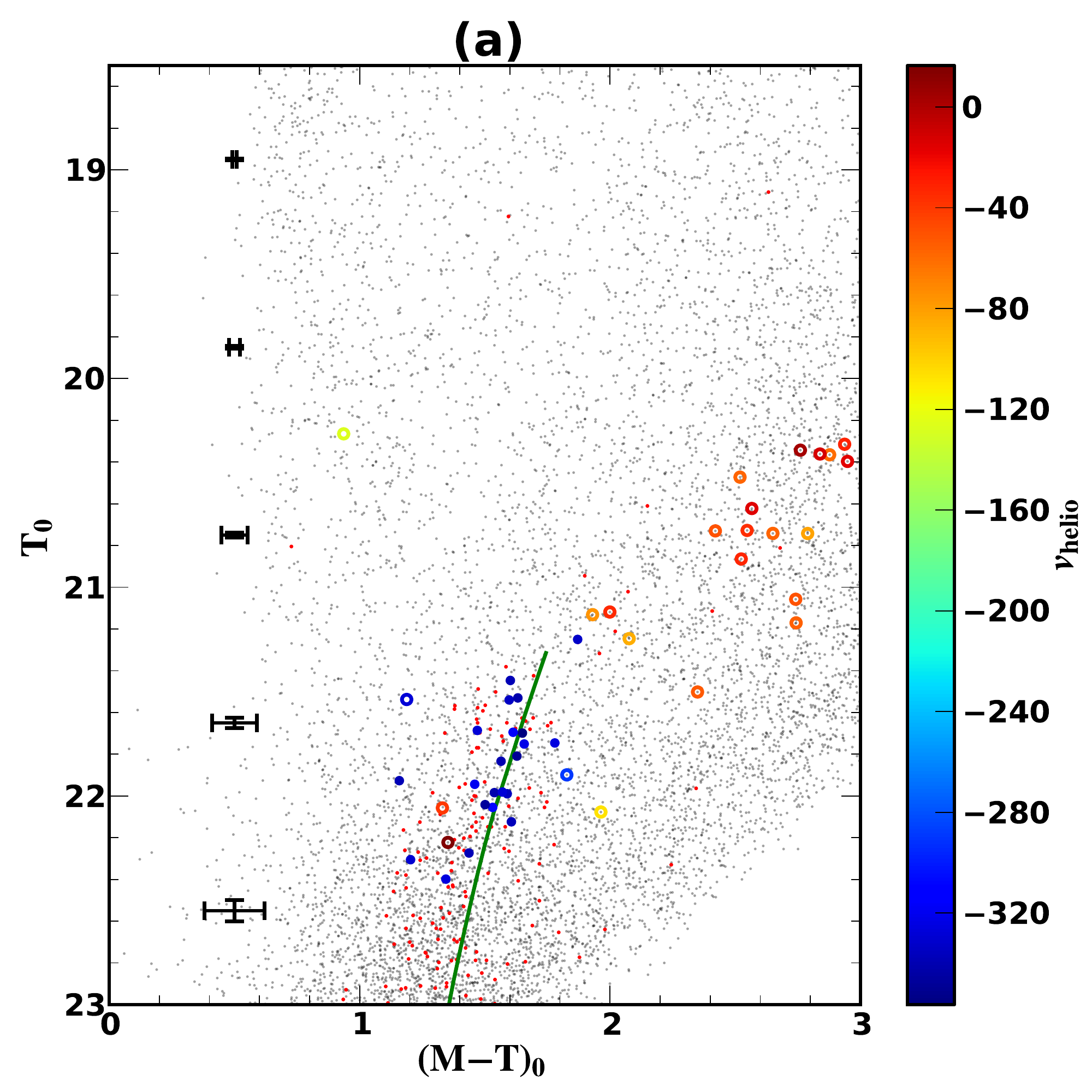}
 \plotone{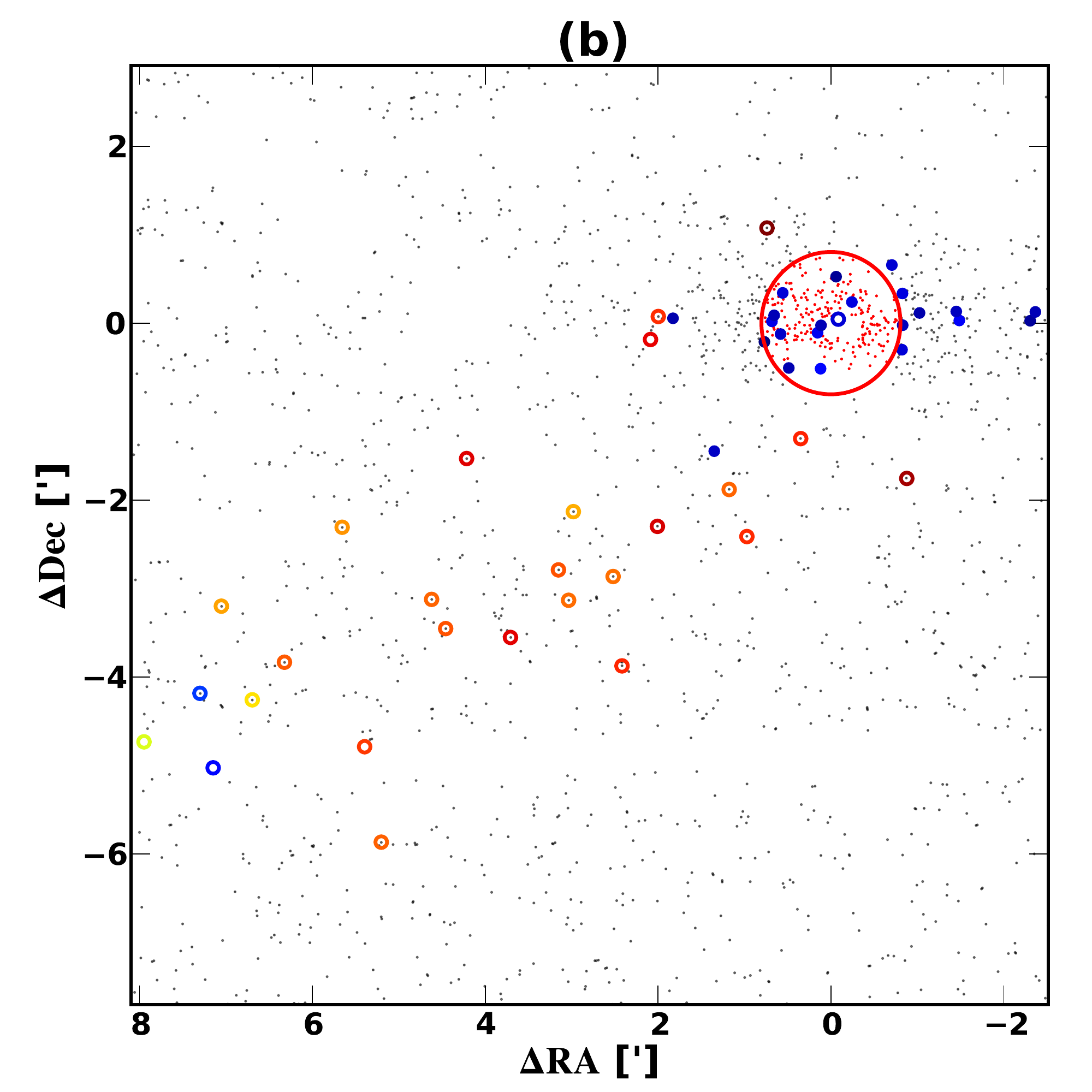}
 \plotone{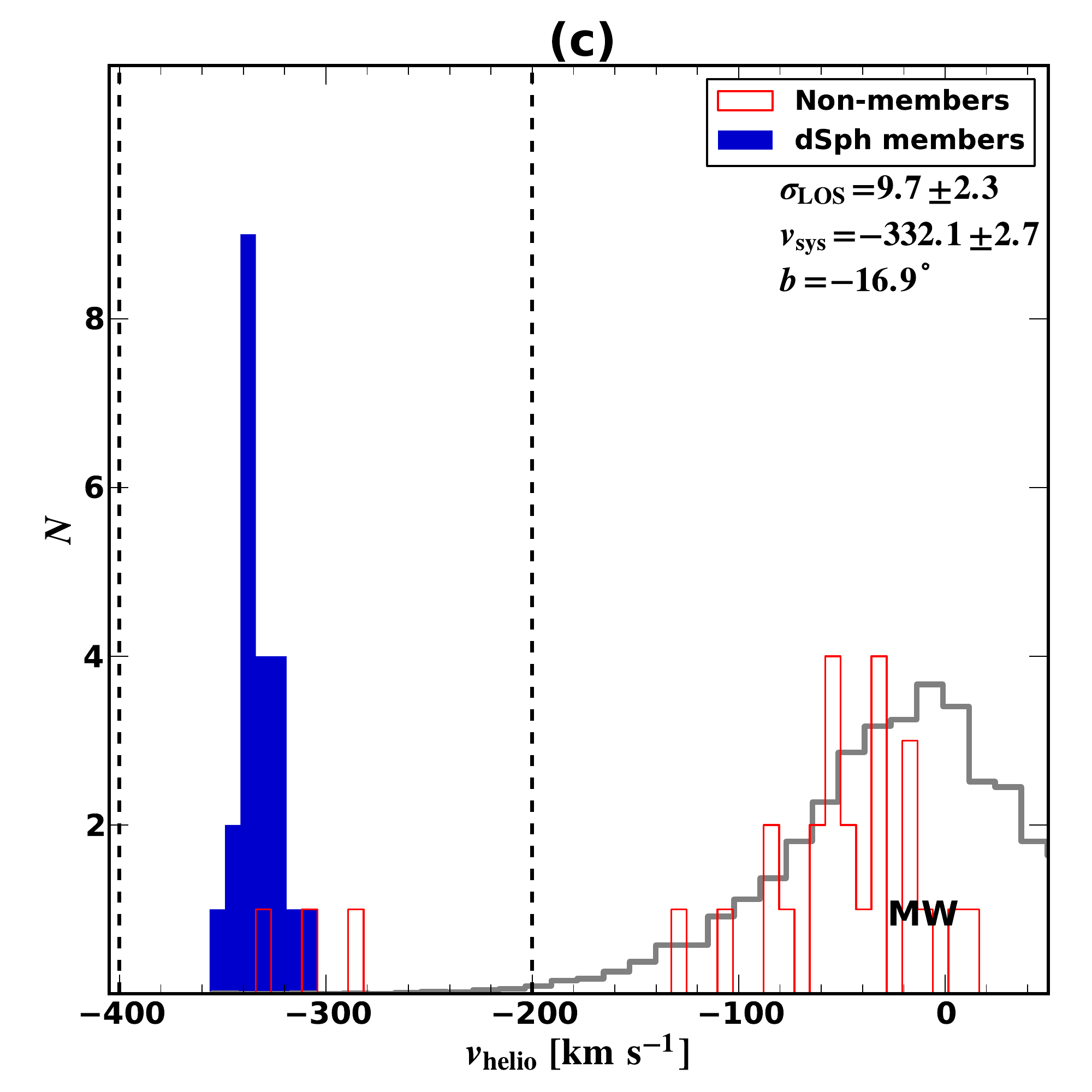}
 \plotone{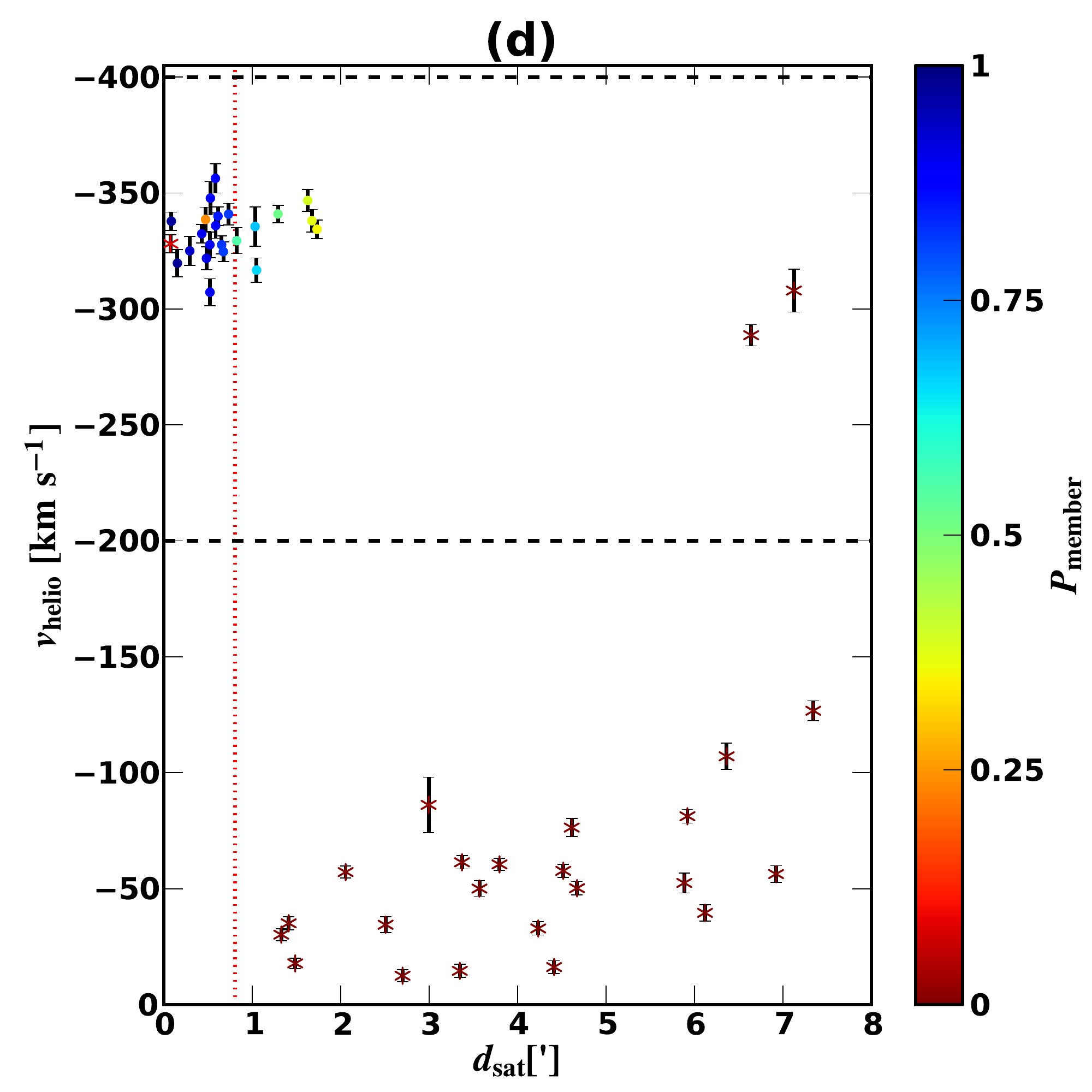}
 \caption{Same as Figure \ref{fig:and1plots}, but for And XVIII.}
 \label{fig:and18plots}
\end{figure*}

And XVIII was discovered by \citet{mc08} and is remarkable in its large line-of-sight distance that formally places it well outside of the M31 system and in fact at the outskirts of the LG.  It is also fairly bright ($M_V=-9.7$, $L_V=6.3 \times 10^5 L_\odot$), and reasonably far from M31 on the sky ($8.3^{\circ}$, or $113$ kpc projected) and therefore contamination from the M31 halo will likely be small.  We present our results for And XVIII in Figure \ref{fig:and18plots}, which shows a clear cold spike near $-330$ \kps.  These stars are centrally concentrated and far from the typical M31 halo star on the CMD.   Hence, we conclude that these observations (the first spectroscopic observations of And XVIII)  definitively confirm it as a kinematically cold satellite galaxy.  We adopt a $\sigma_c=0.3$ to accept the wider-than typical CMD (primarily due to relatively shallow targeting photometry).

The $\vsys$ we measure is very close to M31's $\vsys$.  This is a remarkable coincidence if And XVIII is actually 600 kpc distance from M31 and never interacted.  Hence,  And XVIII is  near its apocenter (and thus nearly at rest relative to M31), and/or closer to M31 (as suggested by our photometry,  \comppapalt).


\subsection{And XXI}

\begin{figure*}[tbp!]

\begin{center}
 \bf \large And XXI
 \end{center}

 \epsscale{.5}
 \plotone{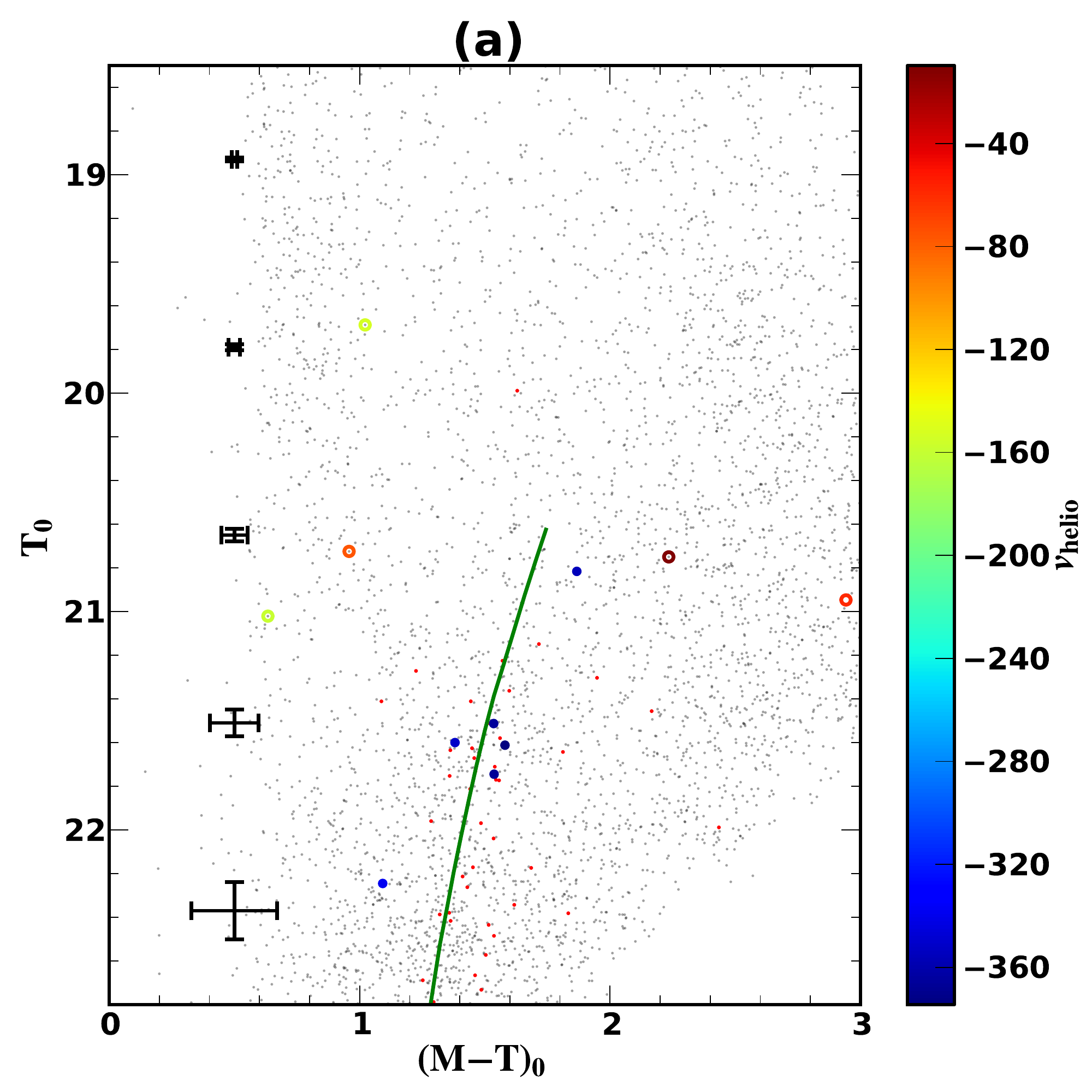}
 \plotone{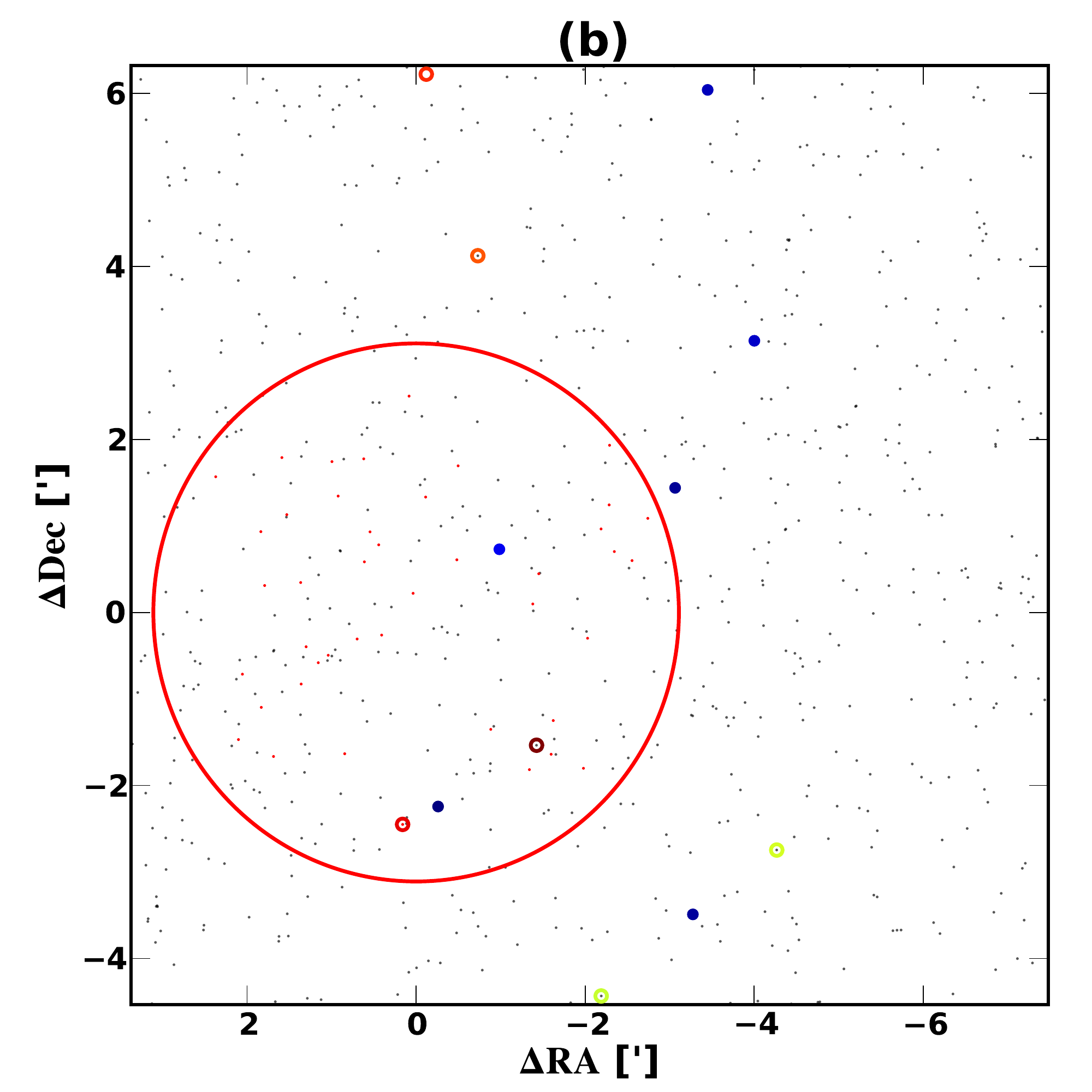}
 \plotone{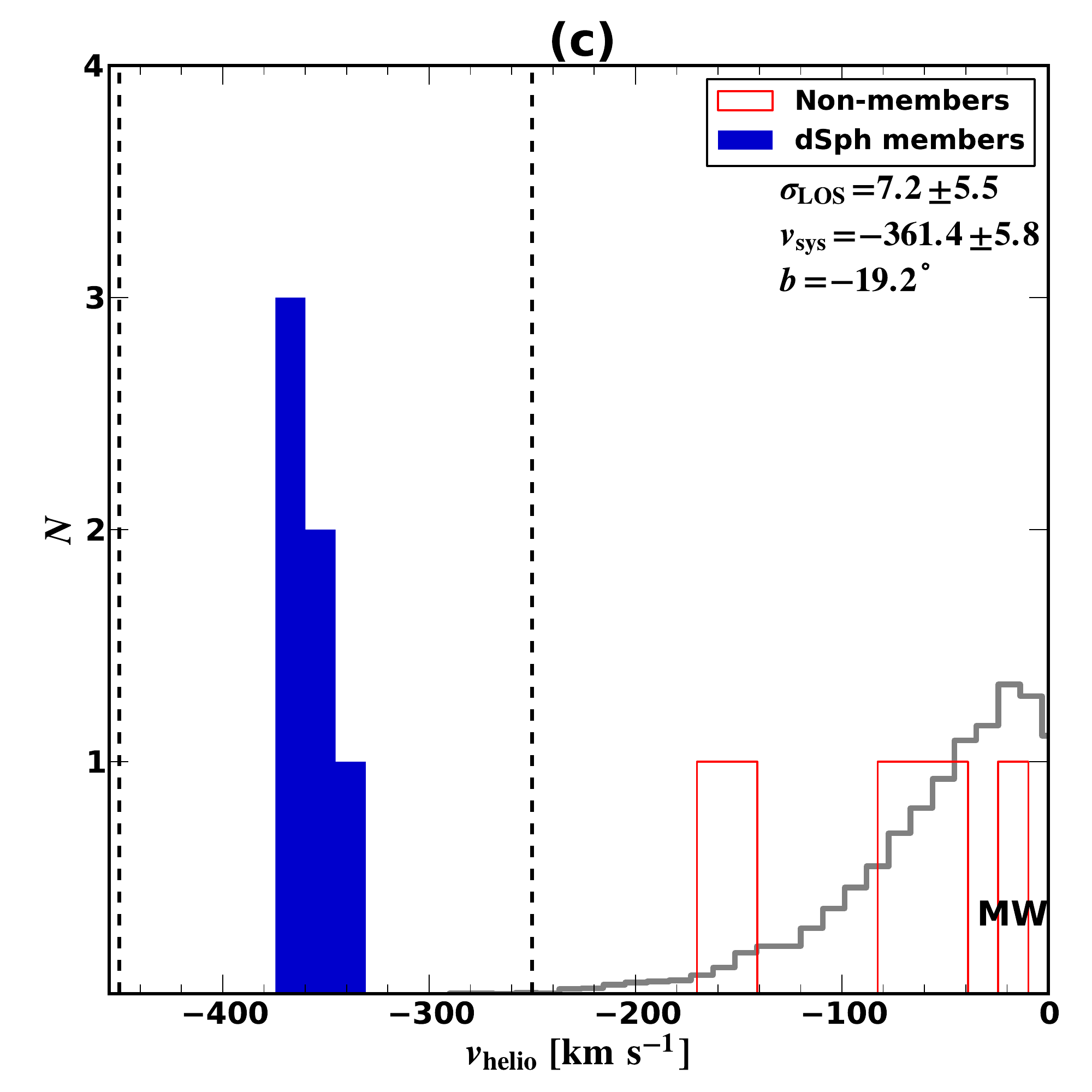}
 \plotone{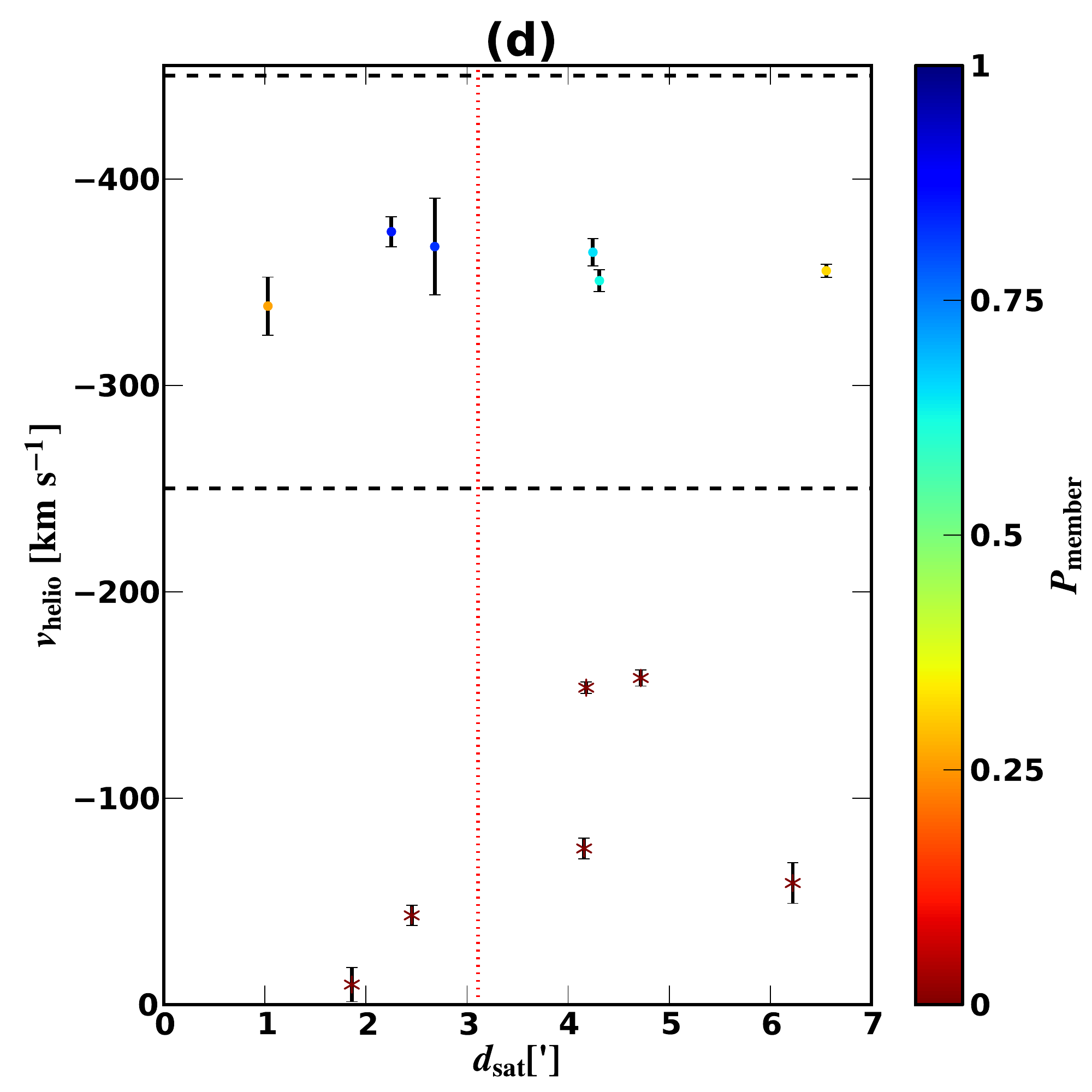}
 \caption{Same as Figure \ref{fig:and1plots}, but for And XXI.}
 \label{fig:and21plots}
\end{figure*}

And XXI was recently discovered as part of the PAndAS survey, and is relatively luminous ($M_V=-9.3$, $L_V=4.6 \times 10^5 L_\odot$) but has an atypically large half-light radius/low surface brightness \citep{martin09}.  It is also quite far from M31 ($9.0^{\circ}$, or $123$ kpc projected), and hence contamination from the M31 halo is quite low.  We present the first spectroscopic observations of And XXI in Figure \ref{fig:and21plots}.  The field is sparse, but there is a definite overdensity of stars with radial velocities near $-360$ \kps.  

For membership criteria, we adopt a larger-than-fiducial $\sigma_c=0.3$ to account for the relatively shallow $T$ exposure.  This mainly serves to include the star near $M-T \sim 1$ (the faintest in the spectroscopic sample), as that star is very near the center of the dSph and hence is plausibly a member.  We also note that the brightest star in the member sample is near (and possibly beyond) the expected tip of the red giant branch (TRGB) for this satellite \citep{martin09}.  It is also the most distant from the center of the dSph in our sample, hence rendering its membership questionable.  It is formally included  following our method here, but removing it from our sample results in changes to our kinematic parameters that are well within errors. 

With this sample, we find a velocity dispersion with large errors but formally inconsistent (at $1\sigma$) with zero. Additionally, given the clear clustering in radial velocity of four of the stars near the identified RGB of  \citet{martin09}, we consider our data to be spectroscopic confirmation that And XXI is a kinematically cold satellite with $\vsys \approx -363$ \kps, although our derived velocity dispersion has large error bars that may be underestimated due to the small sample of only 6 stars.   

\subsection{And XXII}

\begin{figure*}[tbp!]

\begin{center}
 \bf \large And XXII
 \end{center}

 \epsscale{.5}
 \plotone{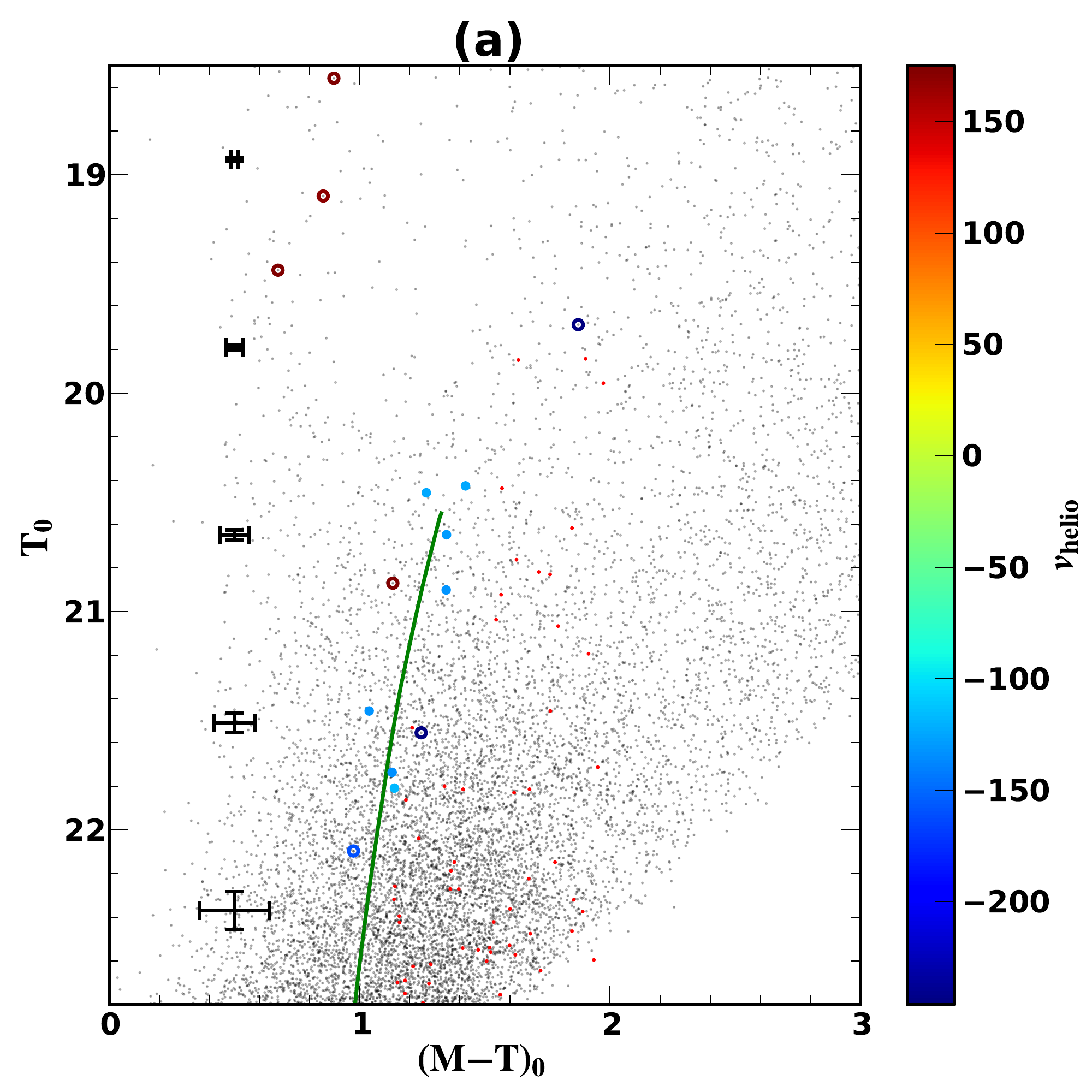}
 \plotone{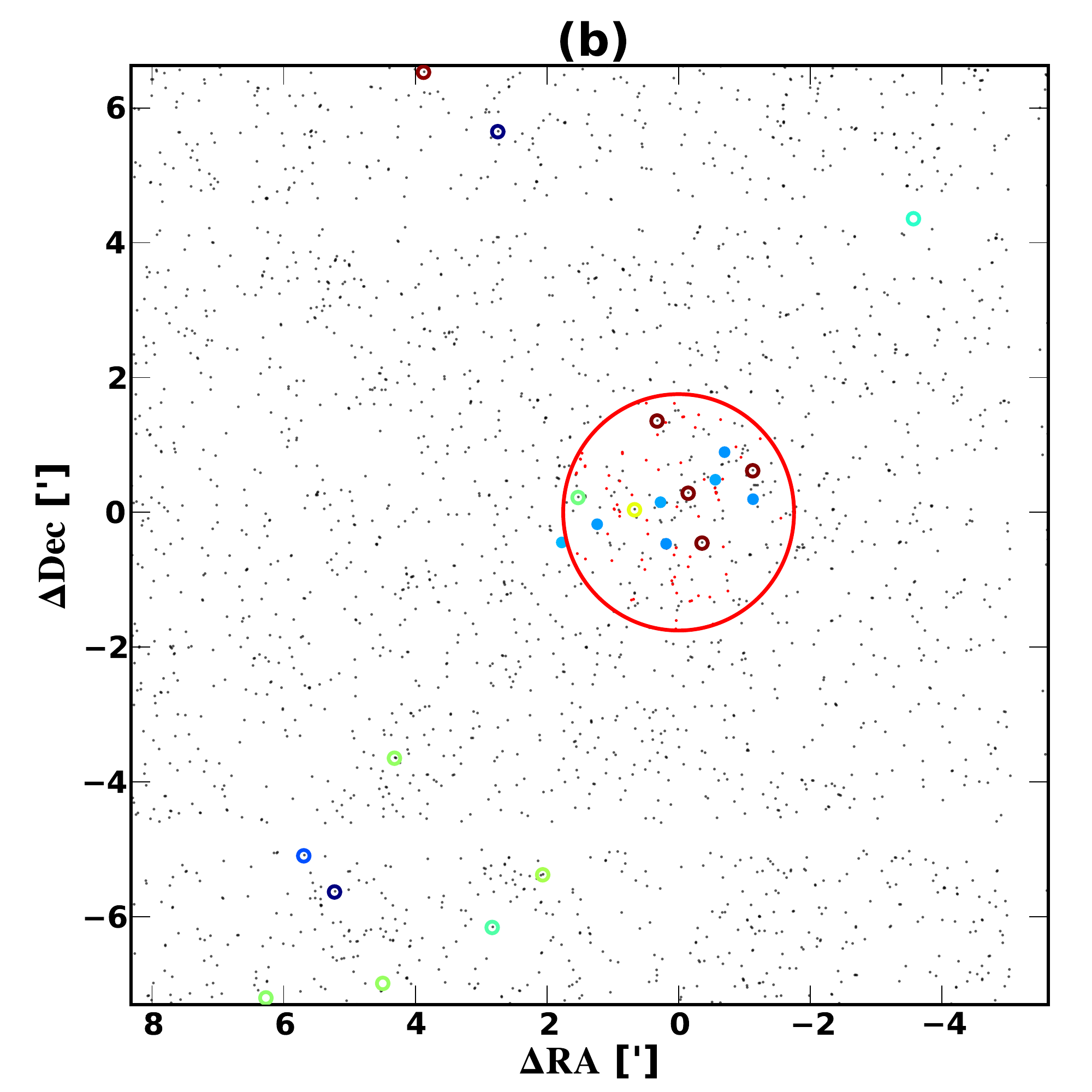}
 \plotone{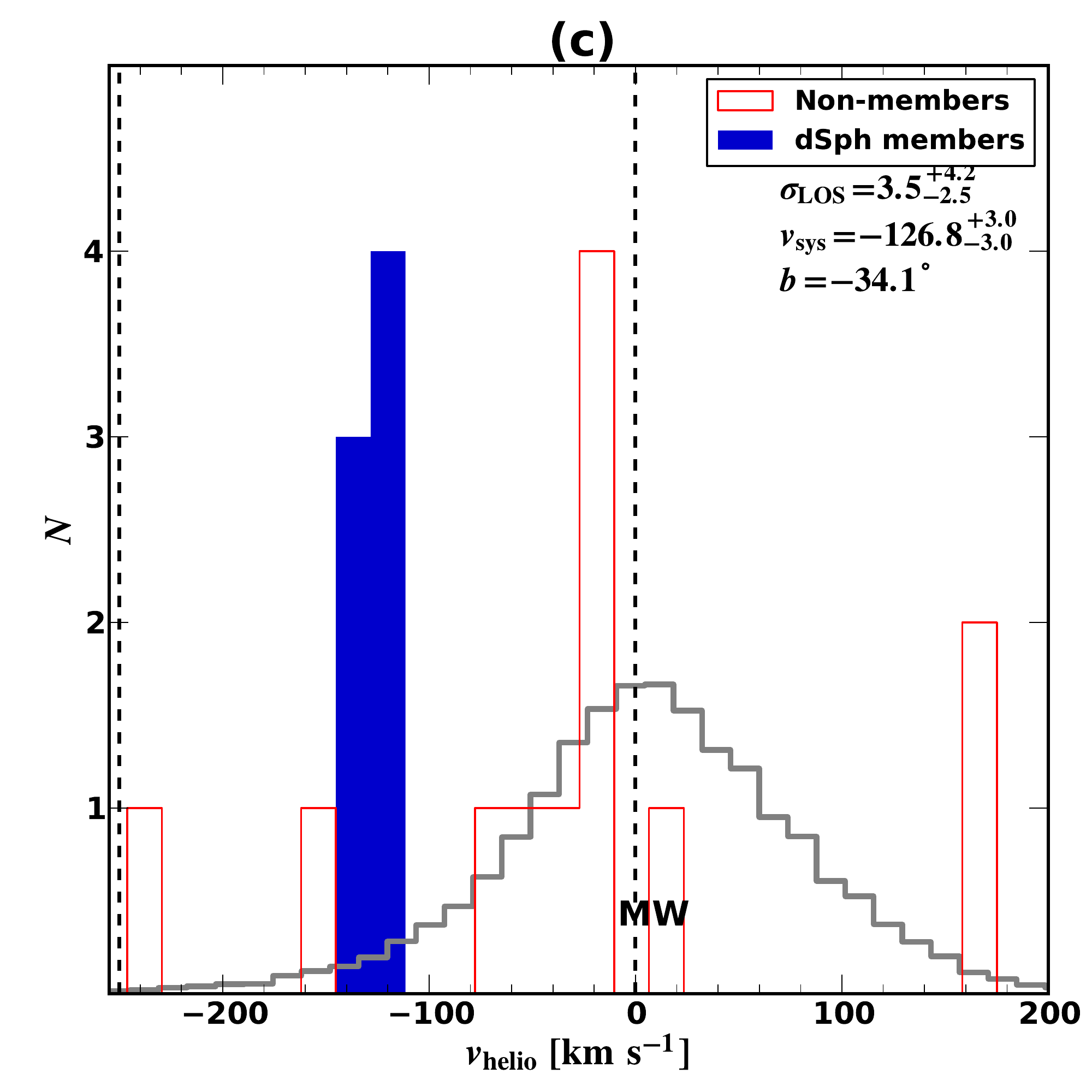}
 \plotone{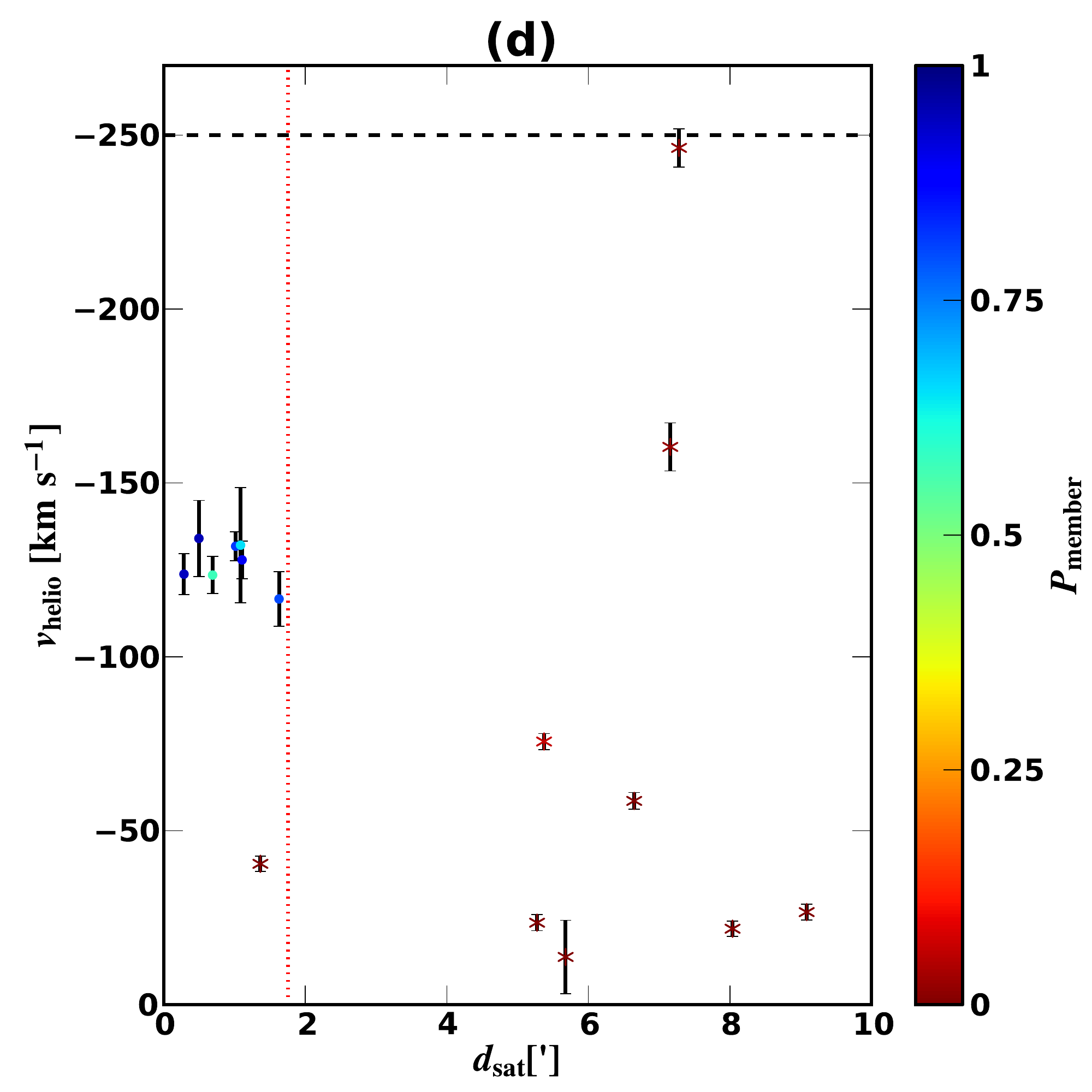}
 \caption{Same as Figure \ref{fig:and1plots}, but for And XXII.}
 \label{fig:and22plots}
\end{figure*}

And XXII, also discovered by \citet{martin09}, is one of the faintest of the M31 dSphs ($M_V=-6.2$, $L_V=2.6 \times 10^4 L_\odot$).  In Figure \ref{fig:and22plots} we present our results for And XXII.  While our success rate at selecting members for spectroscopy was lower for this satellite than others due to a lack of DDO51 pre-selection and the faintness of this galaxy, we do find a concentration of stars within $\Reff$ that are kinematically cold.  These stars lie along a well-defined sequence that matches the CMD of \citet{martin09}, and the kinematically cold sample shows an elliptical distribution consistent with the photometric measurements (and ellipticity was \emph{not} included in the target selection).   Thus, we spectroscopically confirm a cold population consistent with the hypothesis that And XXII is a bound dSph.  

Unlike the other dSphs in this survey, the small number of identified member stars for this dSph results in a likelihood distribution of our velocity dispersion parameter estimate that is clearly non-Gaussian. Hence, our maximum likelihood technique breaks down, because the assumption of normality near the maximum of the likelihood function is invalid.  Thus, for this galaxy, we estimate $\sigma$ and $\vsys$ using a Markov chain Monte Carlo (MCMC) method with uniform priors on both parameters \citep{brad11mcmc}. We have confirmed that the MCMC reproduces the same results as the maximum likelihood method in the case of Gaussian distribution. We  quote the MCMC results in Table \ref{tab:dsphsumm}.

This dSph is also much closer in projected distance to M33 than it is to M31: $2.9^{\circ}$ and $16.1^{\circ}$ to M33 and M31, respectively.  This leads to the suggestion that And XXII may be a satellite of M33 rather than M31 \citep{martin09}.
On the other hand,  the line-of-sight distance is closer to M31 than M33, resulting in three-dimensional (3D) distances of $130$ and $220$ kpc, respectively.  Adopting an assumption for the total mass of M31 and M33 based on \LCDM{} expectations \citep[][Table 1]{guo10}, the Jacobi tidal radius \citep{bandt} of M33 (assuming it is in a circular orbit around M31) is 67 kpc,  suggesting that And XXII cannot be bound to M33.

However, our measurements show that And XXII is significantly closer in systemic velocity $\vsys \approx -127$ \kps{} to M33 \citep[$-180$ \kps, e.g.][]{CFARS} than to M31 ($\sim -300$ \kps).  Experiments with the Via Lactea 2 (VL2) N-body simulation \citep{VL2} suggest that line-of-sight velocities this close for two random subhalos like those M33 and And XXII might inhabit are very unlikely.  Furthermore, the line-of-sight distance to And XXII  is based on TRGB distance measurements \citep{martin09}.  The And XXII RGB is  sparsely populated, and hence mis-identifying even a few RGB stars near the tip as dSph stars when they are actually M31 halo field stars would bias the distance closer.  There is thus a possibility of substantial changes in the TRGB-determined distance with refined stellar membership \citep[e.g.,][]{Letarte09}, in the sense that corrections are likely to push it closer to M33.  If this is indeed the case, our finding that it is also closer in line-of-sight velocity strongly suggests that And XXII is a satellite of M33 rather than M31.  In this case, if M33 is taken to be a satellite of M31, as is plausible in a \LCDM{} context \citep{toll11lmc}, And XXII may be the first detection of a large mass ratio satellite-of-satellite (or sub-subhalo). 


\clearpage

\section{Scaling Relations of M31 dSphs}
\label{sec:scaling}

With velocity dispersions for most of our sample and structural parameters from a variety of previous studies (see Table \ref{tab:dsphsumm}), we are now in a position to consider where M31 dSphs lie on established galaxy scaling relations.  To this end, we consider the scaling relations between mass, luminosity, and size for these satellites.
  In what follows, we exclude both And XI and And XXI from our sample due to the aforementioned problems estimating their velocity dispersions.

\subsection{Empirical MRL Scalings}
\label{ssec:mrlscale}

More specifically, we examine the MRL parameter space explored by \citet{tollerud11a}.  This parameter space consists of deprojected (3D) half-light radius ($r_{1/2}$), the mass within that radius ($M_{1/2}$), and the half-luminosity ($L_{1/2}$).  $L_{1/2}$ is straightforward as simply half the observed luminosity of the galaxy.  $M_{1/2}$ and $r_{1/2}$  are derived from $\slos$ and $R_{\rm eff}$ following the prescriptions of  \citet{wolf10}: $M_{1/2}=3 \slos^2 r_{1/2} G^{-1}$.  $r_{1/2}$ is a simple scaling of $R_{\rm eff}$, at least for light profiles like those of the dSphs we study here: $r_{1/2} = 1.315 \, \Reff$. $M_{1/2}$ is computed using a standard dynamical mass estimator scaled, crucially, to be the mass within $r_{1/2}$, and is \emph{insensitive} to the velocity anisotropy. This only holds in the case of a relatively flat velocity dispersion profile, but the dSphs in our data set satisfy that requirement within the error bars (aside from possibly And I and And III --- see \Ss \ref{ssec:andi} and \ref{ssec:andiii}).  

These scaling relations thus provide a luminosity and mass, as well as a radius inside which they are applicable, suitable for use both with M31 dSphs from this data set and dSphs from similar MW dSph observations. It is important to note, however, that because $M_{1/2}$ is derived from the velocity dispersion, any systematic errors in the velocity dispersion resulting from our method of assigning membership to stars will present themselves in these scalings as errors  in $M_{1/2}$.  Furthermore, MW dSphs are likely to have different systematic errors in membership determination due to the quite different contaminant populations.  Thus, if these errors are present, they could manifest as systematic shifts in $M_{1/2}$ for  M31 dSphs relative to those of the MW dSphs, even if the populations are intrinsically the same.

\begin{figure*}[htbp!]
 \epsscale{.5}
 \plotone{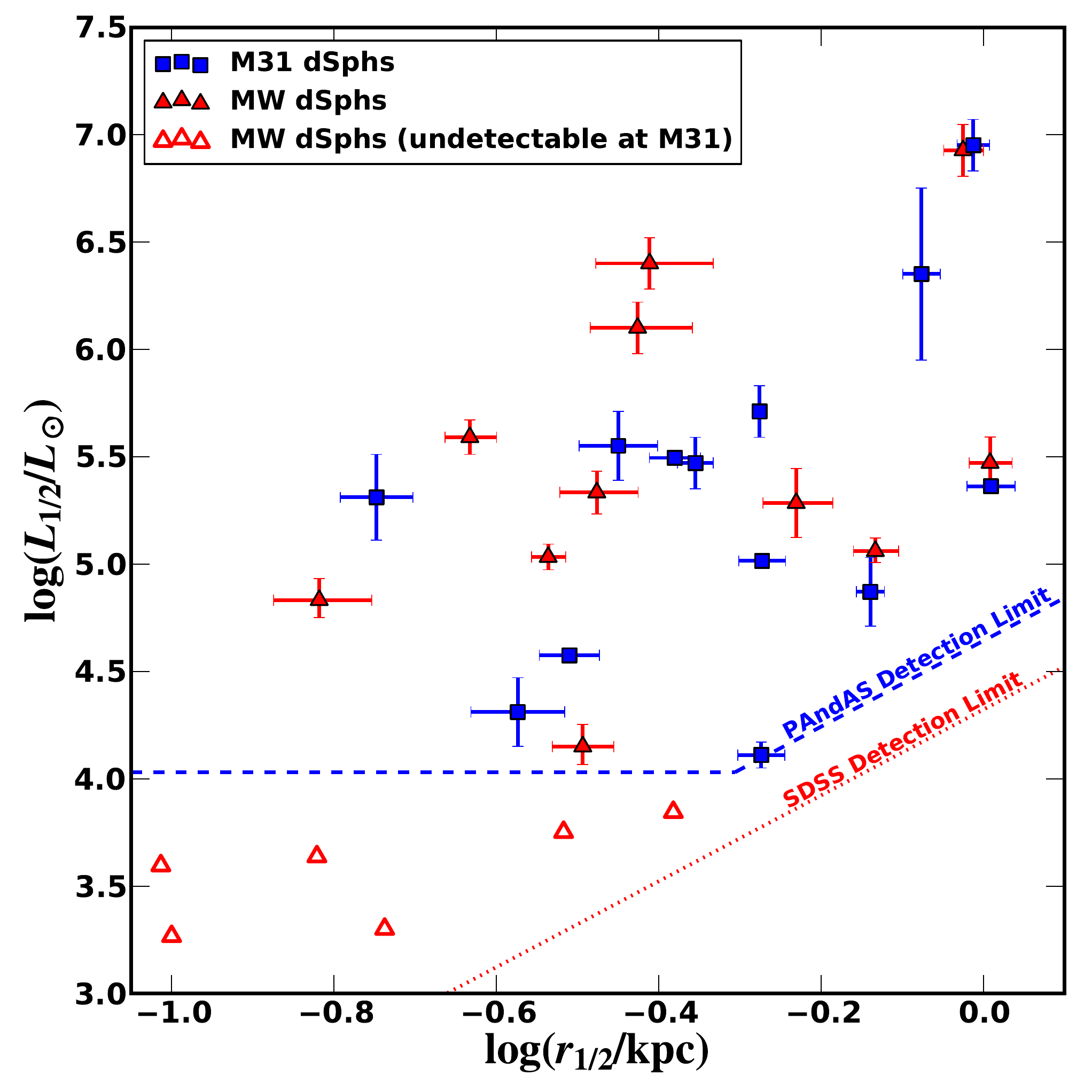}
 
 \epsscale{1}
 \plottwo{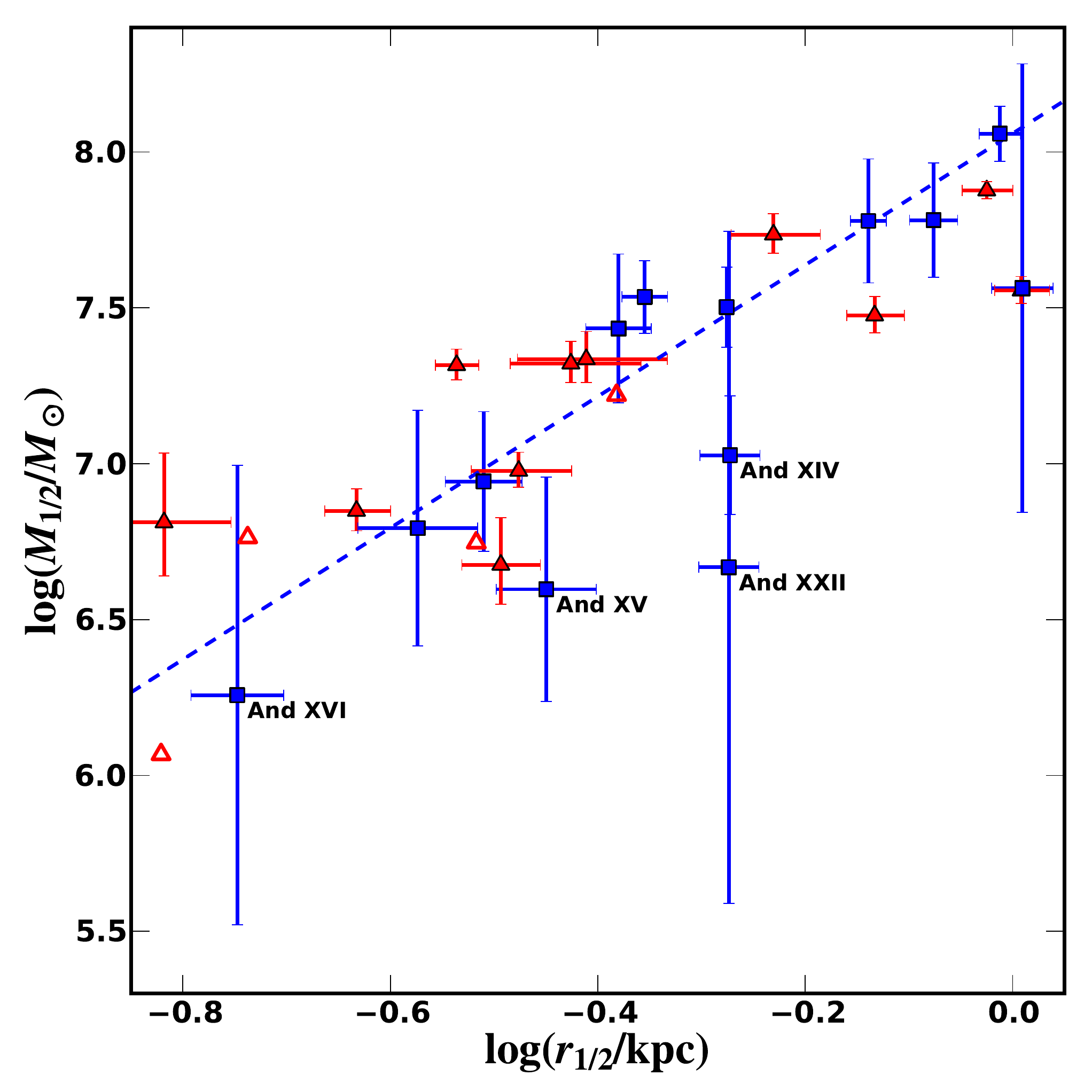}{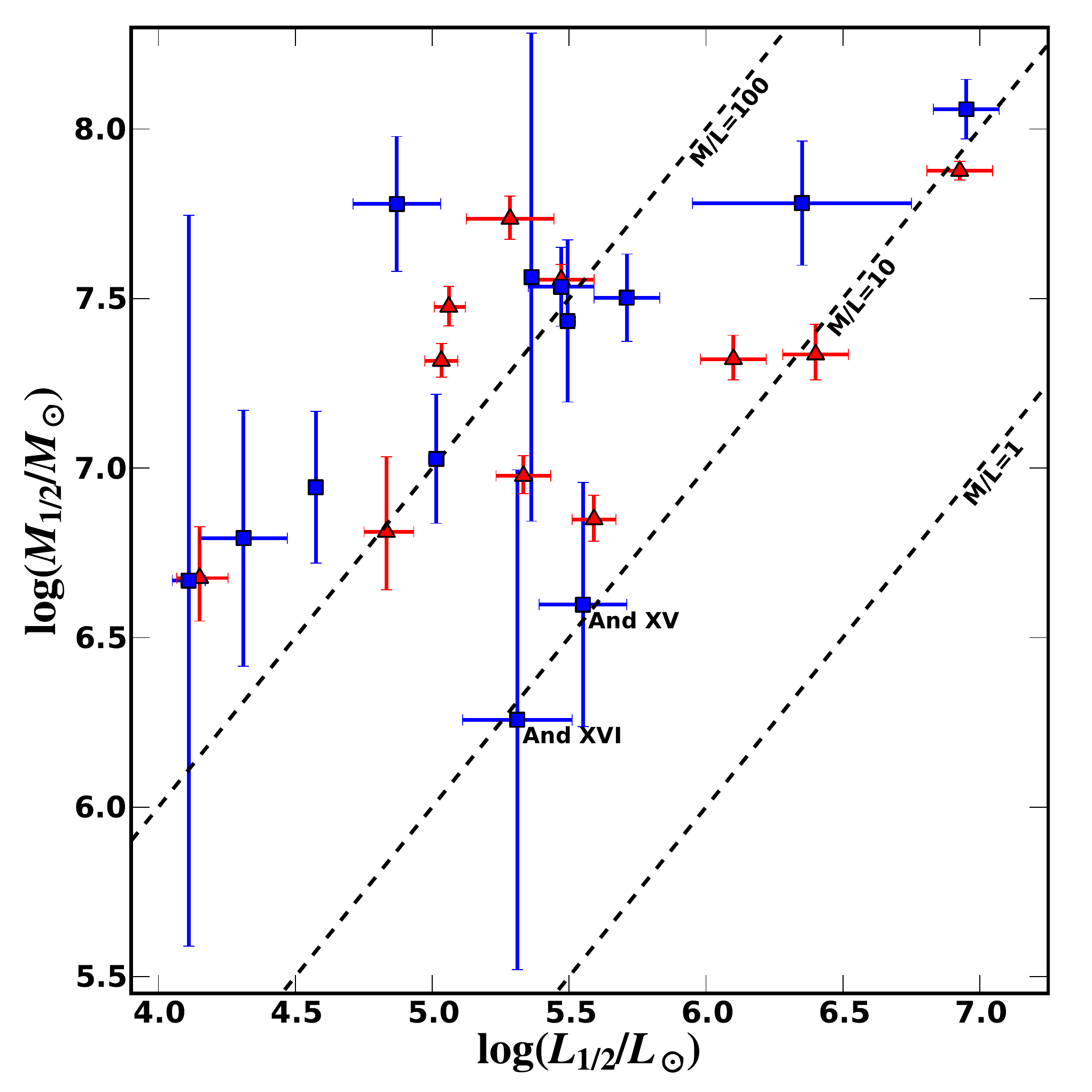}
 \caption{Representation of Local Group dSphs in the MRL space of \citet{tollerud11a}. The upper panel shows the L--R plane (half-luminosity versus deprojected half-light radius).  Points (blue) are M31 dSphs, with associated error bars.    Also shown are detection limits for SDSS searches of the MW \citep{tollerud08} as the red dotted line, and M31 detection limits from the PAndAS survey as reported in \citet{brasseur11}. Note that the PAndAS limits are estimates, rather than true limits from simulations.   Triangles (red) are MW dSphs as tabulated in \citet{wolf10}, where solid triangles are within the M31 detection limit, and unfilled triangles are not.  The same data set is shown in the $R$-$M$ plane (mass within deprojected half-light radius versus deprojected half-light radius)  on lower left.  The dashed (blue) line in this panel indicates the best-fit power law with slope $2.1$.  The lower-right panel shows the $M$-$L$ plane (mass within deprojected half-light radius versus half-luminosity), and the dashed (black) lines in the same panel indicate lines of mass-to-light ratios of 1, 10, and 100 $M_\odot/L_\odot$. }
 \label{fig:mrl}
 \end{figure*}

With this in mind, we present the three projections of the MRL space in Figure \ref{fig:mrl}.  Our M31 dSphs are squares (blue), and we provide the MRL parameters for these satellites in Table \ref{tab:url}.  We also show the MW dSph compilation of \citet{wolf10}, with updates for Bo{\"o}tes I and Segue 1 from \citet{kop11boo} and \citet{simon11seg1}, respectively, as triangles (red). In the $L$-$R$ plane (upper panel), we also plot detection limits for the PAndAS survey \citep[dashed blue,][]{brasseur11} and SDSS \citep[dotted red, ][]{tollerud08}.  Note that the PAndAS detection limits are estimates based on the most marginally detected M31 dSphs, rather than true detection limits from simulations.
Given these limits, a fair comparison requires that MW dSphs that could be detected around M31 be compared to the true M31 satellites.  Thus, we denote MW satellites within the detection limit by filled triangles, and those that could not be detected around M31 as open triangles.

 A careful probabilistic M31/MW comparison for the $L$-$R$ space (upper panel of Figure \ref{fig:mrl}) has already been performed by \citet{brasseur11}, and they conclude that there is no significant evidence that the distributions differ between the MW and M31 dSphs.  In the lower panels of Figure \ref{fig:mrl}, we see similar levels of scatter as in the upper panel, suggesting there is no significant deviation.  We generate 10,000 Monte Carlo samples of the data in each of the three planes, where each resample is generated by assuming Gaussian distributions for both M31 and the MW data points.  We then perform a linear fit, compute the slope and intercept for each resample, and compare the resulting slope/intercept distributions.  For both the $M$-$R$ and $M$-$L$ planes, we find the M31 and MW distributions to be in accord at $<1\sigma$, while for $L$-$R$ there is a slight disagreement at $\sim 1.4\sigma$.  \citet{brasseur11} find closer agreement with a larger sample and a more sophisticated treatment of errors.  In addition, the $M$-$R$ relation shows a detectable slope for the M31 dSphs, and thus we perform an error-weighted fit to a power law (dashed blue line in Figure \ref{fig:mrl}), finding $\log(M_{1/2}) = 2.11 \log(r_{1/2}) + 8.05$.  We conclude that at least our subset of the MW and M31 dSph populations are consistent with lying in the same distribution in major scaling relations.

 Despite overall consistency between the scaling relations of the MW and M31 satellites, there are interesting outliers from the overall trends.  In the $M$-$R$ plane, there are two satellites that show significant deviation from the relation that both MW and M31 satellites seem to follow: And XIV and And XXII.  The error bars for And XXII are large due to the poorly-constrained velocity dispersion, so its interpretation is unclear.  However, And XIV is quite secure and shows a slightly larger $r_{1/2}$ for its mass than the other satellites (or a small mass for its $r_{1/2}$).  And XIV is already unusual in having a large $\vsys$, which has motivated the suggestion that it is only now falling into the M31 system for the first time \citep{maj07and14}.  The fact that it  is distinct from the other satellites in the scaling relations serves to lend further weight to this idea, and supports the notion that environment leaves an observable mark on the structural properties of satellites even in low-density groups like the LG \citep[e.g.,][]{may01stirr,kaz11stirr,toll11lmc}.
 
 In addition, And XV and XVI stand out as outliers from the other satellites in the $M$-$L$ plane, in the sense that they are under-massive (both for luminosity and size). For And XVI, the error bars are very large, admitting a reasonable chance that the satellite lies on the relation with the other satellites, but And XV is more secure.  An offset for And XV is perhaps not surprising however, as it shows hints of tidal interaction, as noted in \S \ref{ssec:andxv}.  This is not entirely satisfying, however, as the presence of tidally stripped or halo contaminant stars would typically \emph{increase} the velocity dispersion.  While its outlier status may be explained by statistical fluctuation, further investigation of this dSph is needed to resolve this oddity.

\subsection{Dark Matter Halo Scalings}
\label{ssec:dmscale}

An additional exercise is suggested by the lower-right panel of Figure \ref{fig:mrl}.  The lowest dashed (black) line indicates the line corresponding to a mass-to-light ratio (with $r_{1/2}$) of 1.  It is clear that the satellites lie far above this, indicating that they are dark matter-dominated galaxies with mass-to-light ratios higher than that expected from any reasonable stellar population.   This warrants considering what dark matter halos they would be expected to inhabit to give the central densities observed here under the assumption that \LCDM{} holds.  

We map galaxies onto their dark matter halos by generating a series of NFW halo profiles \citep{NFW}  and determining the choice of halo that best fits each satellite in the $M$-$R$ plane.  This approach is described in detail by \citet{tollerud11a}, and we only summarize here, highlighting  the differences.  
We take the $M_{1/2}$ and $r_{1/2}$ estimators from the MRL space to deduce the average density within $r_{1/2}$ of each dSph studied here.  Because \LCDM{} dark matter halos are a one-parameter family, we can then map these galaxy scalings onto their dark matter halos by simply matching central densities, and reading off the implied dark matter halo's virial mass.
 
  Because we are examining satellites instead of isolated halos, the appropriate dark matter halos to compare to the satellites are \emph{subhalos} of $M_{\rm vir} \sim 10^{12} M_{\odot}$ hosts.  
For subhalos the concept of  $M_{\rm vir} $ is not always well-defined, as their formal virial radii can reach to radii where the host halo is dominant.  Additionally, tidal stripping alters the shape and total mass of a subhalo, particularly in its outer reaches.  While this stripping does not have as strong of an impact on the central regions where the luminous galaxy sits until the satellite is nearly disrupted \citep{munoz08,penn09}, it does alter the mass of the dark matter subhalo relative to isolated/field halos.  
Thus, we use $\vmax$, the maximum circular velocity of a subhalo (where $V_{\rm circ}(r) = GM(<r)/r$), as a more stable and well-defined parameter for the mass of a subhalo \citep[e.g.,][]{conroy06,moster10,bk10nplh}.   Additionally, to reduce NFW to a one-parameter family of models, we use the $r_{\rm Vmax}$-$\vmax$ relation from the Aquarius project for \emph{subhalos}, rather than a field concentration-mass relation \citep{springel08aqrsubs,neto07}.

It is important to note that this approach depends on the assumption that dissipationless simulations like those of \citet{springel08aqrsubs} are sufficient to explain the central densities of satellite galaxies.   It is possible that baryons can play a role in altering the densities of dSph halos and this would not be included here.  However, as indicated by the lower-right panel of Figure \ref{fig:mrl}, these galaxies have $M/L \lesssim 10$.  For baryon physics to be a major factor, one would then need to evoke a mechanism of baryonic feedback that displaces significantly more dark matter than the mass formed in stars, and it is unclear what mechanism could achieve this \citep[see e.g., the discussion in][]{bkbk12}.   Thus, in this discussion we assume these effects are not significant, as such assumptions have been the backbone of all relevant theory work to date.  This allows for direct interpretation of the data in the context of those models.

\begin{figure}[htbp!]
 \epsscale{1.2}
 \plotone{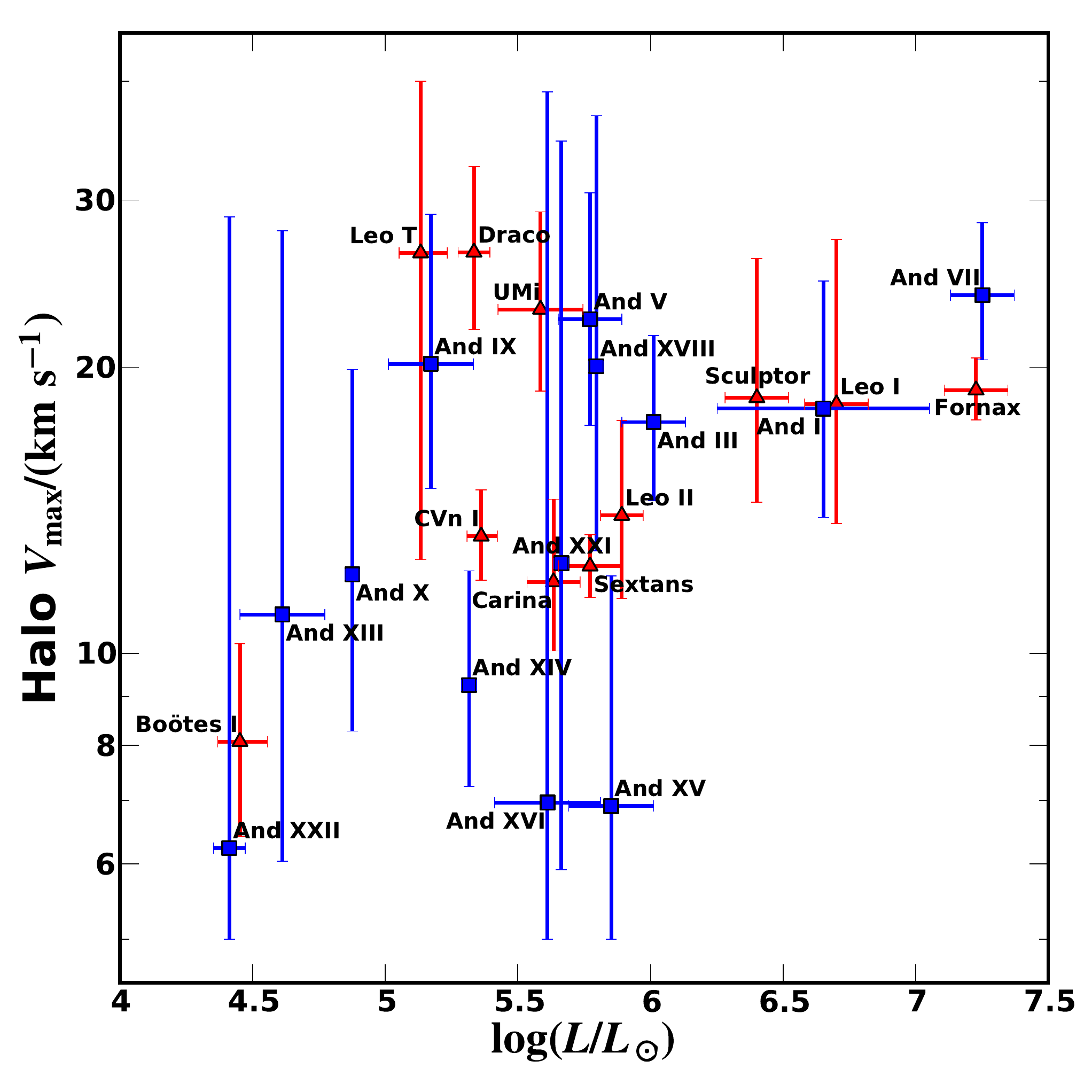}
 \caption{Maximum circular velocity of an NFW halo inferred from dSph scalings as a function of luminosity.  The $\vmax$ values are computed based on the satellite's location in the $R$-$M$ plane as described in the text.  Points (blue) are M31 dSphs, while triangles (red) are MW dSph satellites. }
 \label{fig:vmaxl}
 \end{figure}

With this method used to map from the $R$-$M$ scalings to dark matter halos, we plot the $\vmax$ values for the M31 and MW dSphs in Figure \ref{fig:vmaxl}.  While there is significantly more scatter here than in the MRL relations, this is primarily because the error bars are relatively large, as the $\vmax$ of a subhalo is quite sensitive to the central density around the scales of these dSphs.  We perform the same Monte Carlo simulation method described in \S \ref{ssec:mrlscale}, but for the $\vmax$ versus $L$ relation.  These show that the slope and intercepts for the MW dSphs are consistent with the M31 distribution at the $0.7\sigma$ level.  Hence, there is no definitive sign that the MW and M31 have disjoint dark matter halo scatterings.  Furthermore, the M31 slope distribution is consistent with zero at the $0.8\sigma$ level, providing no clear sign that dark matter halo mass scales appreciably with luminosity at these scales.  \emph{This suggests that the same common mass/halo profile for MW dSphs \citep{stri08commonmass,walker09} holds for M31 dSphs as well.  }

As in the MRL relations, And XIV, And XV, And XVI and And XXII  are  outliers in Figure \ref{fig:vmaxl}. This reveals a possible interpretation of these results: that these dSphs have anamalously small dark matter halos.  In the case of And XV this is not necessarily surprising, as the hints of tidal features suggest it may be heavily stripped. And XXII and And XVI have large error bars that admit masses well above 10 \kps. For And XIV (and the MW dSph Bo\"{o}tes I), the low mass is a puzzle, because it is below 10-15 \kps, the scale below which atomic hydrogen cooling becomes inefficient.  Hence, if And XVI (and possibly XXII, XVI, or Bo\"{o}tes I)  has not been disturbed by interactions, the density implies a dark matter halo that should never have formed galaxies following standard prescriptions of galaxy formation \citep[e.g.,][]{benson02,stringer10,krav10satrev}.    Clearly further investigation is warranted.

 \begin{deluxetable}{cccc}
 \tablecolumns{4}
 \tablecaption{Derived Quantities for M31 dSph.}
 \tablehead{
   \colhead{dSph Name \tablenotemark{1}} &
   \colhead{$\log{(M_{1/2}/M_\odot)}$ \tablenotemark{2}} &
   \colhead{$\log{(r_{1/2}/{\rm kpc})}$ \tablenotemark{3}} &
   \colhead{$\log{(L_{1/2}/L_\odot)}$ \tablenotemark{4}} 
 }
  \startdata

And I & $7.78 \pm 0.18$ & $-0.08 \pm 0.02$ & $6.35 \pm 0.40$ \\
And III & $7.50 \pm 0.13$ & $-0.28 \pm 0.00$ & $5.71 \pm 0.12$ \\
And V & $7.53 \pm 0.12$ & $-0.35 \pm 0.02$ & $5.47 \pm 0.12$ \\
And VII & $8.06 \pm 0.09$ & $-0.01 \pm 0.02$ & $6.95 \pm 0.12$ \\
And IX & $7.78 \pm 0.20$ & $-0.14 \pm 0.02$ & $4.87 \pm 0.16$ \\
And X & $6.94 \pm 0.22$ & $-0.51 \pm 0.04$ & $4.57 \pm 0.03$ \\
And XIII & $6.79 \pm 0.38$ & $-0.57 \pm 0.06$ & $4.31 \pm 0.16$ \\
And XIV & $7.03 \pm 0.19$ & $-0.27 \pm 0.03$ & $5.01 \pm 0.03$ \\
And XV & $6.60 \pm 0.36$ & $-0.45 \pm 0.05$ & $5.55 \pm 0.16$ \\
And XVI & $6.26 \pm 0.74$ & $-0.75 \pm 0.04$ & $5.31 \pm 0.20$ \\
And XVIII & $7.43 \pm 0.24$ & $-0.38 \pm 0.03$ & $5.49 \pm 0.03$ \\
And XXI & $7.56 \pm 0.72$ & $0.01 \pm 0.03$ & $5.36 \pm 0.03$ \\
And XXII & $6.67 \pm 1.08$ & $-0.27 \pm 0.03$ & $4.11 \pm 0.06$ 
  
 \enddata
 
 \tablenotetext{1}{Name of the dSph.}
\tablenotetext{2}{Logarithm of the mass within the 3D half-light radius.}
\tablenotetext{3}{Logarithm of the 3D (i.e., deprojected) half-light radius.}
\tablenotetext{4}{Logarithm of half of the dSphs' total luminosity (i.e., luminosity within the 3D half-light radius).}
 
 \label{tab:dsphmrl}
 \end{deluxetable}
 
\section{M31 Mass Estimate}
\label{sec:M31mass}
In addition to the properties of the dSphs on their own, the kinematics of M31 satellites  as a system can be used to estimate the mass of M31 itself \citep[e.g.,][]{evans00}.  While detailed modeling of such systems can provide in-depth dynamical information \citep[e.g.,]{evans00,klypin02,watkins10}, here we adopt a  simple method, given the relatively small number of tracer particles available here (i.e., the plausible satellites).  

We begin by estimating the distribution of $\vsys$ for the M31 satellites.  We use the $\vsys$ values from Table \ref{tab:dsphsumm}, but with three changes.  First, we do not make use of And XVIII given the fact that its distance may imply it is not a member of the M31 system (see discussion is \S \ref{ssec:and18}).  Second, we add And II, for which we adopt the $\vsys=-193.6 \pm 1.0$ from \citet{kalirai10} and $d_{\rm M31}=185$ kpc from the distance measurement of \citet{mcc06}.  Third, we also add in M31's dE population: NGC 205, NGC 185, NGC 147, and M32.  We obtain $\vsys$ and distances for the first three from \citet{geha06ngc205} and \citet{geha10ngcs}, while for M32 we adopt values from \citet{evans00}.  With those changes, we use the maximum likelihood technique described in \S \ref{ssec:modeling} (Equation \ref{eqn:loglike}) to estimate the parameters of a Gaussian distribution of the dSph's $\vsys$.  This yields a velocity dispersion for the M31 satellite system of $\sigma_{\rm sats}=114 \pm 19$ \kps{}. Furthermore, the resulting mean $v_{\rm dsphs}=-298 \pm 26$ \kps{} of the M31 dSph system \emph{as a whole} is consistent with the mean velocity of M31 and its stellar halo \citep{guhathakurta05,guhathakurta06,chapman06,gilbert07}.

To infer a mass estimate from this distribution, we adopt an empirical approach appropriate for a \LCDM{} context.  We adopt a mass 
estimator proportional to the square of the velocity dispersion , which by dimensional analysis should have the form 
\begin{equation}
\label{eqn:mest}
M_{\rm est}(<r)=c(r) \sigma^2 r/G ,
\end {equation}
\noindent where $\sigma$ is the velocity dispersion, $r$ is a radius, $G$ is the 
gravitational constant, and $c$ is a factor to be empirically determined\footnote{this estimator is inaccurately described in some contexts as a ``virial estimator''; see e.g., \citet{merritt87} Appendix A}.  
To determine this factor, we make use of subhalos with $\vmax>5$ \kps{} in the Via Lactea 2 simulation \citep[VL2,][]{VL2}.  Kinematics of 
subhalos of the VL2 halo ``observed'' from a vantage point as far from the center of the simulated halo as M31 is from the Sun provide a 
plausible sample of proxies for the M31 satellites.  For the radius $r$ within which we measure the mass, we adopt the median (3D) distance from M31 of the satellite sample, \satsrad{} kpc.  The median is used here to reduce the effect of small number statistics, but our final virial mass estimate is not very sensitive to the choice of distance, as the correction factor adjusts for different measurement distances.  We then determine the correction factor as $c(r)= M_{\rm true}(<\satsrad \; {\rm kpc})/M_{\rm est}(c=1)$ for 1000 random orientations of the VL2 halo.  The median and $68 \%$ tails of the $c$ distribution are $c(r) = 1.96^{+0.37}_{-0.27}$.    We note that this coefficient is somewhat different from the analytically-derived  \citet{wolf10} mass estimator of the
  same form.  This difference is a result of the fact that the \citet{wolf10} estimator requires that the full set of tracers be available for computing the half-light radius and velocity dispersion, which is not the case here because the satellite population's completeness is only well-defined to the limits of the PAndAS survey.  
  
With this correction factor in hand, we use Equation \ref{eqn:mest} with this value of $c$ and propagate errors for $c$ and $\sigma$ to provide 
a mass estimate for M31 within \satsrad{} kpc of $\andmass$.
With a mass within a fixed radius, we are in a position to estimate the virial mass of M31's dark matter halo.  We follow the approach of \citet{tollerud11a}, and use a grid of NFW halos following the \emph{field} $c-\mvir$ relation of \citet{klypin10bolshoi}, choosing one based on the $M_{\rm M31}(r<\satsrad \;{\rm kpc})$ determined above.   This results in a virial mass for M31 of $\andmvir$.
This corrected result is comparable to the results of \citet{evans00} and \citet{watkins10}, and thus suggests a moderately lower mass than expected based 
on assuming  abundance matching, i.e., a monotonic halo mass to galaxy luminosity mapping  \citep[$\sim 3 \times 10^{12} M_\odot$ based on][]{guo10}.

\section{Conclusions}
\label{sec:conc}

In this paper, we have described spectroscopy of M31 dSph satellites as part of the SPLASH Survey.  We  filter out MW foreground and M31 halo field contamination to identify M31 dSph member stars, and use these data to determine $\vsys$ and $\slos$ for the satellites.  Based on these kinematics, we determine for each dSph the implied mass within the half light radius, and (under the assumption that these objects are dark matter-dominated) we estimate their dark halo properties.  This paper can be summarized as follows:

\begin{enumerate}

\item We provide a homogenous spectroscopic survey of \ndwarfs{} M31 dSphs and provide radial velocities of resolved stars in these galaxies.

\item We confirm that And XVIII, XXI, and XXII are kinematically cold and hence likely true satellite galaxies.  

\item We find that And XXII has a $\vsys$ close to with M33,  suggesting it is associated with M33 rather than M31. If so, this is likely the first large mass ratio sub-subhalo (or satellite of a satellite) known.

\item We find that the M31 dSphs obey very similar mass-size-luminosity scalings to those of MW satellites.  This suggests that the MW satellite population is not particularly unique and may be typical of a starforming $L_*$ galaxy. 

\item We use the scalings of the M31 dSphs to infer properties of their dark matter halos.  The masses of these halos show no sign of scaling with luminosity, similar to the MW dSphs \citep{stri08commonmass,walker09}.  

\item The density of And XIV, as well as perhaps And XV and And XVI, is consistent with dark matter halos with $\vmax<10$ \kps{} (although consistent with higher masses at $\sim 1\sigma$).  If the most-likely masses for these systems are correct, these (along with the MW satellite Bo\"{o}tes I) are the lowest-mass dark matter halos hosting stars, with potential well depths indicative of field halos that are below the atomic hydrogen cooling limit.

\item Using the systemic velocities of M31 dSphs as tracer particles and adopting an empirical mass estimator suggested by n-body 
simulations, we estimate the mass of M31 within \satsrad{} kpc to be $\andmass$.
This corresponds to a virial mass for M31's dark matter halo of $\andmvir$.

\end{enumerate}

This analysis of M31 dSphs thus represent a major step forward in understanding the faintest known class of galaxies. The M31 satellites present
an opportunity to understand these galaxies in a new way, as a system, along with their host halo, providing
 a rich  set of opportunities for examining  galaxy formation and \LCDM.  Their similarity to MW dSphs also confirm of the Copernican
 principle, affirming that the MW may be a typical galaxy with typical satellites, albeit in an extraordinary universe.

\acknowledgements{We wish to acknowledge Nhung Ho, Greg Martinez, and Ricardo Munoz for helpful discussions, as well as Stacy McGaugh, Mark Fardal, Alan McConnachie, and the anonymous referee for helpful suggestions regarding the manuscript.  

EJT acknowledges support from a Graduate Assistance in Areas of National Need (GAANN) Fellowship and a Fletcher Jones Fellowship. RLB acknowledges receipt of the the Mark C. Pirrung Family Graduate Fellowship from the Jefferson Scholars Foundation and a Fellowship Enhancement for Outstanding Doctoral Candidates from the Office of the Vice President of Research at the University of Virginia. MG acknowledges support from NSF grant
AST-0908752 and the Alfred P.~Sloan Foundation. 
PG, JSB, and  SRM acknowledge support from collaborative NSF grants AST-1010039, AST-1009973, AST-1009882, and AST-0607726.
AC thanks the UC Santa Cruz Science Internship Program for support. Additional support for this work was provided by NASA through Hubble Fellowship grants 51256.01 and 51273.01 awarded to ENK and KMG by the Space Telescope Science Institute, which is operated by the Association of Universities for Research in Astronomy, Inc., for NASA, under contract NAS 5-26555.

The spec2d pipeline used to reduce the DEIMOS data was developed at UC Berkeley with support from NSF grant AST-0071048. 

This work made extensive use of code developed for the Astropysics\footnote{http://packages.python.org/Astropysics/} and Pymodelfit\footnote{http://packages.python.org/PyModelFit/} \citep{pymfascl} open-source projects.

The authors wish to recognize and acknowledge the very significant cultural role and reverence that the summit of Mauna Kea has always had within the indigenous Hawaiian community.  We are most fortunate to have the opportunity to conduct observations from this mountain.

Some slitmasks were designed based on data acquired using the Large Binocular Telescope (LBT). The LBT is an international collaboration among institutions in the United States, Italy and Germany. LBT Corporation partners are: The University of Arizona on behalf of the Arizona university system; Istituto Nazionale di Astrofisica, Italy; LBT Beteiligungsgesellschaft, Germany, representing the Max-Planck Society, the Astrophysical Institute Potsdam, and Heidelberg University; The Ohio State University, and The Research Corporation, on behalf of The University of Notre Dame, University of Minnesota and University of Virginia.

This research used the facilities of the Canadian Astronomy Data Centre operated by the 
National Research Council of Canada with the support of the Canadian Space Agency.}

{\it Facilities:}  \facility{Keck:II (DEIMOS)},  \facility{Mayall (Mosaic)}, \facility{LBT (LBC)}

\bibliography{../splashdwarfs}{}
\bibliographystyle{hapj}

\end{document}